\documentclass{elsart}
\usepackage{epsfig}
\usepackage{graphicx}
\usepackage{amsmath}
\usepackage{amssymb}
\usepackage{amsfonts}
\include{Acknowledge}

\begin{document}

\begin{frontmatter}
\title{Measurement of the background in the NEMO~3 double beta decay experiment}
\author[LAL]{J.~Argyriades},
\author[IPHC]{R.~Arnold},
\author[LAL]{C.~Augier},
\author[INEL]{J.~Baker},
\author[ITEP]{A.S.~Barabash},
\author[LAL]{M.~Bongrand},
\author[LAL]{G.~Broudin-Bay},
\author[JINR]{V.B.~Brudanin},
\author[INEL]{A.J.~Caffrey},
\author[LPC]{A.~Chapon}, 
\author[UB,CENBG]{E.~Chauveau}, 
\author[UCL]{Z.~Daraktchieva},
\author[LPC]{D.~Durand}, 
\author[JINR]{V.G.~Egorov}, 
\author[UM]{N.~Fatemi-Ghomi},
\author[UCL]{R.~Flack},
\author[UCL]{A.~Freshville},
\author[LPC]{B.~Guillon}, 
\author[UB,CENBG]{Ph.~Hubert}, 
\author[LAL]{S.~Jullian},
\author[UCL]{M.~Kauer},
\author[UCL]{S.~King},
\author[JINR]{O.I.~Kochetov},
\author[ITEP]{S.I.~Konovalov},
\author[IPHC,JINR]{V.E.~Kovalenko},
\author[LAL]{D.~Lalanne},
\author[UT]{K.~Lang},
\author[LPC]{Y.~Lemi\`{e}re},
\author[UB,CENBG]{G.~Lutter},
\author[CTU]{F.~Mamedov},
\author[UB,CENBG]{Ch.~Marquet},
\author[IFIC]{J.~Martin-Albo},
\author[LPC]{F.~Mauger},
\author[UB,CENBG]{A.~Nachab},
\author[UM]{I.~Nasteva},
\author[JINR]{I.B.~Nemchenok},
\author[Barc]{F.~Nova},
\author[IFIC]{P.~Novella},
\author[Saga]{H.~Ohsumi},
\author[UT]{R.B.~Pahlka},
\author[UB,CENBG]{F.~Perrot},
\author[UB,CENBG]{F.~Piquemal},
\author[LSCE]{J.L.~Reyss},
\author[UB,CENBG]{J.S.~Ricol},
\author[UCL]{R.~Saakyan},
\author[LAL]{X.~Sarazin},
\author[LAL]{L.~Simard},
\author[JINR]{Yu.A.~Shitov},
\author[JINR]{A.A.~Smolnikov},
\author[UM]{S.~Snow},
\author[UM]{S.~S\"{o}ldner-Rembold},
\author[CTU]{I.~\v{S}tekl},
\author[MHC]{C.S.~Sutton},
\author[LAL]{G.~Szklarz},
\author[UCL]{J.~Thomas},
\author[JINR]{V.V.~Timkin},
\author[IPHC,JINR]{V.I.~Tretyak\corauthref{CA}} \ead{tretyak@jinr.ru},
\author[KINR]{Vl.I.~Tretyak},
\author[ITEP]{V.I.~Umatov},
\author[CTU]{L.~V\'{a}la},
\author[ITEP]{I.A.~Vanyushin},
\author[UCL]{V.A.~Vasiliev},
\author[Charles]{V.~Vorobel},
\author[JINR]{Ts.~Vylov}
\collab{NEMO Collaboration}

\address [LAL] {LAL, Universit\'e Paris-Sud, CNRS/IN2P3, 
  F-91405 Orsay, France}
\address [IPHC] {IPHC, Universit\'e de Strasbourg, CNRS/IN2P3,
  F-67037 Strasbourg, France}
\address [INEL] {INL, Idaho Falls, ID83415, USA}
\address [ITEP] {Institute of Theoretical and Experimental Physics, 
  117259 Moscow, Russia}
\address [UB]{Universit\'e de Bordeaux, 
Centre d'Etudes Nucl\'eaires de Bordeaux Gradignan, UMR 5797, 
 F-33175 Gradignan, France}
\address [CENBG]{CNRS/IN2P3,  
Centre d'Etudes Nucl\'eaires de Bordeaux Gradignan, UMR 5797, 
 F-33175 Gradignan, France}
\address [JINR] {Joint Institute for Nuclear Research, 
  141980 Dubna, Russia}
\address [UM] {University of Manchester, M13 9PL Manchester,
7  United Kingdom}
\address [CTU] {IEAP, Czech Technical University in Prague, 
  CZ-12800 Prague, Czech Republic}
\address [UCL] {University College London, 
  WC1E 6BT London, United Kingdom}
\address [UT] {University of Texas at Austin, Austin, Texas 78712-0264, USA}
\address [LPC] {LPC Caen, ENSICAEN, Universit\'e de Caen, CNRS/IN2P3, F-14032 Caen, France}
\address [Barc] {Universitat Aut\`onoma de Barcelona, Spain}
\address [IFIC] {IFIC, CSIS - Universidad de Valencia, Valencia, Spain}
\address [Saga] {Saga University, Saga 840-8502, Japan}
\address [LSCE] {LSCE, CNRS, F-91190 Gif-sur-Yvette, France}
\address [MHC] {MHC, South Hadley, Massachusetts, MA01075, USA}
\address [Charles] {Charles University in Prague, Faculty of Mathematics and 
Physics, CZ-12116 Prague, Czech Republic}
\address [KINR] {INR, MSP 03680 Kyiv, Ukraine}
\corauth [CA] {Corresponding author.}

\begin{abstract}
In the double beta decay experiment NEMO~3 
a precise knowledge of the background in the
signal region is of outstanding importance.
This article presents the methods used in NEMO~3 to evaluate the backgrounds 
resulting from most if not all possible origins. 
It also illustrates the power of the combined
tracking-calorimetry technique used in the experiment.
\end{abstract}

\end{frontmatter}
\newpage

\section{Introduction}
NEMO~3 is a currently running experiment located in the Laboratoire
Souterrain de Modane (LSM) searching for neutrinoless double beta decay 
($\beta\beta0\nu$). 
This decay is a uniquely sensitive probe of
the mass and charge conjugation properties of the neutrino. Should it be
observed, it would demonstrate that at least one neutrino is a massive
Majorana particle.
NEMO~3 is also able to detect the rare second order weak double beta
decay with its two accompanying neutrinos ($\beta\beta2\nu$), as well as the 
non-standard neutrinoless decay with Majoron emission ($\beta\beta\chi$). The three
decay modes are distinguishable experimentally by the energy sum
distribution of the two beta particles. 
Although the observation of neutrinoless double beta decay is the goal of NEMO~3, its aim is
also to measure (or give limits on) half-lives of the other double beta 
decay processes. 

\begin{figure}
\centerline{\parbox[t]{14.0cm}{\fbox{\epsfxsize14.cm\epsffile{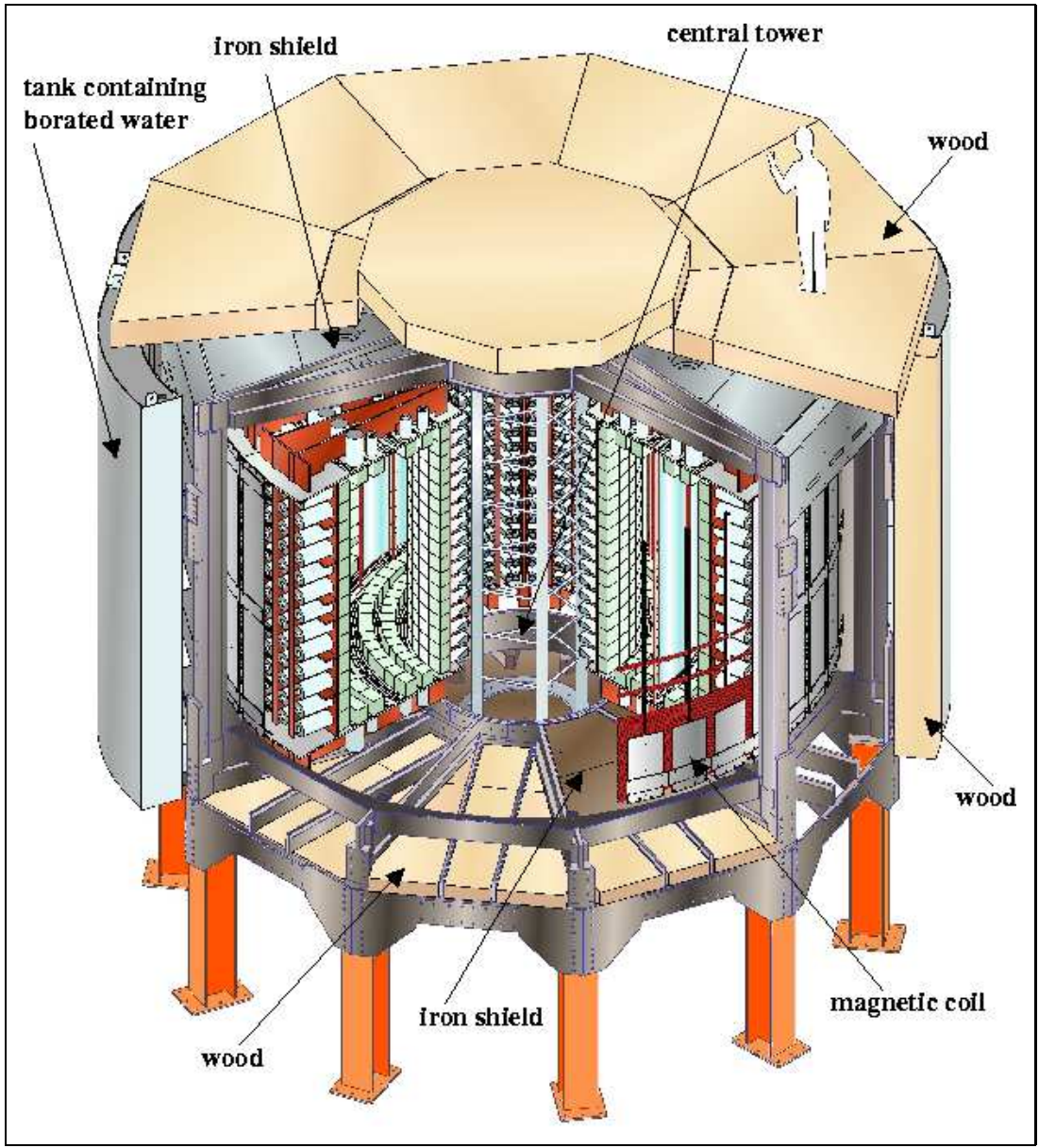}}}}
\caption{Schematical view of the NEMO~3 detector.}
\label{fig:NEMO3}
\end{figure}

The NEMO~3 detector provides the direct detection of two electrons
from the decay by the use of a tracking device and a calorimeter 
(see Fig.~\ref{fig:NEMO3}). It has three integrated components: 
\begin {description}
  \setlength{\itemsep}{0cm}%
 \setlength{\parskip}{0cm}%
 \item {-}
a foil consisting of different sources of double beta
emitters \{$^{100}$Mo (6914~g), $^{82}$Se (932~g), $^{116}$Cd (405~g), $^{130}$Te (454~g), 
natural Te (614~g of TeO$_2$), $^{150}$Nd (37~g), $^{96}$Zr (9~g), $^{48}$Ca (7~g)\} and pure 
copper (621~g); 
 \item {-}
a tracking volume based on open Geiger cells; 
 \item {-}
a calorimeter made of plastic scintillator blocks with photomultiplier 
tubes. 
\end {description}
A magnetic field created by a solenoidal coil surrounding the detector
provides  
identification of electrons by the curvature of their tracks. 
Besides the electron  and photon identification, the calorimeter measures the energy and
the arrival time of these particles while the tracking chamber can measure 
the time of delayed tracks associated with the initial event 
for up to 700~$\mu$s. A full description of the NEMO~3 detector and its 
performance can be found in ~\cite{ARN05}. The first results were 
published in ~\cite{NEMO0nu}-\cite{NEMOexc}. 

The most significant concern in double beta decay experiments is the
background. Although the candidates for double beta decay are
selected as events with two electron  (2$e^-$) tracks with a common
origin on the source foil, there are certain non $\beta\beta$
processes that can  mimic the 2$e^-$ topology.
According to its origin, the background in NEMO~3 is divided into two
categories:
\begin{description}
\item{-} the "internal background" having its origin inside the double beta
decay source foil;
\item{-} the "external background" coming from all radioactive sources
  located outside of the foil.
\end{description}
  
\begin{figure}[htb]
\begin{center}
\includegraphics[width=0.85\textwidth]{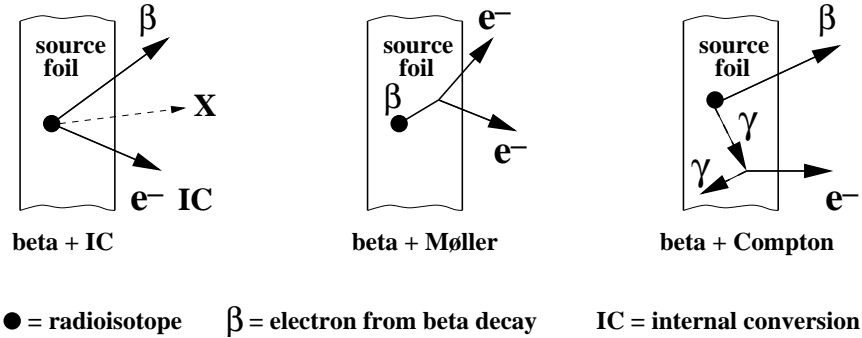}
\hfill
\caption{Internal background production in the source foil.}
\label{fig:bdf_int}
\end{center}
\end{figure}

The "internal background" is mainly due to the presence of radioactive
isotopes from the $^{238}$U and $^{232}$Th decay chains. The dominant mechanisms
leading to the 2$e^-$ topology are (see Fig.~\ref{fig:bdf_int}):
\begin{description}
\item{-} $\beta$ decay accompanied by an internal conversion;
\item{-} $\beta$ decay followed by M{\o}ller scattering;
\item{-} $\beta$-$\gamma$ cascades in which a $\gamma$-ray undergoes
  Compton scattering.   
\end{description}
The first mechanism is the dominant one. Particularly
troublesome are the isotopes with large $Q_\beta$ values such as $^{208}$Tl
($Q_\beta$=4.99~MeV) and $^{ 214}$Bi ($Q_\beta$=3.27~MeV). Fortunately, 
for these  
isotopes there are good estimates of the activities obtained from 
identifiable topologies ($e\alpha$ for $^{ 214}$Bi, $e\gamma\gamma$ 
and $e\gamma\gamma\gamma$ for $^{208}$Tl).

Great care was taken in the production and subsequent purification of 
the enriched materials, as well as during the source foil production
and the mounting of foils
in the detector so as to keep any contamination to a minimum given the
strict radioactivity limits.

In the region where a signal of neutrinoless $\beta\beta$ decay is
expected, the allowed $\beta\beta2\nu$ decay can be an important
fraction of the background. 
Its contribution depends upon the $\beta\beta2\nu$ half-life and the energy
resolution of the detector. It is therefore important to carefully study 
the background in a large energy region where the $\beta\beta2\nu$
decay takes place in order to obtain a good measurement of the
half-life of this process.

A component of the external background producing events similar to the
internal background is caused by the presence of radon and thoron
inside the detector. These elements are 
highly diffusive radioactive gases. They are outgased in the air from
the rock walls of the experimental hall and can enter the detector
either through tiny gaps 
between sectors or through gas pipe joints.
The progeny of radon and thoron produces $\gamma$-rays and $\beta$-decays 
accompanied by internal conversion (IC), 
M{\o}ller or Compton scattering. If such an event occurs on or near a
foil and appears with a 2$e^-$ topology it becomes indistinguishable from a
double beta decay candidate.
\begin{figure}[htb]
\begin{center}
\includegraphics[width=0.90\textwidth]{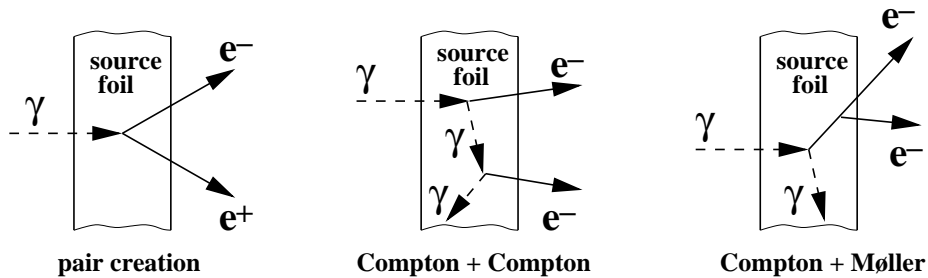}
\hfill
\caption{External background production in the source foil.}
\label{fig:bdf_ext}
\end{center}
\end{figure}

Another component of this background is due to external $\gamma$-rays
interacting inside the foil. These $\gamma$-rays are of different origins:
\begin{description}
\item{-} $\gamma$-rays inside the laboratory, mostly coming from the rock
  walls;
\item{-} neutron interactions in the shield and material of the detector;
\item{-} radioactive isotopes present in the detector materials despite the
  rigorous selection done during the detector construction;
\item{-} presence of radon in the air surrounding the detector.
\end{description}
The interaction of $\gamma$-rays in the foil can appear like 2$e^-$ events by
$e^+e^-$ pair creation with misidentification of the charge, double Compton
scattering or Compton scattering followed by M{\o}ller scattering (see 
Fig.~\ref{fig:bdf_ext}).

In this article the methods used to evaluate
the various backgrounds are presented. The very pure copper 
foils (OFHC) are used to prove their validity. 
The experimental data of the NEMO~3 detector were used 
to perform the background measurements.

\section{Natural radioactivity inside the tracking volume}
\subsection{$^{222}$Rn measurement inside the tracking chamber}
\label{sect:Radon}
The most bothersome external background comes from radon.
\begin{figure}
\centerline{\parbox[t]{14.0cm}{\fbox{\epsfxsize14.cm\epsffile{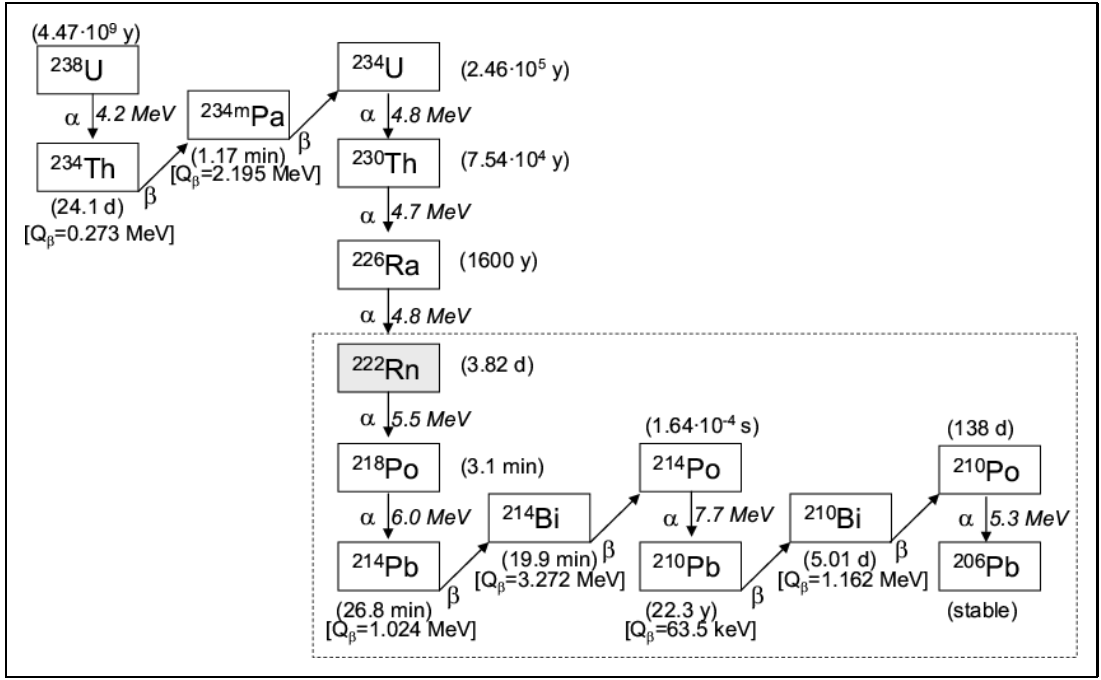}}}}
\caption{Decay chain of the radioactive family of
  $^{238}$U. The half-lives   
  and decay energies are taken from~\cite{TOI}.}
\label{fig:u238scheme}
\end{figure}
As shown in Fig.~\ref{fig:u238scheme}, $^{214}$Bi is one of the
descendents of $^{222}$Rn. 
The  $\beta^-$decay of \nuc{214}{Bi} to $^{214}$Po is generally
accompanied by several photons which can mimic a
$\beta\beta0\nu$ event given its large $Q_\beta$ value. The 
$^{214}$Po has a half-life of 164~$\mu$s and it
disintegrates to $^{210}$Pb via $\alpha$-decay. 

The ejected alpha particle from the decay of $^{222}$Rn can free
several electrons from the $^{218}$Po atom transforming it to a positively
charged ion. Diffusing through the gas of the
tracking chamber this ion may be
neutralized by  different processes, such as recombining with negative
ions in the gas, a charge transfer by neutral molecules with a small
ionization potential or the capture of an electron created during
the gas discharge near the open Geiger cell wires~\cite{Frey1981,Busigin1981}. It is
difficult to predict the proportions of neutral and charged atoms of
$^{218}$Po in the gas of the NEMO~3 tracking chamber. However,
following some earlier
studies~\cite{Wellisch1913,Renoux1965,Porstendorfer1979} one can 
suppose that the proportion of  
neutral atoms of $^{218}$Po which are in the gas 
is small and that the majority of the charged $^{218}$Po is
deposited on the surfaces of the  
cathode wires of the Geiger cells.
Here the measurement of the radon activity in the tracking chamber 
is done under the assumption that the descendents of $^{222}$Rn are deposited
on the wires. 
It should be noted that the final background resulting from the presence of 
$^{222}$Rn does not depend critically upon this assumption. If one supposes that 
the descendants of this gas are uniformly distributed in the tracking chamber 
the background estimation remains unchanched within the errors.
A possible deposition
on the foil surfaces is also taken into account in the analysis performed
to measure internal foil contamination by $^{214}$Bi (see section~\ref{sect:inbg}).

\subsubsection{Event selection}
\label{sect:rnmodel}

\begin{figure}
\parbox[t]{13.0cm}{\fbox{\epsfxsize6.cm\epsffile{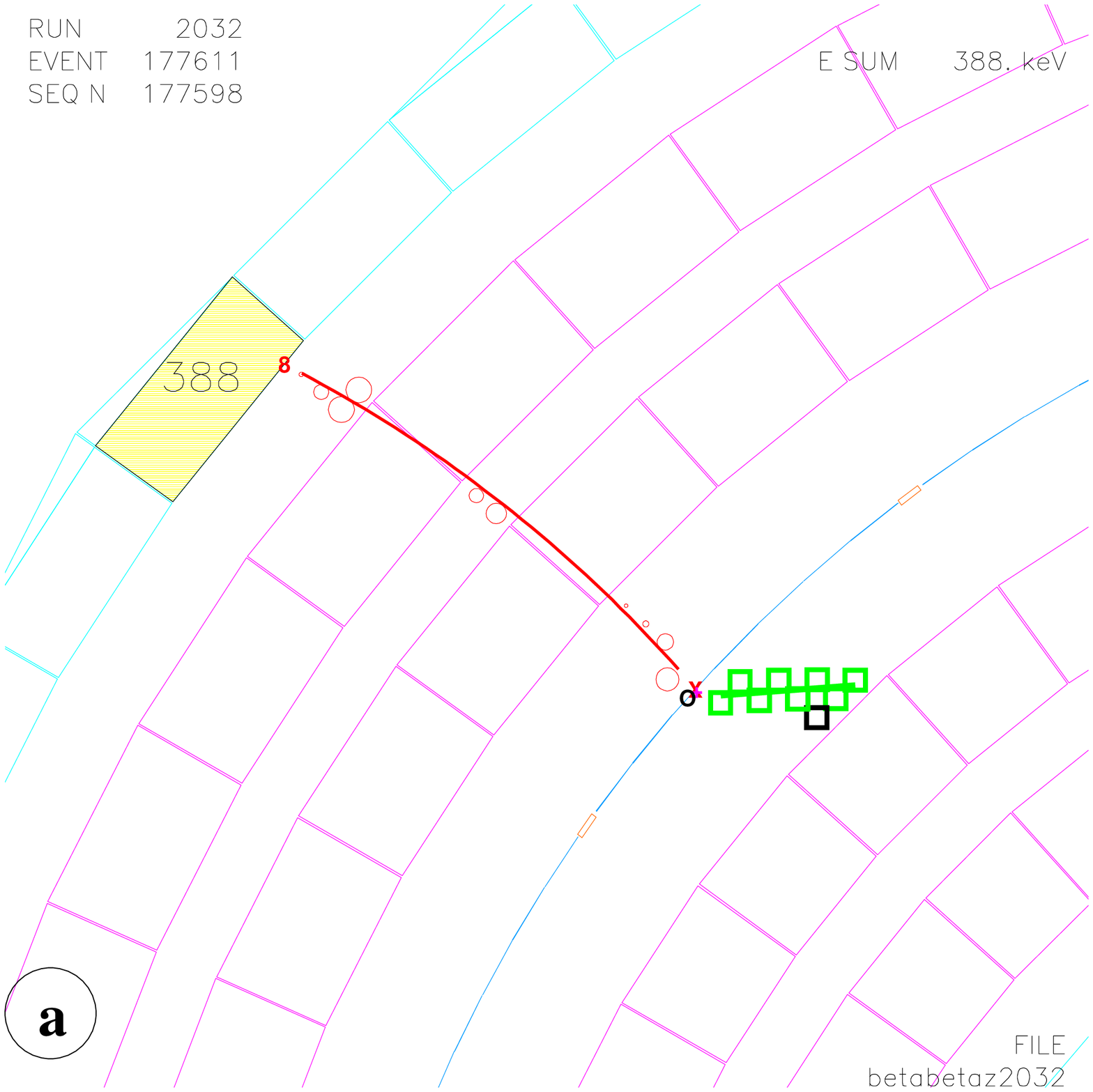}}
\hfil\fbox{\epsfxsize6.cm\epsffile{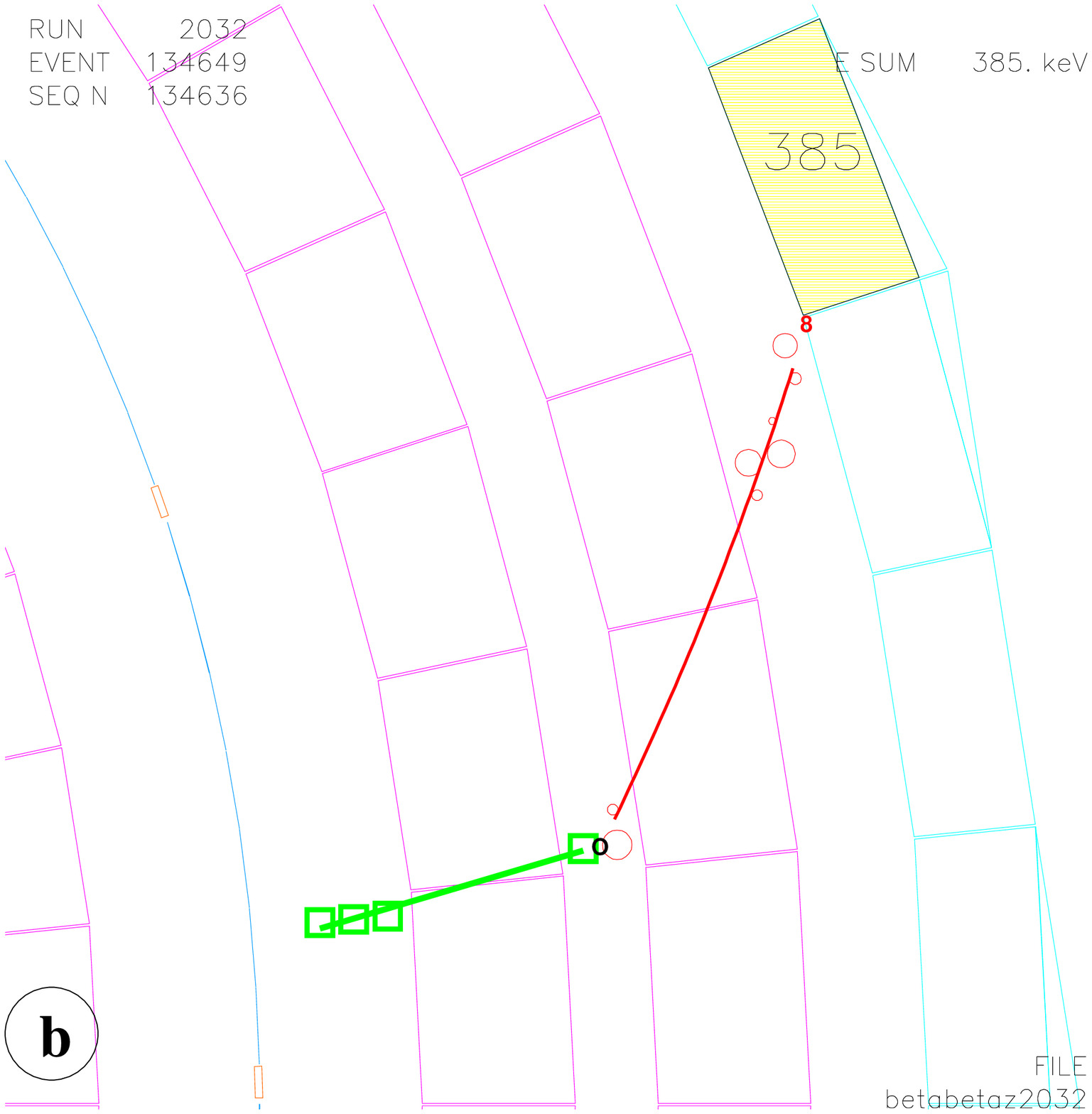}}}
\caption{Examples of BiPo event candidates, viewed from the top.
In each event the prompt track is shown in red (electron), delayed track -
in green ($\alpha$-particle). The two tracks have a common vertex: a) on the source foil, 
b) inside the tracking volume. 
Scintillator blocks located at top and bottom (wall) are shown as magenta (blue) boxes.}
\label{fig:ev_examples}
\end{figure}

In the NEMO~3 experiment, the data acquisition allows one to readout
signals from delayed tracks 
in order to tag the $\alpha$ particles from $^{214}$Po decays associated with
electrons from  $^{214}$Bi decays. 
The $^{214}$Bi decays followed by $^{214}$Po decays
(hereafter referred to as BiPo events)
are used for the measurement of the radon activity.
Typical examples of such events with detected electron and $\alpha$-particle tracks
are shown in Fig.~\ref{fig:ev_examples}.
Two types of spurious events can appear in a sample of BiPo candidates.
They are due to:	
\begin{itemize}
\item[--] random coincidence of two independent events closely 
  localized in space and occurring inside the 700~$\mu$s time window.
  The delay time distribution is flat in this case.
\item[--] a single event accompanied by one or more delayed signals caused by 
  refiring of neighbouring Geiger cells.
  The number of this kind of event decreases sharply with the delay time, 
  so their distribution is nearly exponential.   
\end{itemize}

In order to reduce the contribution from refirings, the
following cuts are applied:
\begin{itemize}
\item[-] for events with only one delayed signal, the delay
  must be greater than 90~$\mu$s;
\item[-] for events with more than one delayed signal, the
  delay must be greater than 30~$\mu$s.
\end{itemize}
It is required that the delayed signals 
have to be located close to the ``prompt'' (i.e. electron) track
or to the event vertex on the source foil.
The  group of selected delayed signals must be  within
2.1~$\mu$s (corresponding to the maximum transversal drift time in the 
Geiger cells) and follow a straight line in space
in order to exclude random
coincidences  of delayed 
events of different origins.
In the tracking chamber the maximum range of the 
7.7~MeV $\alpha$-particle emitted by $^{214}$Po is 36 cm.
This value is also taken into account in the selection of events.

Applying these criteria 
the mean efficiency to select a BiPo event produced on a wire surface 
has been estimated by a Monte Carlo (MC) simulation to be 16.5\%.
In this work all simulations~\cite{nemos} are based on 
GEANT3~\cite{GEANT3}  using DECAY0~\cite{genbb} as event generator.

\begin{figure}
\centerline{
\parbox[t]{14.cm}{\epsfxsize6.5cm\epsffile{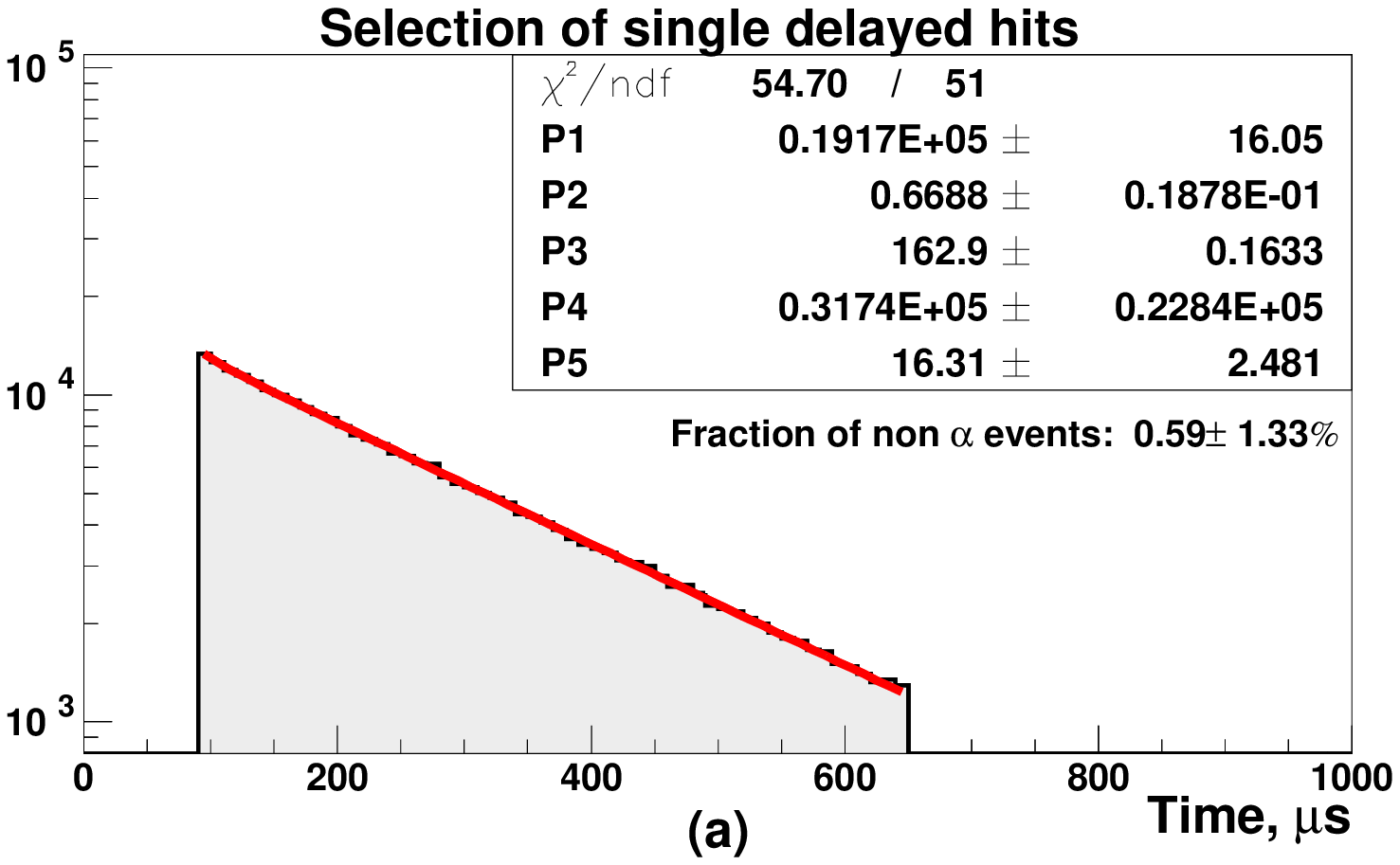}
\hfil{\epsfxsize6.5cm\epsffile{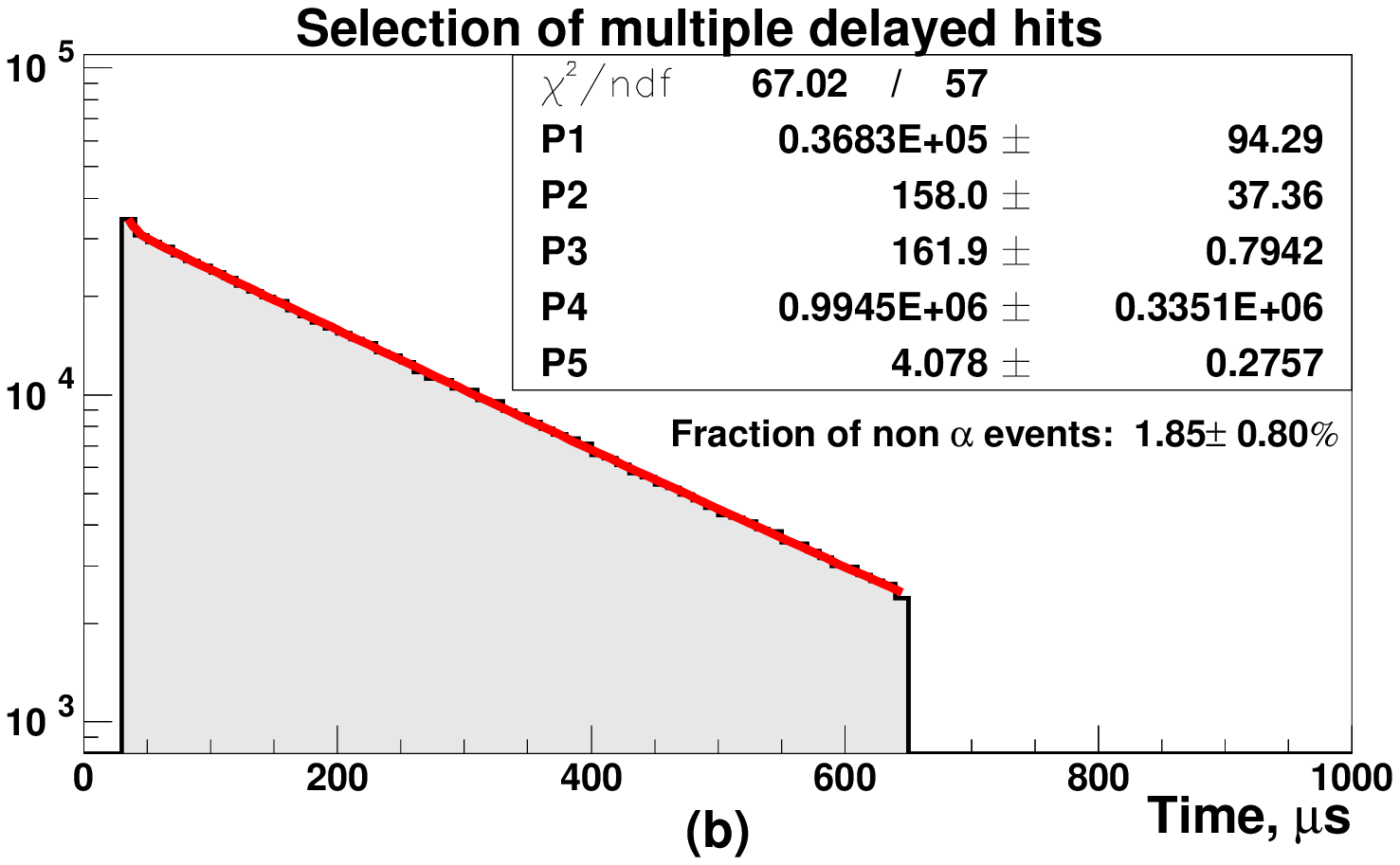}}}
}
\caption{The time distributions of events selected with (a) single and
  (b) multiple delayed signals.  
  Each distribution was fit to the function: $f(time) =
  P1\cdot e^{-\frac{time}{P3/\ln2}} +
  P2 + P4\cdot e^{-\frac{time}{P5/\ln2}}$ where  P1 and P4 are scaling
  constants,  P2 is the amplitude of random coincidences,
  P3 the $^{214}$Po half-life in $\mu$s  and P5 the time constant
  of the refirings.}
\label{fig:time_alpha}
\end{figure}

The time distribution of the delayed tracks (see Fig.~\ref{fig:time_alpha}) 
provides an efficient way to validate the quality of the event selection. 
The fit to the distributions  allows the half-life of $^{214}$Po 
to be evaluated. The proportions of events due to refirings and random 
coincidences are also determined from the fit.
They are 
found to be negligibly small with  
0.6$\pm$1.3\% for the single delayed signal and 1.9$\pm$0.8\% 
for the multiple delayed signals as shown in
Fig.~\ref{fig:time_alpha}. In spite of the systematic
uncertainty of $\sim$1~$\mu$s on the result, the
half-life of $^{214}$Po for the single delayed signal events is 
$T_{1/2}$($^{214}$Po) = 162.9$\pm$0.2(stat.only)~$\mu$s and for the
multiple delayed 
signal events is $T_{1/2}$($^{214}$Po) = 161.9$\pm$0.8(stat.only)~$\mu$s,
a comparison  with the table value of $T_{1/2}$($^{214}$Po) =
164.3$\pm$2.0~$\mu$s~\cite{Akovali1995}  confirms
that the delayed tracks are due to the $\alpha$-particles of $^{214}$Po. 

\subsubsection{$^{222}$Rn monitoring}

\begin{figure}[htb]
\centerline{
\parbox[t]{10.5cm}{\epsfxsize10.5cm\epsffile{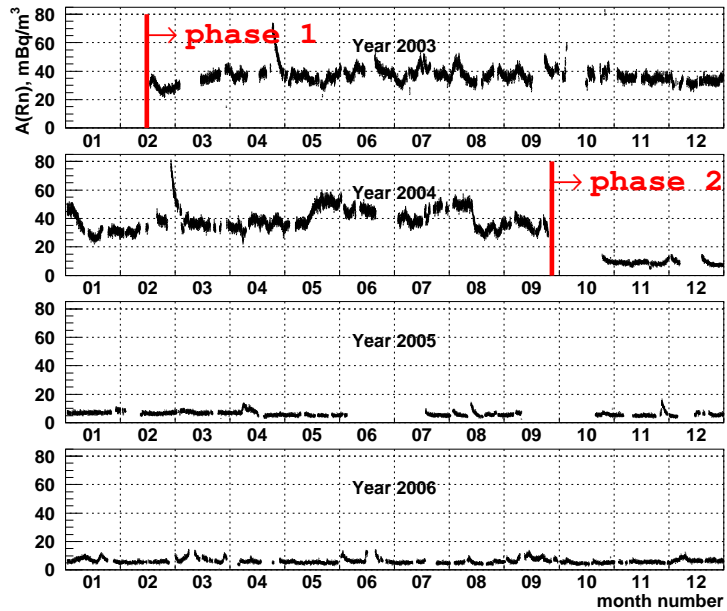}}
}
\caption{The $^{222}$Rn activity in mBq/m$^3$ inside the tracking
  chamber measured on  an hourly basis.}
\label{fig:rnmonitoring}
\end{figure}

Using the method described in section~\ref{sect:rnmodel}
the mean $^{222}$Rn level in the tracking volume was analysed.
The results of the radon monitoring are shown in
Fig.~\ref{fig:rnmonitoring}. 

In order to decrease the level of $^{222}$Rn, a radon reduction
factory has been installed in the underground laboratory to inject
nearly radon free air into a tent built around the detector. It became
operational at the beginning of October 2004. Therefore the NEMO~3
data has been 
divided into two parts according to the collection dates. 
The first part of the data corresponds to acquisition done
from the beginning of the experiment in February 2003 up to the
end of September 2004 (``Phase~1''). The second part of the data
presented here includes the runs from October 2004 up to the end of
2006 (``Phase~2''). Data collection continues under the 
conditions of Phase~2.
The mean $^{222}$Rn level was measured to be 37.7$\pm$0.1~mBq/m$^3$ for Phase~1,
and 6.46$\pm$0.02~mBq/m$^3$ for Phase~2 inside the tracking volume
(only statistical errors are given). 

According to the goal of the analysis
the $^{222}$Rn monitoring provides the possibility to select data
with different radon levels.

\subsubsection{Spatial distribution of the $^{214}$Bi}

In the preceding paragraph the results  of the mean $^{214}$Bi
activity measurements in the tracking volume were given.
For double beta decay studies it is important to know the $^{214}$Bi distribution 
close to the source foils.

\begin{figure}[h]
\centerline{\parbox[t]{13.0cm}{\fbox{\epsfxsize13.cm\epsffile{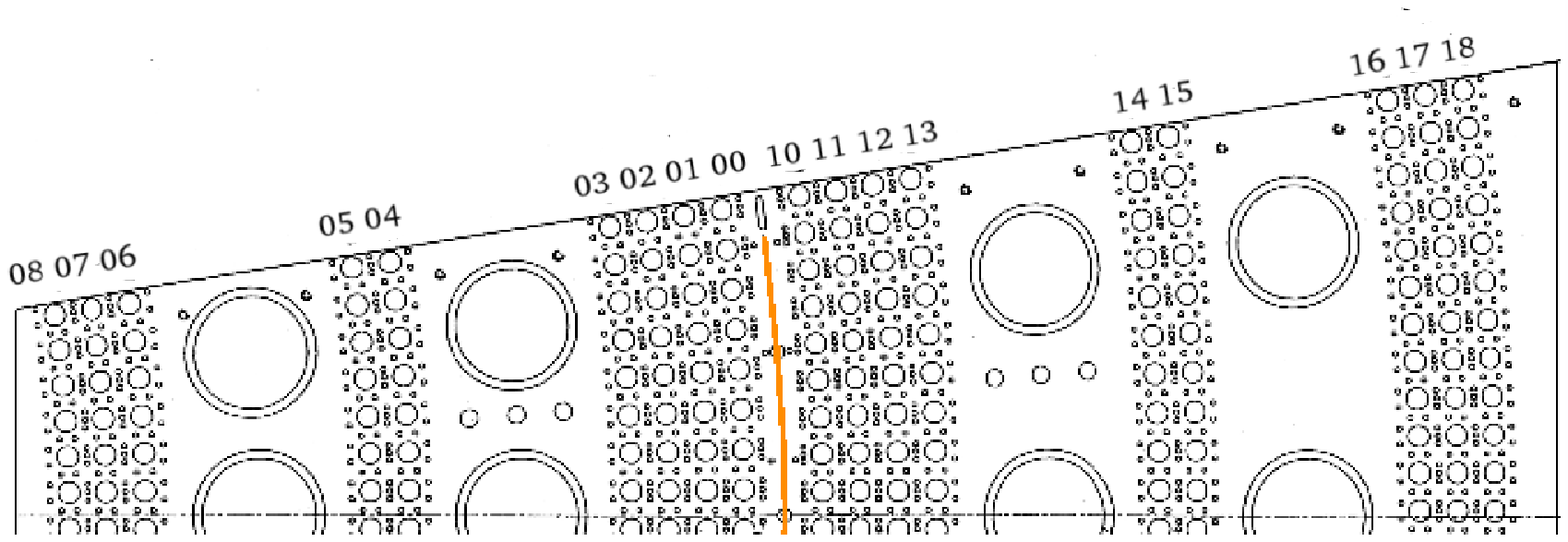}}}}
\caption{Lay out of the end-cap showing the Geiger cell layer
  configuration, top view. The inner layers 
  are numerated 00 to 08 starting from the source foil shown in red. The outer layers are 
  10 to 18 as indicated.} 
\label{fig:gg}
\end{figure}

The starting point of the electron track closest to the delayed track is
used to localize the decay point of a BiPo event. 
The position of such a point is 
defined by the Geiger cell layer number, running from 08 to 00 for
the inner Geiger cell layers and from 10 to 18 for the outer Geiger cell
layers~(Fig.~\ref{fig:gg}).   
The vertical position is the measured Z coordinate\footnote{Origin Z =
  0 is in the middle plane of the detector} of the electron's
origin.  Its azimuthal position is given by the sector number 
from 0 to 19.
\begin{figure}[!p]
\centerline{
\parbox[t]{10.cm}{\epsfxsize10.cm\epsffile{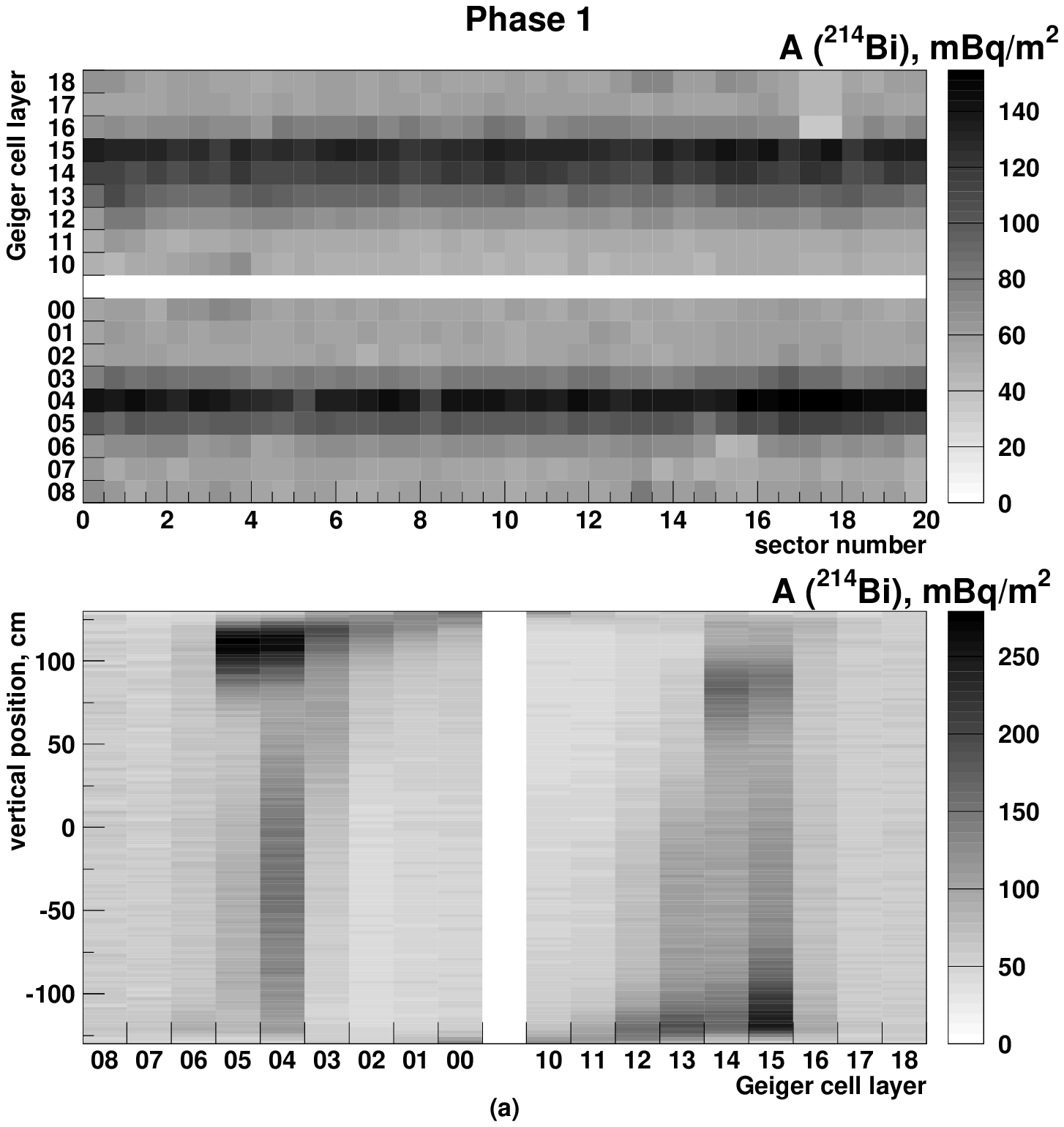}}}
\centerline{
\parbox[t]{10.cm}{\epsfxsize10.cm\epsffile{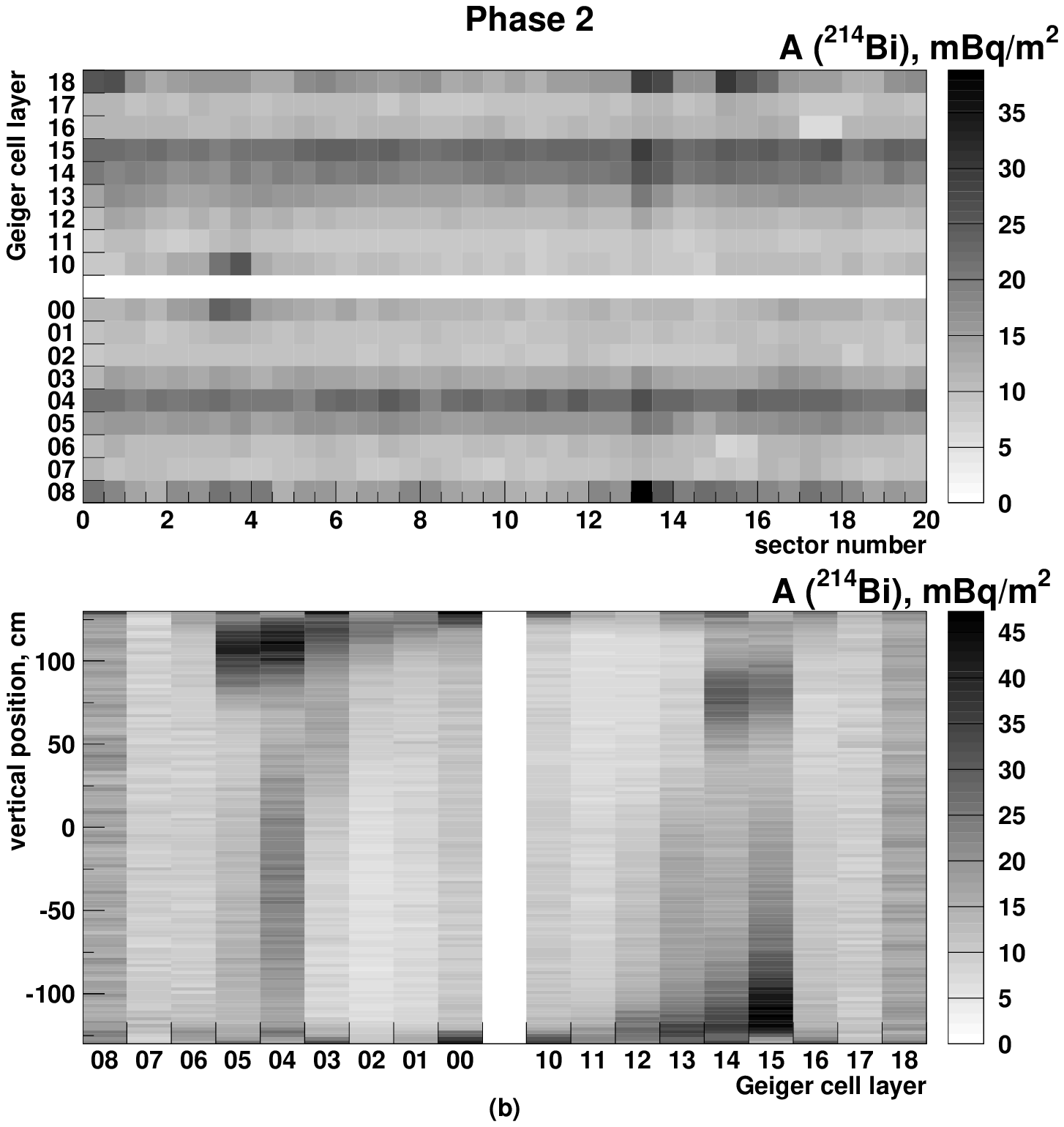}}}
\caption{Plots of the $^{214}$Bi activity as a function of 
  Geiger cell layer, sector number and vertical position inside the
  tracking chamber for data of (a) Phase~1~ and (b) Phase~2.  }  
\label{fig:rn_2d}
\end{figure}

The $^{214}$Bi distribution inside the tracking chamber for Phase~1 and Phase~2 is
shown in Fig.~\ref{fig:rn_2d}.
For both phases there is a consistent pattern that is
scaled to the radon activity. Greater activity is
observed in layers 03 to 06 and 13 to 16 and is explained by the large
gap between Geiger layers 03-04, 13-14, 05-06 and 15-16~(Fig.~\ref{fig:gg}).

\begin{figure}[htb]
\centerline{
\parbox[t]{10.0cm}{\epsfxsize10.cm\epsffile{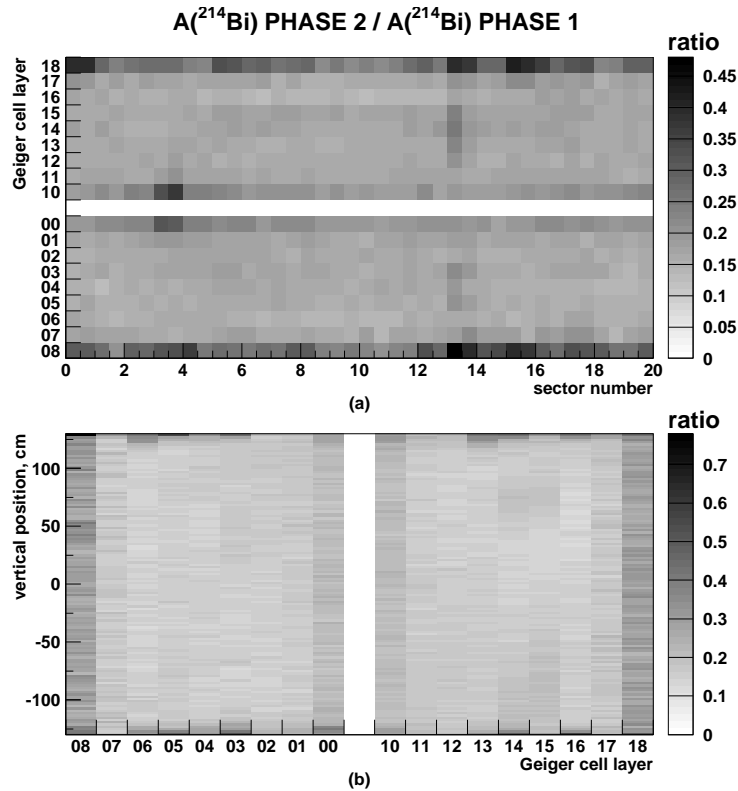}}
}
\caption{Ratio of $^{214}$Bi activity for two data samples
  (a) as a
  function of Geiger cell layer and
  sector number and (b) vertical position in the Geiger cell layers.}
\label{fig:rnratio}
\end{figure}

In order to detect possible sources of $^{222}$Rn outgasing or
$^{214}$Bi pollution, the spatial distribution of $^{214}$Bi in Phase~2 has
been compared with Phase~1.
The results are shown in Fig.~\ref{fig:rnratio}
and indicate that the distributions of $^{214}$Bi for the two phases are
similar in the whole tracking volume except at the borders. There is a 
residual $^{214}$Bi activity near the scintillator walls and end caps.
The enhanced $^{214}$Bi activity near the foil extremities
(especially in sector~3 with molybdenum foils)   probably
originates from the foil holders and/or
the scintillator surfaces.
In order to exclude the top and bottom parts with the ``extra'' $^{214}$Bi
activity, only events with vertices 
of $|$Z$| <$ 120~cm  are taken into account in the following. 

In the $\beta\beta$ analysis
most of the background due to
$^{222}$Rn appears to come from the regions close to the source foils. 
Therefore the observed non-uniformity of $^{214}$Bi distribution along the vertical 
direction for distant Geiger cell layers
is neglected. While the variations in the azimuthal direction from sector-to-sector
are accounted for.
The mean activities measured for each sector layer-by-layer are used.
The test of the background model (see section~\ref{sect:testmodel})
is performed with copper foils located in sector 0.
For this sector the results of the $^{214}$Bi activity
measurements for the layers closest to the foil are given 
in Table~\ref{tab:Tswbysectors}.
 
\begin{table}
\caption{The $^{214}$Bi activity in mBq  for the Geiger cell layer closest to the
  copper foils of sector 0. The
  average activity for the 20 sectors is also presented.
  Statistical errors are given.}  
\label{tab:Tswbysectors}
\label{page:swbysectors}

\vspace*{0.1cm}
{\small
\begin{center}
\begin{tabular}{|l|c|c|}
\hline \hline
&\multicolumn{2}{|c|}{A($^{214}$Bi),mBq} \\ \cline{2-3}
&Phase 1& Phase 2\\ \hline
Inner foil side, sector 0&724$\pm$10 &134$\pm$4 \\
Outer foil side, sector 0&598$\pm$8  &101$\pm$3\\ \hline
20 sectors average & 700$\pm$1 & 140$\pm$1 \\ \hline \hline 
\end{tabular}
\end{center}
}
\end{table}

\subsubsection{$^{222}$Rn activity measurement using $e\gamma$ events}
It is also possible to detect $^{214}$Bi using $e\gamma $ events.
A large fraction of  the $^{214}$Bi decays is accompanied by a high energy
$\gamma$-ray. However, the $e\gamma$ events are contaminated by 
external $\gamma$-rays which Compton scatter on the wires of the Geiger
cells (see section~\ref{sect:extbg}). 

To suppress this background contribution, only
events with a $\gamma$-ray of energy greater than 1~MeV are
used.  In order to select events originating from the tracking volume
and not from the source foils, only electrons with their starting point on the
Geiger cell layers 01, 11, 02 and 12 are analyzed.
The results for the mean $^{214}$Bi activity for the second and
third Geiger cell layers using $e\alpha$ and $e\gamma$ topologies
are shown in Table~\ref{tab:rn2}. 
\begin{table}[ht]
\caption{Results of the $^{222}$Rn measurements on the second and
  third Geiger cell layers of the tracking chamber using two event
  topologies are shown. Statistical errors are given.} 
\label{tab:rn2}

\vspace*{1mm}
\begin{center}
\begin{tabular}{|c|ccccc|}
\hline
&\multicolumn{5}{|c|}{A($^{214}$Bi), mBq}\\ \hline
Geiger cell layer&02&01&&11&12 \\ \hline
 $e\alpha$ & 598$\pm$3& 701$\pm$3 &&706$\pm$3&800$\pm$3 \\ \hline
$e\gamma$ &688$\pm$7&624$\pm$7&&645$\pm$5&727$\pm$6 \\
\hline
\end{tabular}
\end{center}
\end{table}

The method involving  $e\gamma$
events has a larger background with approximately three times smaller detection 
efficiency compared to  the method using the delayed tracks.
However, the $e\gamma$ events allows one to address the measurements 
of delayed tracks and to estimate the systematic error on 
$^{222}$Rn activity inside the tracking chamber to within 10\%.

\subsection{$^{220}$Rn ($^{208}$Tl) activity measurements inside the
  tracking chamber}
\label{sect:thoron}
If $^{220}$Rn is present in the gas of the tracking chamber, 
it constitutes a source of $^{208}$Tl.
Given the high $Q_\beta$ value, $^{208}$Tl  is a serious concern for
neutrinoless double beta decay search.
The beta decay of
$^{208}$Tl is almost always accompanied by a $\gamma$-ray of 2615~keV
from the first excited state of $^{208}$Pb~\cite{TOI}.  In the case of
the de-excitation by IC electron emission one can
observe two electrons with a total energy of approximately 3~MeV which
can mimic $^{100}$Mo and $^{82}$Se neutrinoless double beta decay. 
Therefore the thoron content of the gas mixture has been 
studied 
by means of the  $^{208}$Tl beta decay detection.

\subsubsection{Selection criteria}
\label{sect:thoron_sel}
The beta decay of $^{208}$Tl is mainly
accompanied by two or three $\gamma$-rays. Therefore $e\gamma\gamma$
and $e\gamma\gamma\gamma$ topologies  are used 
in the analysis. The event vertex is defined by the origin of
the electron.
In order to reject events coming from the source foil
only events with vertices on the Geiger cell layers 01$-$04 and
11$-$14 are analyzed.  
Tracks starting in planes 05 to 08 and 15 to 18 are too short to 
allow accurate event selection by time-of-flight.

For both topologies the electron has an energy greater
than 200~keV and each $\gamma$-ray an energy greater than 150~keV.
The time-of-flight method is used to ensure the common origin of 
all the particles involved.
In order to reduce backgrounds the 
$\gamma$-ray with the highest measured energy (E$_\gamma$) is required 
to have E$_\gamma$~$>$~1700~keV  
and the energy of the electron must satisfy the condition 
E$_e$~$>$~(4200~-~$\sum$E$_{\gamma}$)~keV.
\subsubsection{Results of the measurement}
The data has been analyzed separately for the two phases and the two
topologies.
The results of the measurements are shown in Table~\ref{tab:tl208ch}.
There is good agreement between the results obtained for the
$e\gamma\gamma$ and $e\gamma\gamma\gamma$ topologies.
The average activities resulting from them are
3.5$\pm$0.4 mBq for Phase~1 and 2.9$\pm$0.4 mBq for Phase~2. 
The systematic uncertainty is estimated to be
less than 10\% and comes mostly from the $\gamma$-ray detection efficiency.\begin{table}[tbh]
\caption{The number of observed events, number of
  estimated background events, signal efficiency and the results of the
  measurements of $^{208}$Tl activity inside the 
  tracking chamber found for both phases via
  $e\gamma\gamma$ and $e\gamma\gamma\gamma$ topologies. }
\label{tab:tl208ch}
\vspace*{0.1cm}
\begin{center}
\scriptsize{
\begin{tabular}{|l|c|c|c|r|}
\hline
Topology & N (observed) & N (estimated bgr)& Eff, \% &
\multicolumn{1}{|c|}{A($^{208}$Tl), mBq} \\
\hline
\multicolumn{5}{|l|}{Phase~1 }\\ \hline
$e\gamma\gamma$                      &342 	&22.4 	&0.26&3.5$\pm$0.4 	\\
$e\gamma\gamma\gamma$                &63 	&1.8 	&0.05&3.3$\pm$0.5 	\\
\hline                     
\multicolumn{5}{|l|}{Phase~2 }\\ \hline               
$e\gamma\gamma$                      &322 	&6.6 	&0.24&2.8$\pm$0.3 	\\
$e\gamma\gamma\gamma$                &79 	&1.2 	&0.05&3.5$\pm$0.5 	\\
\hline
\end{tabular}
}
\end{center}
\end{table}
A comparison of the results for the two phases shows 
no strong difference due to the detector environment but indicates a possible 
weak outgasing of thoron inside the
tracking volume. Taking into account the 35.94\% branching fraction of 
$^{208}$Tl in the $^{232}$Th chain, 
the measured $^{208}$Tl activity of $\sim$3~mBq corresponds to $\sim$8~mBq of $^{220}$Rn.
MC simulations  show that this low level of thoron
inside the tracking device 
is much less of a concern for 
neutrinoless double beta decay than the presence of radon.
For both phases the background from thoron in the 2$e^-$ channel 
is more than one order of magnitude 
lower than the background originating from radon.
\subsection{$^{210}$Pb ($^{210}$Bi) activity inside the tracking chamber}
One probable source of background with its origin on the Geiger
cell wires is the $\beta$-decay of $^{210}$Bi  from $^{210}$Pb ($T_{1/2}$ = 22.3~y) in
the $^{238}$U decay chain (see Fig.~\ref{fig:u238scheme}).
Given its $Q_\beta$ of 1.16~MeV, this radioactive isotope is of no concern
for this neutrinoless double beta 
decay search, but it must be  considered in the precise measurement of the
two neutrino double beta decay spectra. 

\begin{figure}[hb]
\centerline{
\parbox[t]{14.0cm}{\epsfxsize14.cm\epsffile{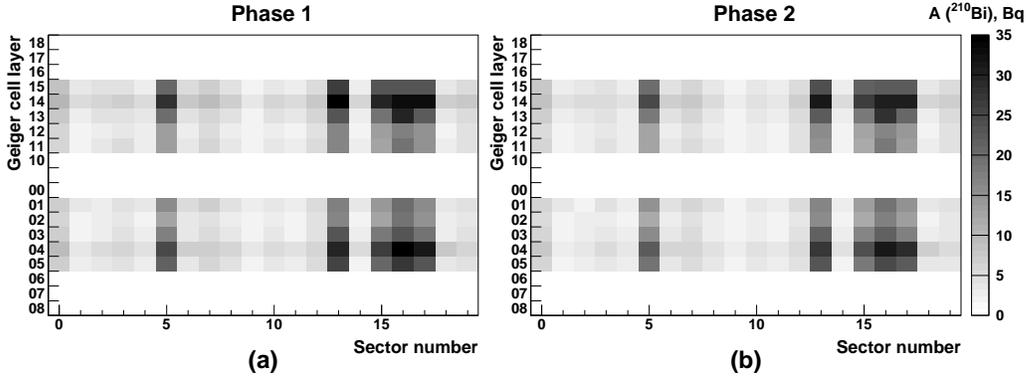}}
}
\caption{Plot of $^{210}$Bi
  activity  in Bq on the Geiger cell wire surfaces as a function of the
  event vertex position for (a) Phase~1 and (b) Phase~2. }
\label{fig:bi210sw}
\end{figure}

One electron events with an energy greater than 600~keV and
their vertices associated with Geiger cells  are
selected to  measure the $^{210}$Bi activity on the wire surfaces.
The results for both phases are in Fig.~\ref{fig:bi210sw} and 
Table~\ref{tab:bi210swres}. 

\begin{table}
\caption{Activity of $^{210}$Bi
  in Bq on wire surfaces of Geiger cell layers which are next to the
  foils. } 
\label{tab:bi210swres}
\vspace*{0.1cm}
\scriptsize{
\begin{center}
\begin{tabular}{|r|c|c|c|c|c|c|c|c|c|c|c|}
\hline
Sector&0&1&2&3&4\\ \hline
Phase~1&  6.21$\pm$0.37& 2.20$\pm$0.11& 2.95$\pm$0.28& 3.52$\pm$0.21
& 2.60$\pm$0.17\\
Phase~2&  5.83$\pm$0.35& 2.10$\pm$0.12& 2.76$\pm$0.25& 3.31$\pm$0.20& 2.41$\pm$0.13
\\ \hline
Sector&5&6&7&8&9\\ \hline
Phase~1&12.62$\pm$0.73& 3.26$\pm$0.17& 5.09$\pm$0.52& 3.29$\pm$0.25& 1.89$\pm$0.24\\
Phase2&11.84$\pm$0.63& 3.17$\pm$0.16& 4.71$\pm$0.42& 3.13$\pm$0.19&
1.85$\pm$0.14 \\ \hline
Sector&10&11&12&13&14\\ \hline
Phase~1&2.86$\pm$0.23& 2.10$\pm$0.24& 4.51$\pm$0.49&16.72$\pm$0.93& 2.21$\pm$0.14\\
Phase~2& 2.71$\pm$0.18& 2.00$\pm$0.13& 4.28$\pm$0.37&17.00$\pm$1.16&
2.11$\pm$0.14 \\ \hline
Sector&15&16&17&18&19\\ \hline
Phase~1&12.91$\pm$0.70&18.37$\pm$0.99&14.48$\pm$0.73& 2.48$\pm$0.12& 3.87$\pm$0.44\\
Phase~2&12.17$\pm$0.63&17.43$\pm$0.93&13.80$\pm$0.70& 2.40$\pm$0.12&
3.84$\pm$0.41\\ \hline
\end{tabular}
\end{center}
}
\end{table}

Approximately the same activity values and spatial distribution are measured in 
both data samples, while
a large variation of $^{210}$Bi activity from one sector to
another is observed.
The origin of the non-uniformity
in $^{210}$Pb deposition on the wires is most probably due to the different 
histories of the wires and conditions during the wiring of the sectors.

\section{External $\gamma$-ray flux}
\label{sect:extbg}

The external $\gamma$-ray flux is one of the sources of 2e$^-$ events and
therefore a background for double beta decay.
With the NEMO~3 data it is possible to measure this flux
using the events resulting from single or double Compton scattering. 

\begin{figure}[htp]
\begin{center}
\fbox{\includegraphics[width=0.41\textwidth]{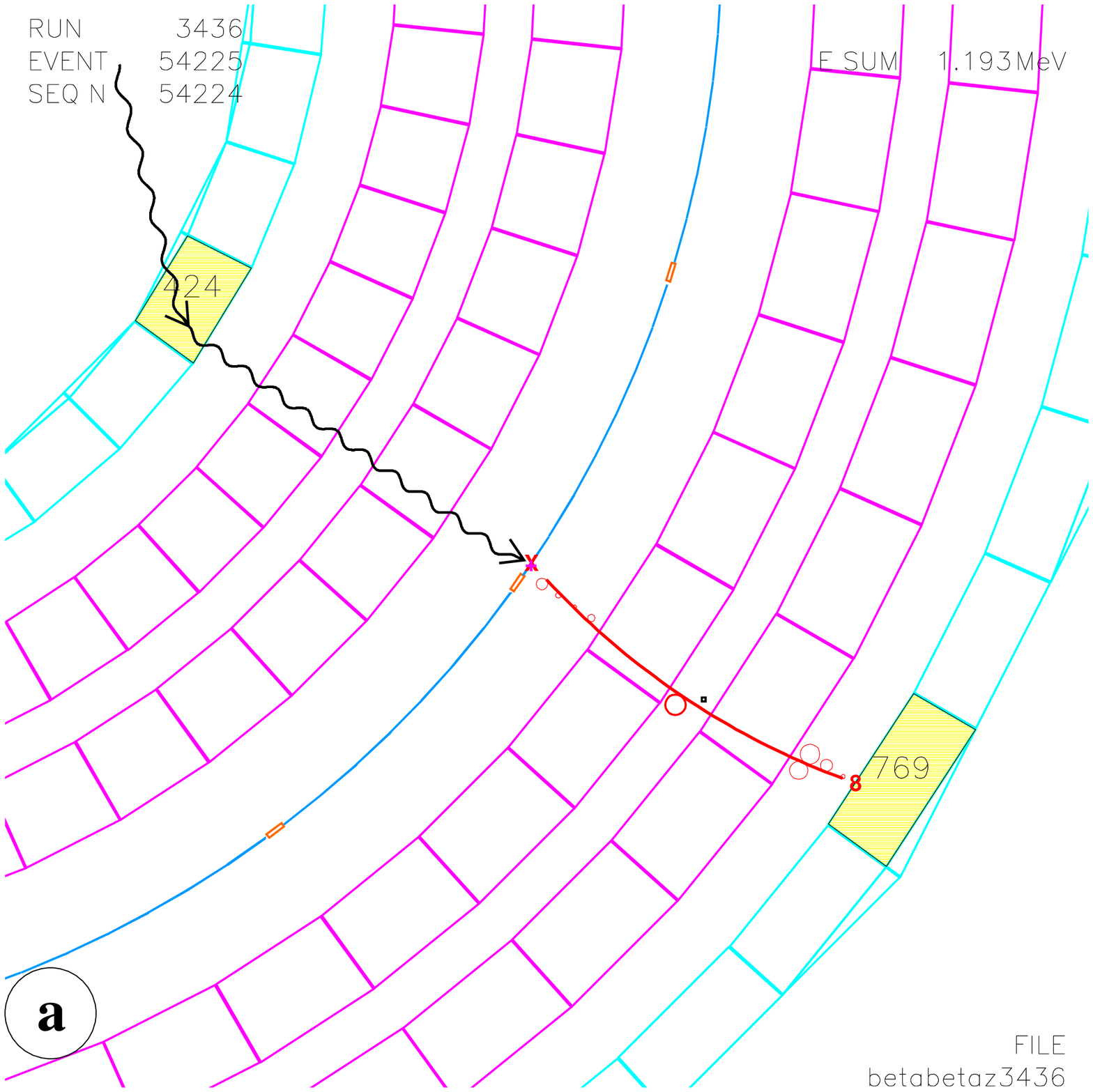}}
\fbox{\includegraphics[width=0.41\textwidth]{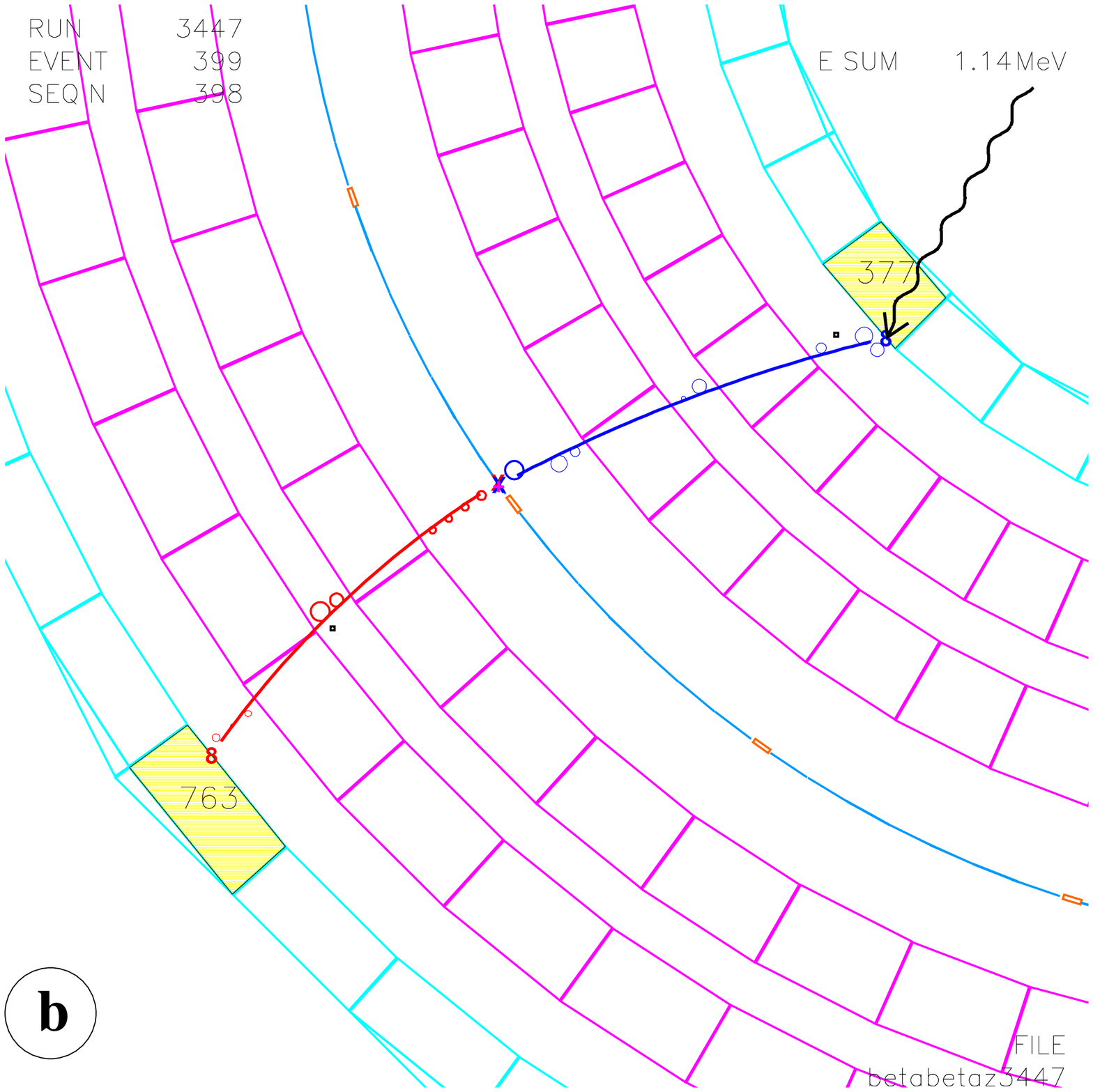}}
\end{center}
\caption[Tl]{Examples of (a) $e\gamma$-external  and (b) crossing electron
 events. Presumed $\gamma$-rays shown by wavy lines are superimposed to the event display.}
\label{fig:egee_example}
\end{figure}

If an incoming $\gamma$-ray undergoes Compton scattering inside a
scintillator block leaving the scattered electron unseen in the tracking chamber and
subsequently rescatters in the foil ejecting an electron which hits a
scintillator block, an event of $e\gamma$ topology is observed, 
see Fig.~\ref{fig:egee_example}a. 
Such an event from a double Compton scattering is
identified as an "$e\gamma$-external" event as opposed to 
an "$e\gamma$-internal" event with both  particles coming
from a decay or an interaction produced inside the foil.
Using time-of-flight measurements and the timing properties of the
detector it is easy to distinguish an $e\gamma$-external event from an 
$e\gamma$-internal event.

When a single Compton scattering of the incoming
$\gamma$-ray occurs very close to the surface of the scintillator block, the
scattered electron, provided it has sufficient energy,  crosses the tracking chamber. 
The topology of such an event is a pair
of tracks of opposite curvature sign coming from the foil. 
Using the  track curvatures and time-of-flight information 
it is easy to select such "crossing electron" events, see Fig.~\ref{fig:egee_example}b.
A possible background for this type of event is 
from $\beta$-emitters at the detector's inner surface.

\subsection{$\gamma$-ray flux from surrounding rocks}
In the LSM underground laboratory there is a significant flux of $\gamma$-rays
 coming from natural radioactivity in the surrounding rocks. This mostly includes  
gamma radiation from $^{40}$K, $^{214}$Bi and $^{208}$Tl decays. A passive shield,
surrounding the detector, 
has been constructed to reduce this source of background. 
The shield consists of low radioactivity iron plates and tanks filled with borated water,
see Fig.~\ref{fig:NEMO3}.

The $\gamma$-ray energy spectrum in the LSM laboratory has
been measured 
using a NaI detector.
In Table 3 of reference \cite{OHSUMI2002} one can find intensities of the 2.61~MeV  and 
1.46~MeV $\gamma$-ray lines of $^{208}$Tl and $^{40}$K decays, respectively.
The intensities of the lines corresponding to $^{214}$Bi decay have not been 
evaluated in that work, however, with the published spectrum, one can estimate the strength
of the 1.76~MeV $\gamma$-ray line and use it as a reference. Relative ratios between 
different $^{214}$Bi lines can be found in an earlier work \cite{ARP92}. 
Although these measurements were done for the Gran Sasso underground laboratory, 
one can assume that
the rock composition of the two sites is similar, and therefore the 
attenuation is similar too. The result of this compilation for the $\gamma$-ray flux 
is summarized in Table~\ref{tbflux}.

\begin{table}
\caption{
Simplified model of $\gamma$-ray flux in the LSM underground laboratory.}
\vspace{0.5cm}
\begin{center}
\begin{tabular}{|c|c|c|}
\hline
Isotope
 &
$\gamma$-ray energy, keV& Flux, $\rm cm^{-2}s^{-1}$  \\
\hline
$^{40}$K & 1461 & 0.1 \\
\hline
$^{208}$Tl & 2615 & 0.04 \\
\hline
$^{214}$Bi & 1764 & 0.05 \\
           & 1600 & 0.026 \\
& 1300 & 0.041 \\
& 1120 & 0.046 \\
&  609 & 0.109 \\
\hline
Total & & 0.411\\
\hline
\end{tabular}
\label{tbflux}
\end{center}
\end{table}

In order to evaluate the corresponding background in NEMO~3 a two stage 
MC simulation has been done. First,  a 
simple MC simulation was performed to estimate the 
attenuation of the NEMO~3 shielding (iron plates and water tanks) for the spectrum from Table~\ref{tbflux}.  
The MC provided energy and angular spectra ($\frac{\rm dN}{\rm dE \; dcos(\theta)}$) 
of the $\gamma$-rays after the shielding 
with a total attenuation factor of  
$3.5\cdot 10^{-5}$. 

\begin{figure}[hb]
\begin{center}
\includegraphics[width=0.45\textwidth]{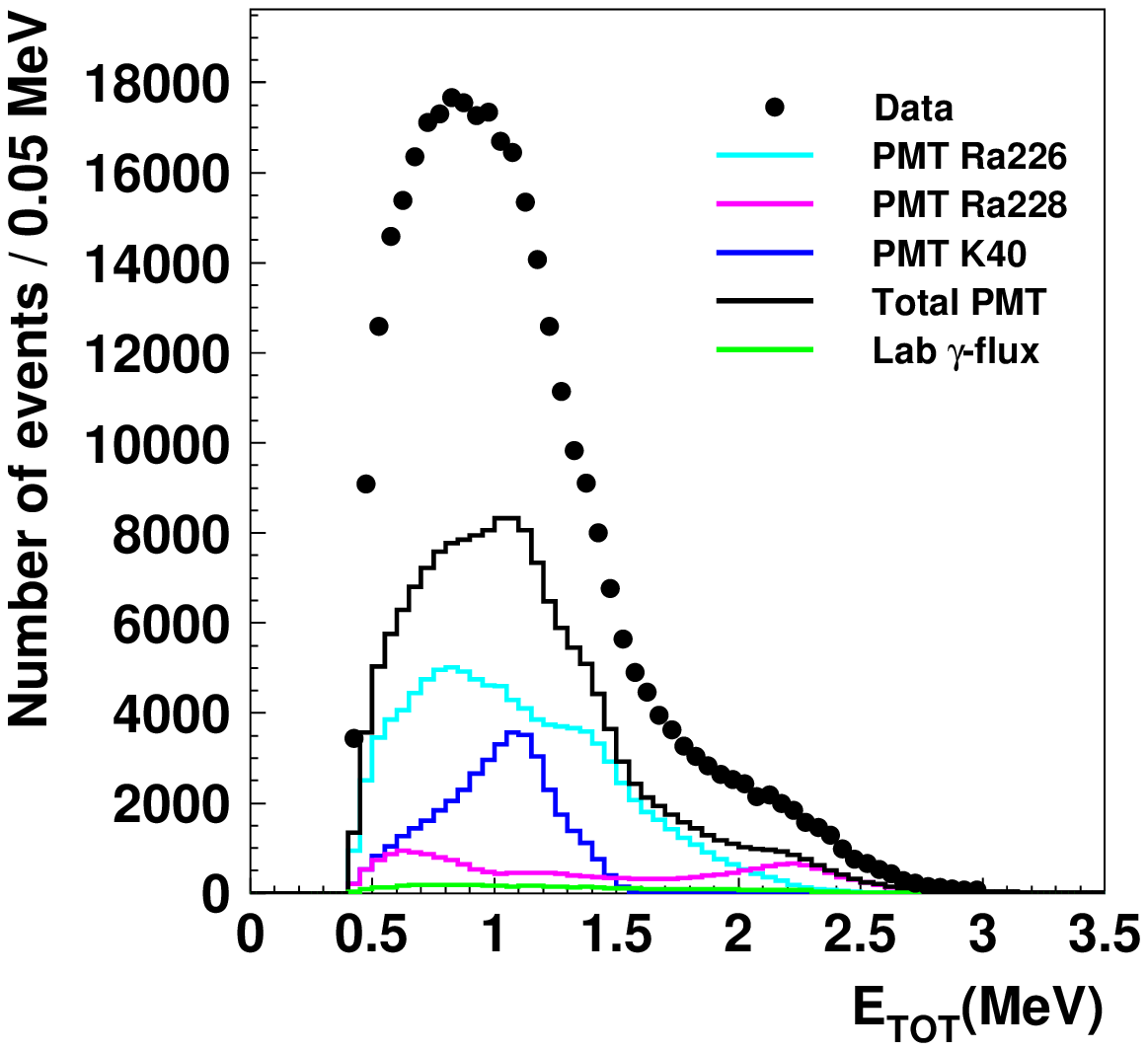}
\includegraphics[width=0.45\textwidth]{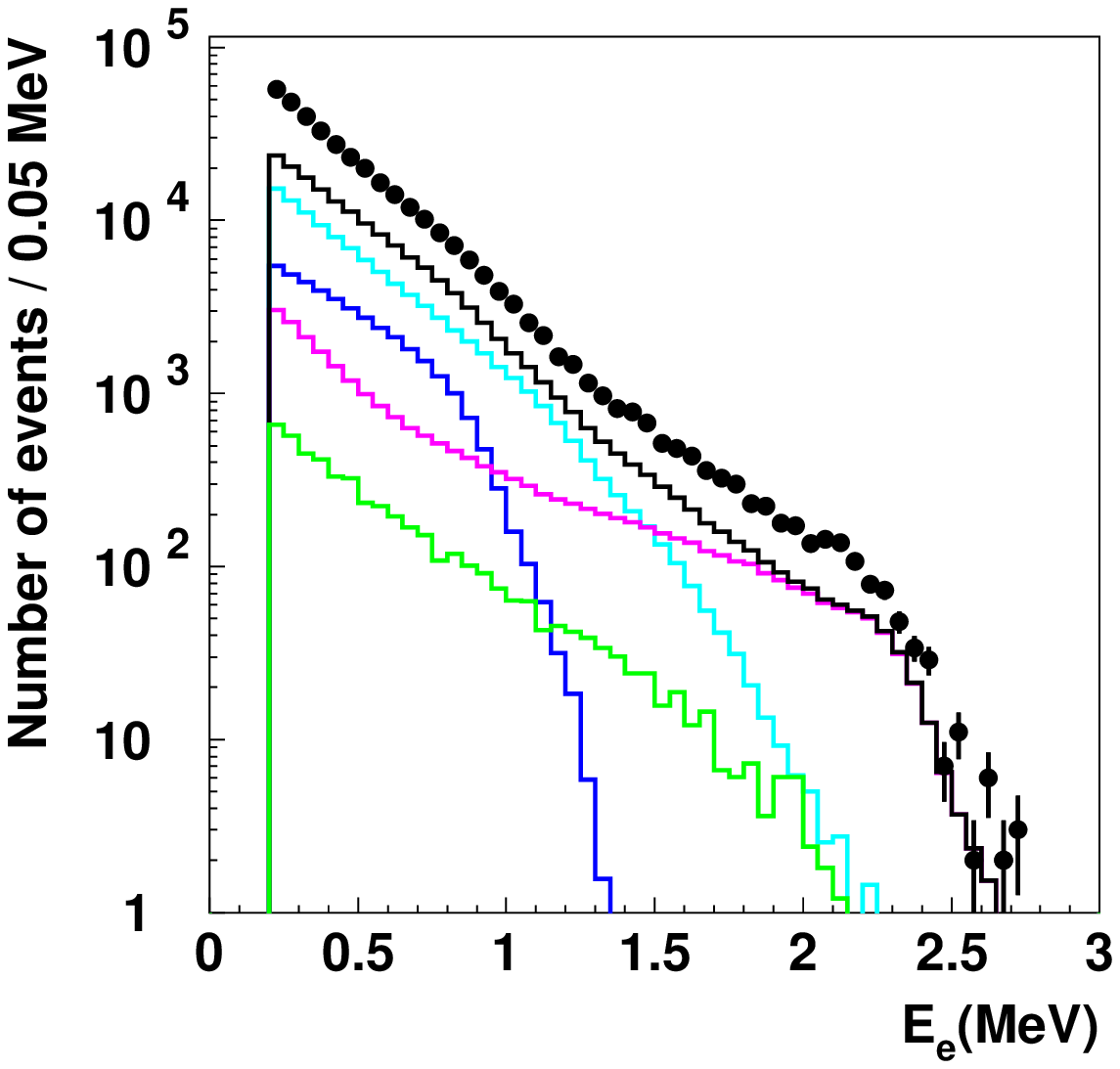}
\end{center}
\caption{
The sum  of the $\gamma$-ray  and electron energy ($E_{TOT})$, and the electron energy
($E_{e}$) of the $e\gamma$-external events in Phase~2 data. 
The expected contributions due only to the $\gamma$-ray flux from the laboratory 
(see Table~\ref{tbflux}) and from the photomultiplier
tubes are also shown.}
\label{fgstd}
\end{figure}

The  energy and angular 
spectra were then parametrized for a MC simulation with all the components
of NEMO~3. 
This simulation 
showed that, after cuts were applied, the $\gamma$-ray flux from the laboratory 
accounts for $\sim$2\% of the total measured external background, see Fig.~\ref{fgstd}.
Consequently it can be neglected and taken into account by a slight adjustment 
of other components in the external background model. The most important contribution 
expected due to the PMT radioactivity measured with the HPGe detectors~\cite{ARN05} 
is also demonstrated in this figure.

\subsection{Neutrons}
Neutrons can contribute to the external background via the neutron capture process 
resulting in emission of $\gamma$-rays.
The neutron flux in the LSM has been measured \cite{neutron_nemo},\cite{neutron_edelweis}
and originates from spontaneous fission 
and ($\alpha$,$n$) reactions due to trace amounts of uranium in the rocks.
The NEMO~3 neutron shield thermalizes fast neutrons with energies of a few MeV 
and suppresses thermal and epithermal neutrons ~\cite{ARN05}.

\begin{figure}[hb]
\begin{center}
\includegraphics[height=6.cm]{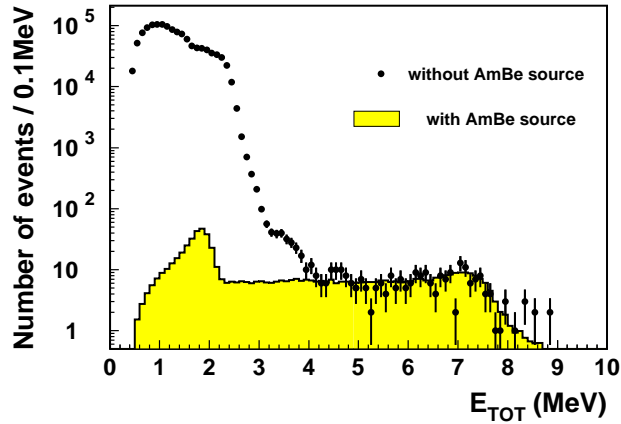}
\caption{
Energy sum distribution of crossing electron events
for 404 days of data (Phase~1).
The superimposed histogram represents the distribution obtained
for 25 hours of data  with an AmBe neutron 
source and scaled by the factor $6.9 \cdot 10^{-3}$.}
\label{fgneutron}
\end{center}
\end{figure}

A series of calibration runs with an AmBe neutron source in the vicinity of the 
detector has been done to check the shield's efficiency. These runs may also be used  to 
evaluate the  neutron
background for $\beta\beta2\nu$ measurements. 
The energy sum distribution of crossing electron events for Phase~1 data is shown 
in Fig.~\ref{fgneutron}.
The distribution for a run with the AmBe source is superimposed on it.
A pronounced tail at  energies up to 8~MeV is seen in both 
distributions and is a characteristic feature of
neutron captures in the detector walls.
This tail above 4.5~MeV 
is used for normalization of the neutron source distribution (Fig.~\ref{fgneutron}).
Attributing all the data in this energy region to neutrons
one can see that the neutron contribution to the low energy part of the spectrum 
measured without the AmBe source is very small.
In this way one finds that 
the neutron background can amount to 0.03\% of the total at energies below 4~MeV.
Therefore this background can be neglected for $\beta\beta2\nu$ analysis.

\subsection{External $\gamma$-ray flux model}
The principal source of the external background is the natural radioactivity  of the
detector components. 
The dominant one is due to the PMTs contamination by  $^{226}$Ra, $^{228}$Ra and $^{40}$K
as known from the results of HPGe detector measurements \cite{ARN05}.
It is addressed in the MC simulation with decays of $^{214}$Bi, $^{208}$Tl, $^{228}$Ac
and $^{40}$K inside the PMT glass.
However, the use of the PMT activities allows one to reproduce roughly only half of 
the observed experimental 
$e\gamma$-external events, see Fig.~\ref{fgstd}.
The small fraction of each component of NEMO~3 was measured extrapolating the results to the 
total amount of the material.
Therefore these activities are used as free parameters in order to fit the experimental data.
The presence of these isotopes and of $^{60}$Co in other parts of the detector have 
also been taken into account.
The list of potential isotopes providing the low energy $\gamma$-rays is long. 
It includes for example 
$^{54}$Mn, $^{58}$Co, $^{65}$Zn, $^{137}$Cs, and $^{234m}$Pa. 
In order to take them into account and to improve the description of the low energy 
part of the gamma spectrum, a flux of 1~MeV $\gamma$-rays 
was simulated at the external surface of the calorimeter.

Not all components of the apparatus or sources of background 
have been considered so the results of the fit described below
should not be interpreted as a  
measurement of the internal contamination of the corresponding elements of the detector. 
The purpose of this study is to provide an effective model able
to reproduce the external $\gamma$-ray flux.

\begin{table}[htb]
\caption{Components of the external background model.}
\bigskip
\begin{center}
\begin{tabular}{|l|r|r|r|r|r|}  
\hline
Components   & \multicolumn {4} {c|}{Activity (Bq)}\\ \cline{2-5}
of NEMO~3    & $^{40}$K  & $^{214}$Bi & $^{208}$Tl & $^{60}$Co \\
\hline 
Photomultiplier tubes   & $1078. \pm 32.$   & $324. \pm 1.$   & $27.0 \pm 0.6$   &        \\ 
Plastic scintillators   & $21.5 \pm 0.9 $  &         &         &        \\
$\mu$-metal PMT shield  &          &         &         & $14.6 \pm 2.6$ \\
Iron petals             & $100. \pm 4.$    & $9.1 \pm 1.0$   & $3.1 \pm 0.5$  & $ 6.1 \pm 1.8$ \\
Copper on petals        &          &         &         & $47.6 \pm 7.8$ \\
Internal tower          &          &         &         & $18.4 \pm 0.8$ \\
Iron shield             &          & $7360. \pm 200.$  & $484. \pm 24. $ &        \\
\hline 
\end{tabular}
\label{table:exbg_model}
\end{center}
\end{table}
\vskip 0.5cm
The external background was evaluated  using two types of events, 
$e\gamma$-external and crossing electrons.
A global fit of the following distributions was performed: the total energy released 
in the calorimeter, 
the energy deposited by the electron, the cosine of the angle between the
directions of the electron and $\gamma$-ray for the $e\gamma$-external events, 
the total energy and finally the energy of the electron after crossing the tracking volume
for the crossing electron events.
In order to describe the external $\gamma$-ray flux coming into the detector
in more detail
three sets of histograms were produced. These are for the incoming $\gamma$-ray signals 
detected by a counter at the internal wall, external wall and end-caps.
 
\begin{figure}[htb]
\begin{center}
\includegraphics[width=0.32\textwidth]{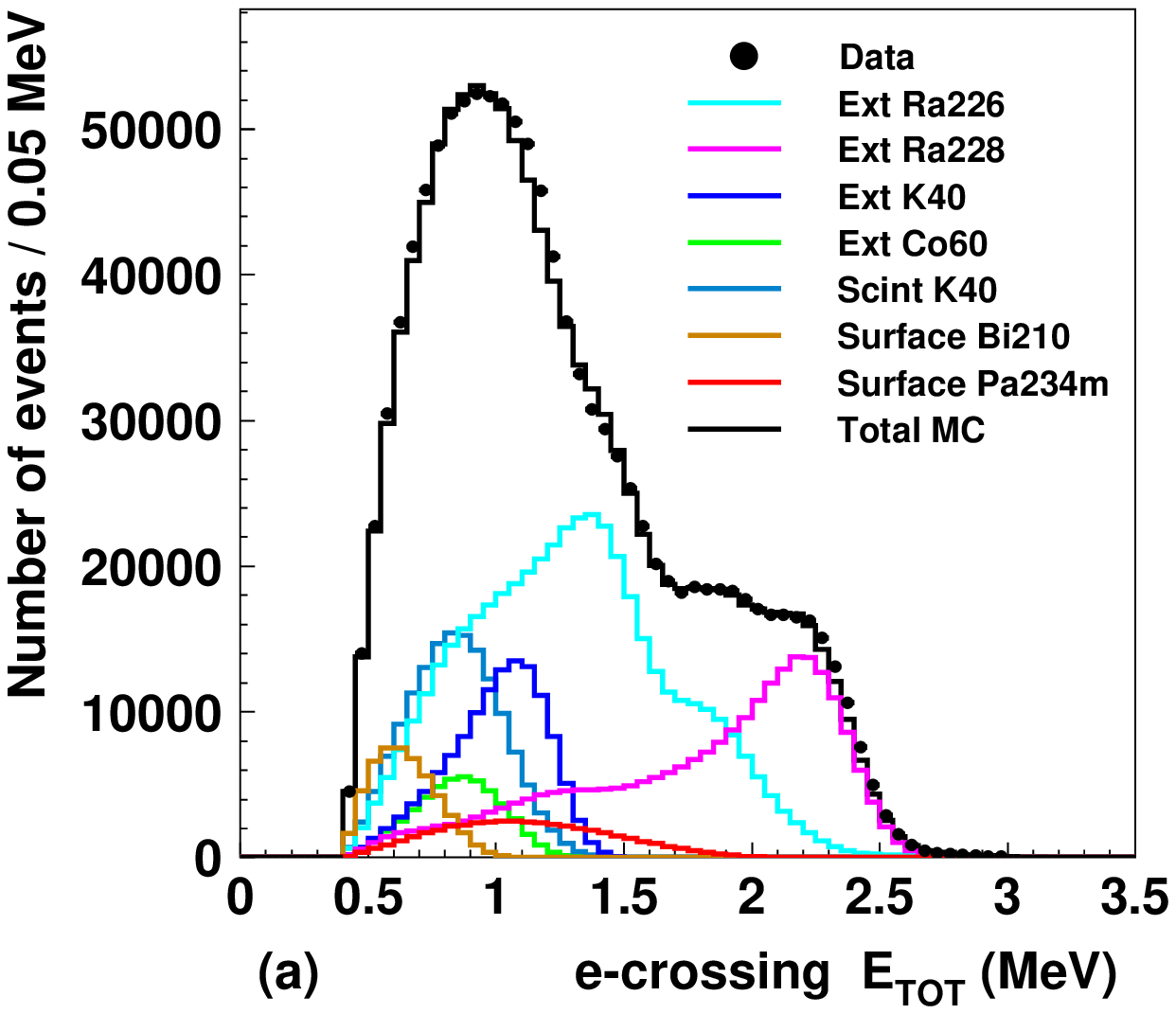}
\includegraphics[width=0.32\textwidth]{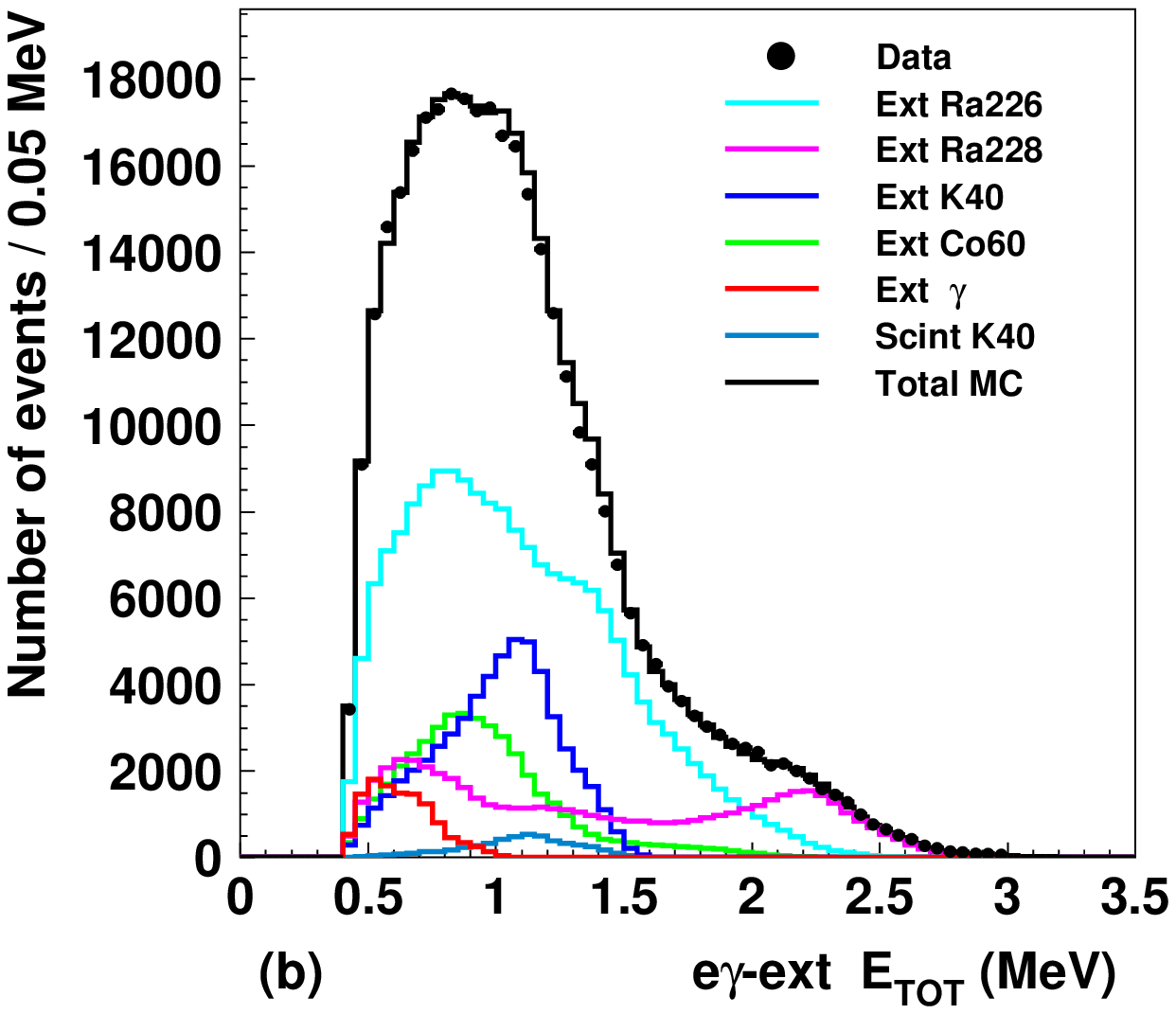}
\includegraphics[width=0.32\textwidth]{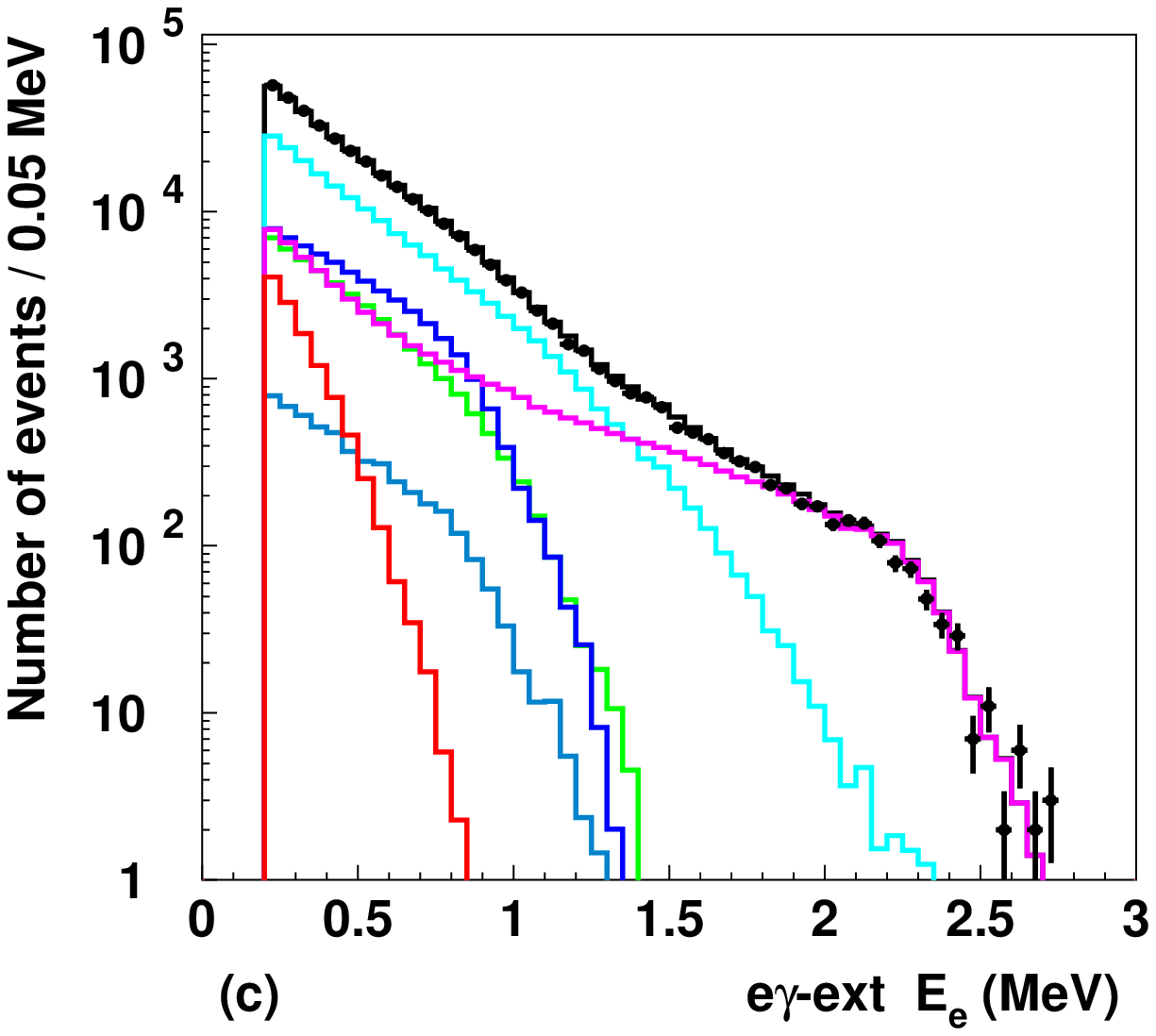}\\
\end{center}
\caption[Tl]{ Results of the Phase~2 fit for the whole detector: a) energy sum of crossing 
electron events,  b) the energy sum of the electron and $\gamma$-ray for 
$e\gamma$-external events, c) the detected electron energy for $e\gamma$-external events.}
\label{fig:fit_all_low} 
\end{figure}

First the external background model parameters are fixed from the 
Phase~2 data where one can 
neglect the radon in the 
air surrounding the detector given its very low level ($\sim$~0.1~Bq/m$^3$).
The activities of the background model presented 
in Table~\ref{table:exbg_model} and the 1~MeV $\gamma$-ray flux of
0.446~m$^{-2}$s$^{-1}$ reflect 
the data for the incoming $\gamma$-rays detected
by the calorimeter of NEMO~3.
\begin{figure}[b]
\begin{center}
\includegraphics[width=0.32\textwidth]{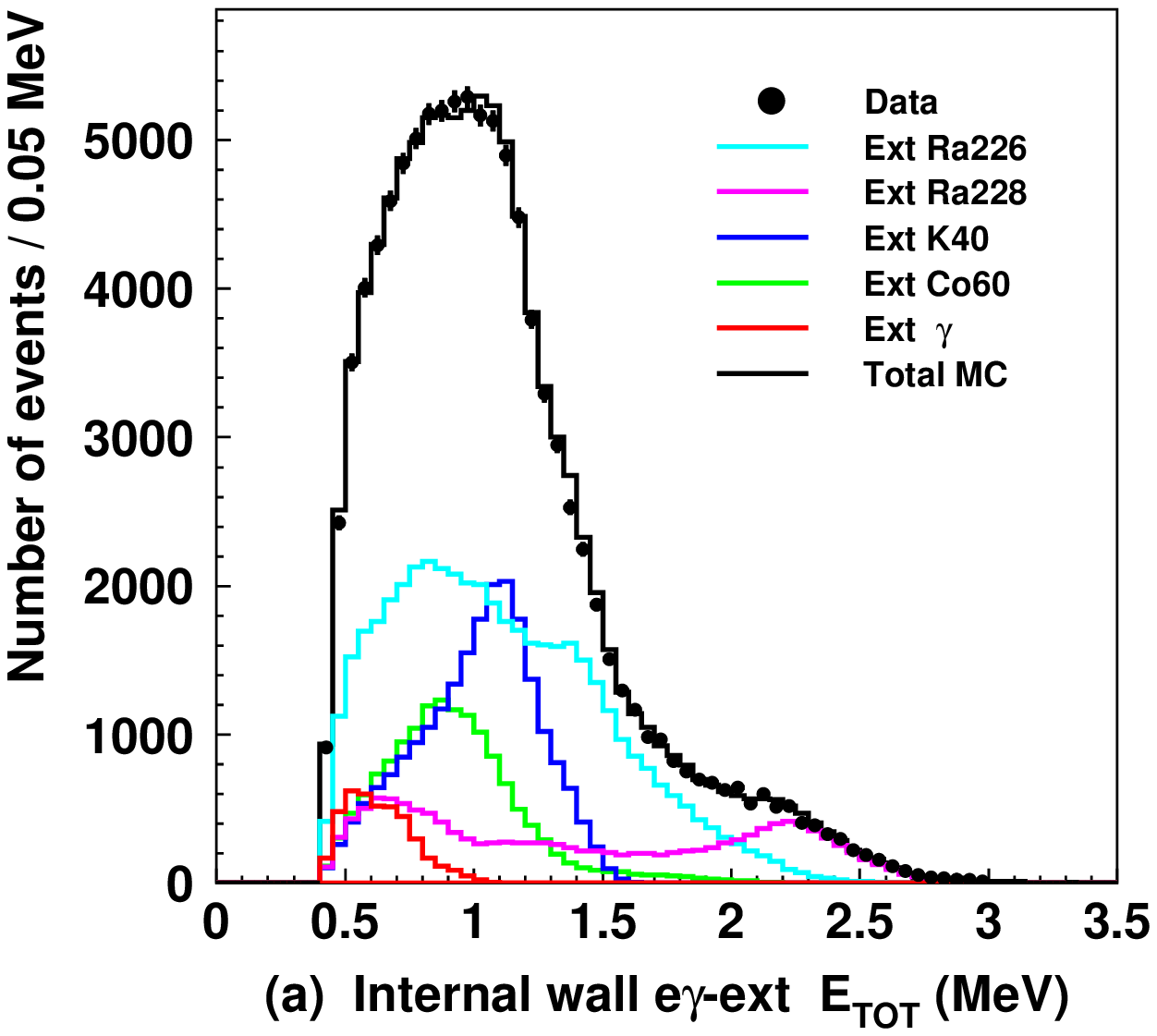}
\includegraphics[width=0.32\textwidth]{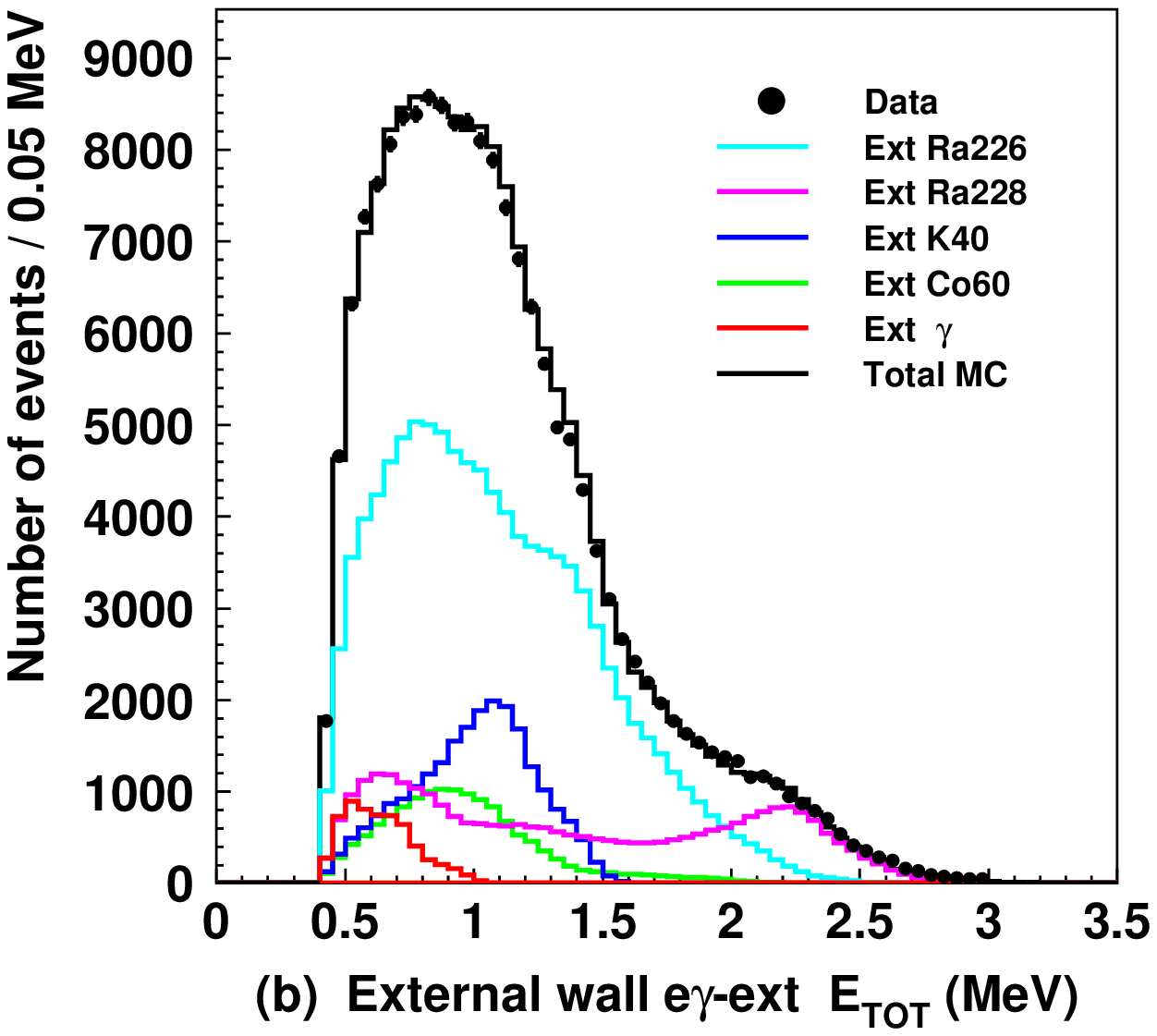}
\includegraphics[width=0.32\textwidth]{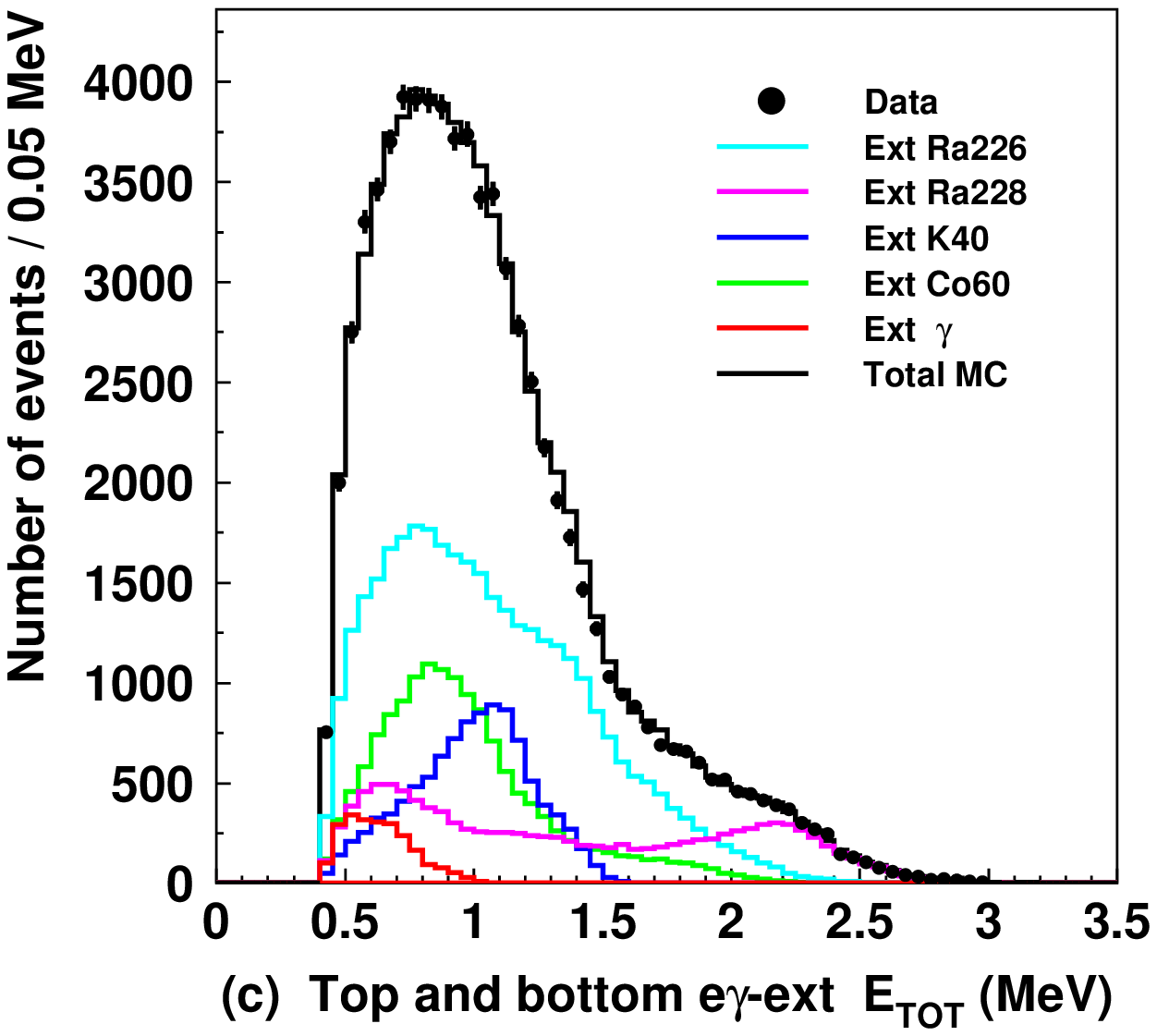}\\
\end{center}
\caption[Tl]{Phase~2 data. The energy sum distribution of $e\gamma$-external 
events with incoming $\gamma$-rays detected at (a) the internal wall, (b) the external wall
and (c) the end-caps at the top and bottom of the detector.}
\label{fig:fit_ori}
\end{figure} 
In Fig.~\ref{fig:fit_ori}  the
energy sum ($E_{\rm TOT}$) distributions of the  $e\gamma$-external events are shown. 
The crossing electron channel requires an additional source of electrons
that was found to be possible to reproduce
with 7.3~mBq/m$^2$ of $^{234m}$Pa and 120 mBq/m$^2$ of $^{210}$Bi on the surface of 
the scintillator blocks.
The results of the fit for the whole detector for Phase~2 data are presented 
in Fig.~\ref{fig:fit_all_low}.
In the Fig.~\ref{fig:fit_all_low}(a) one can see the distribution of the total energy 
measured in the calorimeter for crossing electron events, where the beta emitters 
provide a noticeable contribution.
However, they are negligible in the case of  $e\gamma$-external events for which one can see 
the distributions of the energy sum of the electron and $\gamma$-ray 
in Fig.~\ref{fig:fit_all_low}(b)
and of the detected electron energy 
in Fig.~\ref{fig:fit_all_low}(c).

\begin{figure}[htb]
\begin{center}
\includegraphics[width=0.32\textwidth]{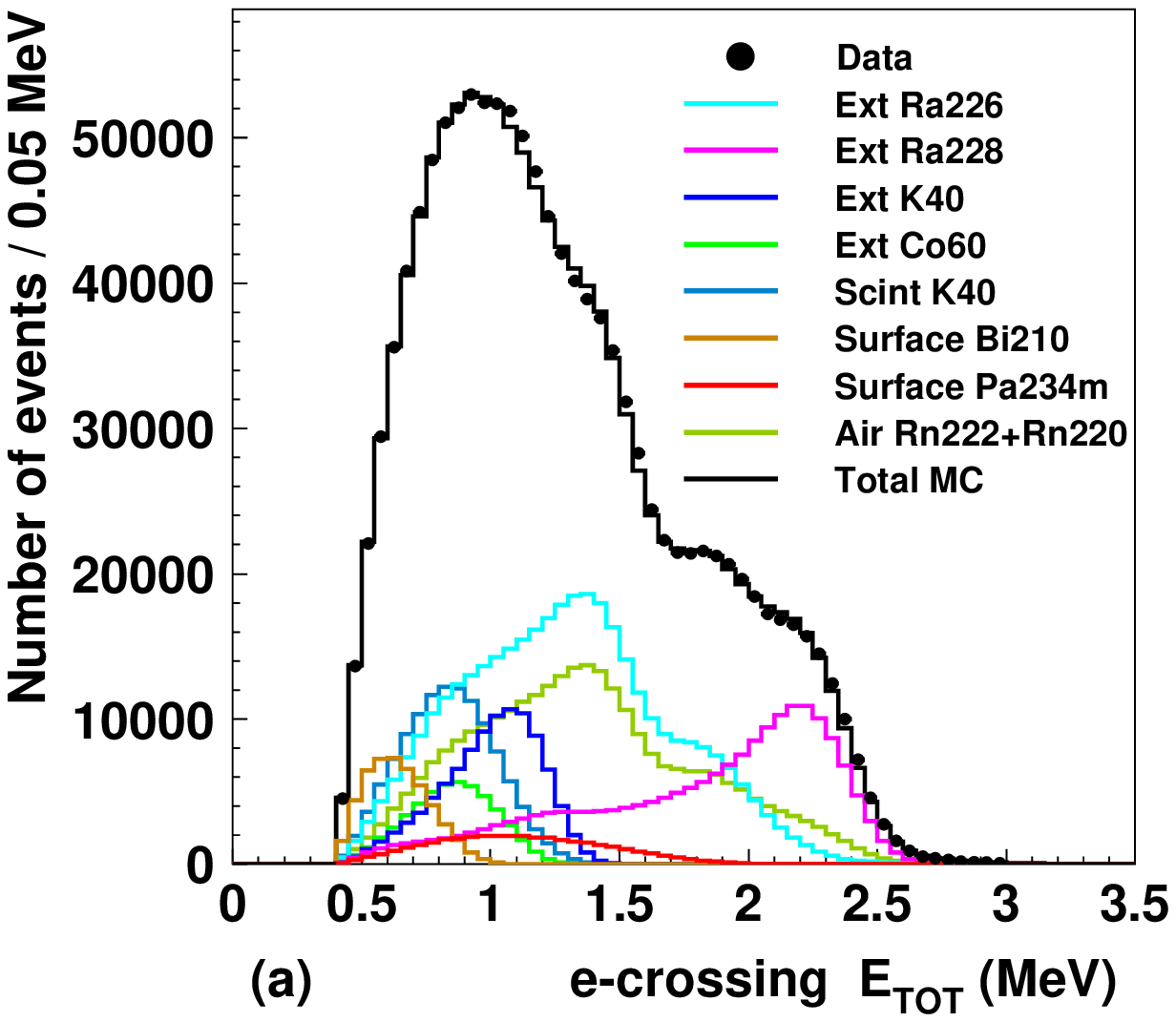}
\includegraphics[width=0.32\textwidth]{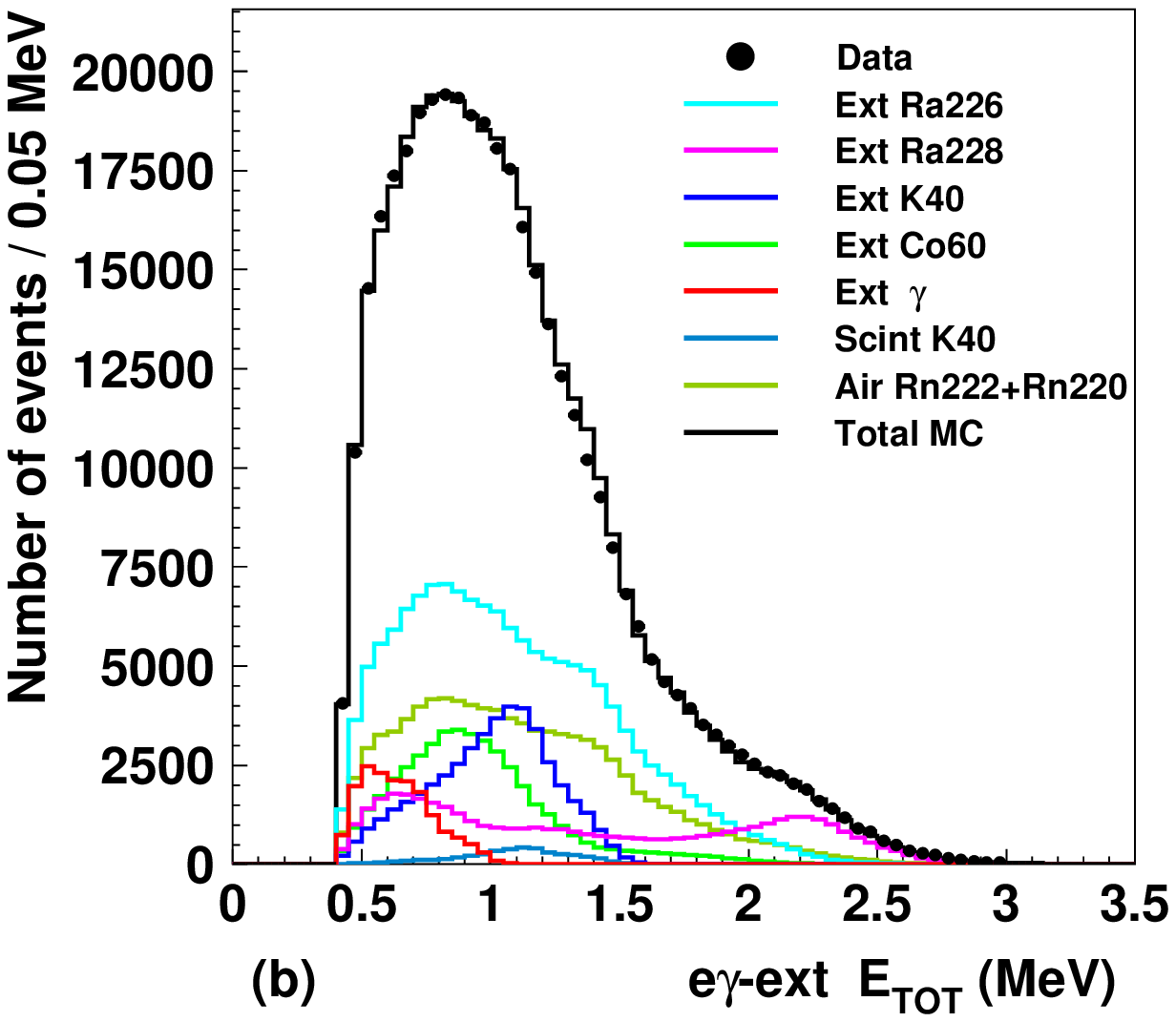}
\includegraphics[width=0.32\textwidth]{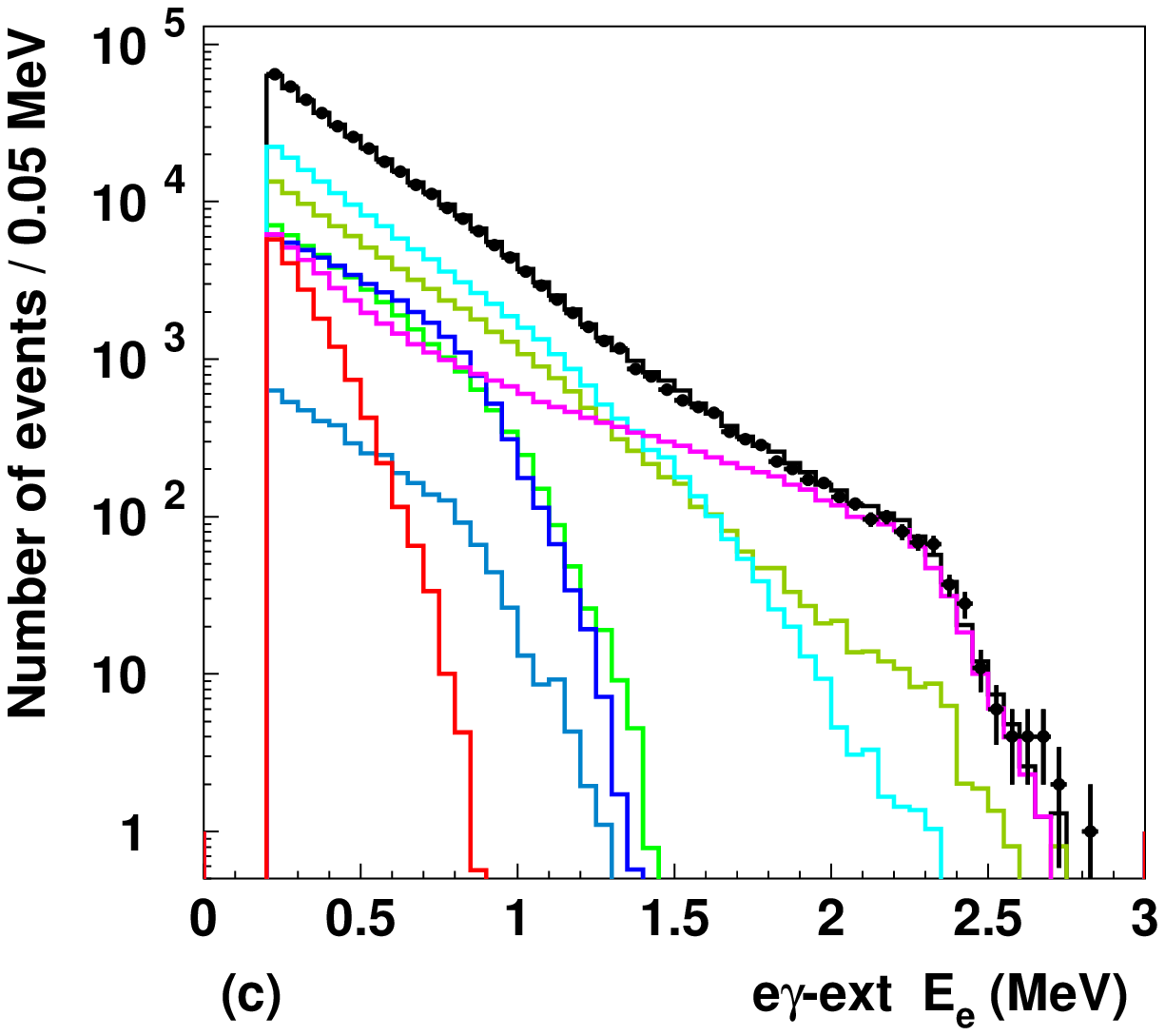}
\caption{Results of the fit for the Phase~1 data: a) energy sum of crossing 
electron events,  b) the energy sum of the electron and $\gamma$-ray for 
$e\gamma$-external events, c) the detected electron energy for $e\gamma$-external events.}
\label{fig:bestfitHigh}
\end{center}
\end{figure}

Then the Phase~1 data were fit with the model including the contribution of radon 
and thoron 
in the air thus simulating $^{214}$Bi and $^{208}$Tl decays in the gap between 
the iron shield and the tracking chamber walls. 
For radon  the activity of 11~Bq/m$^3$ is obtained in  good
agreement with the results of the radon monitoring in the LSM.
The value of the thoron activity which agrees with the NEMO~3 data is 
0.22~Bq/m$^3$.
The mean activity of $^{60}$Co for Phase~1 is higher by a factor of 1.3 when
compared to Phase~2. This is reasonable considering the $^{60}$Co half-life of $T_{1/2}$=5.2~y.
The low-energy $\gamma$-ray flux for Phase~1 was found to be twice as high
than in Phase~2. 
As one can see in Fig.~\ref{fig:bestfitHigh} the background model fits well the 
Phase~1 data too.

\begin{figure}[htp]
\begin{center}
\includegraphics[width=0.41\textwidth]{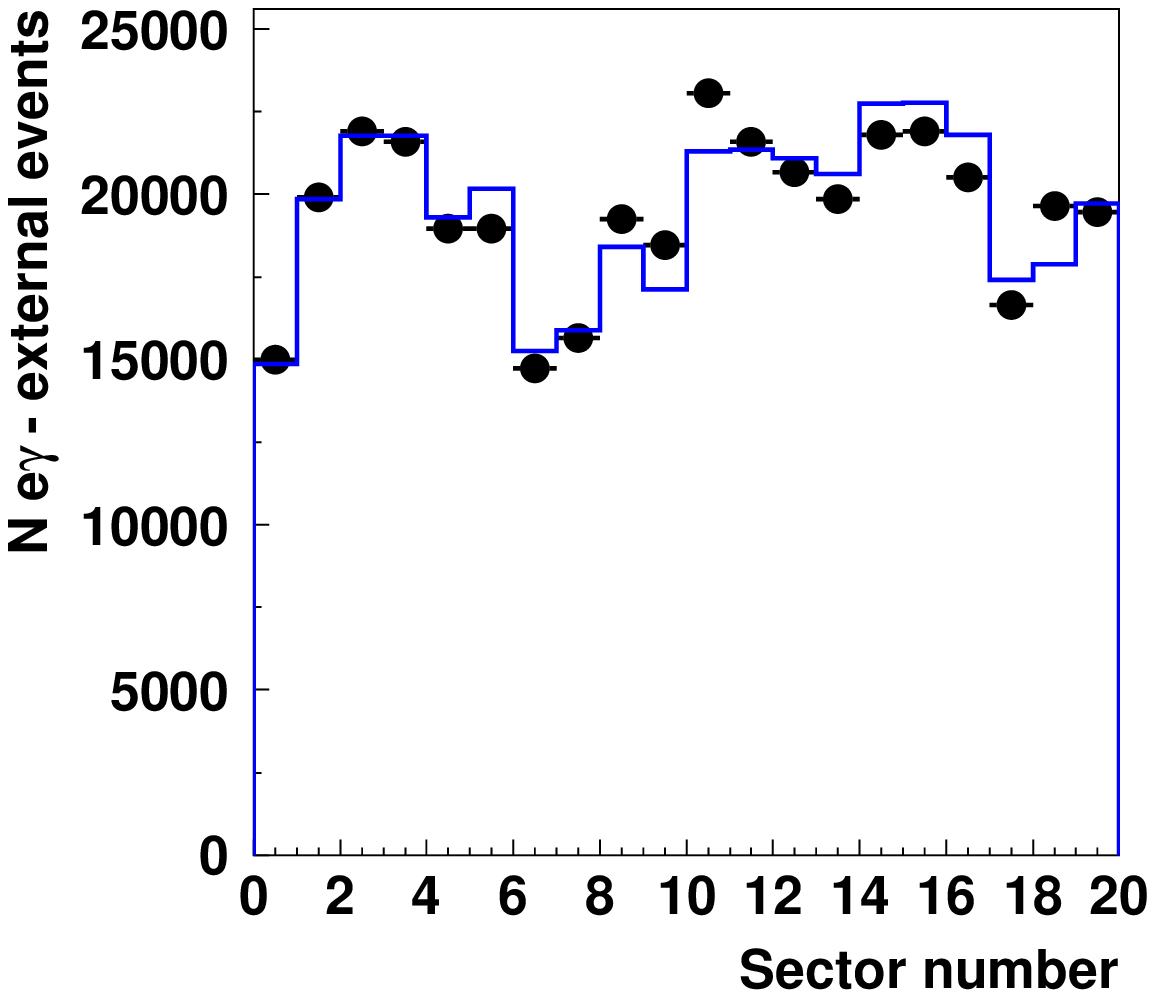}
\includegraphics[width=0.41\textwidth]{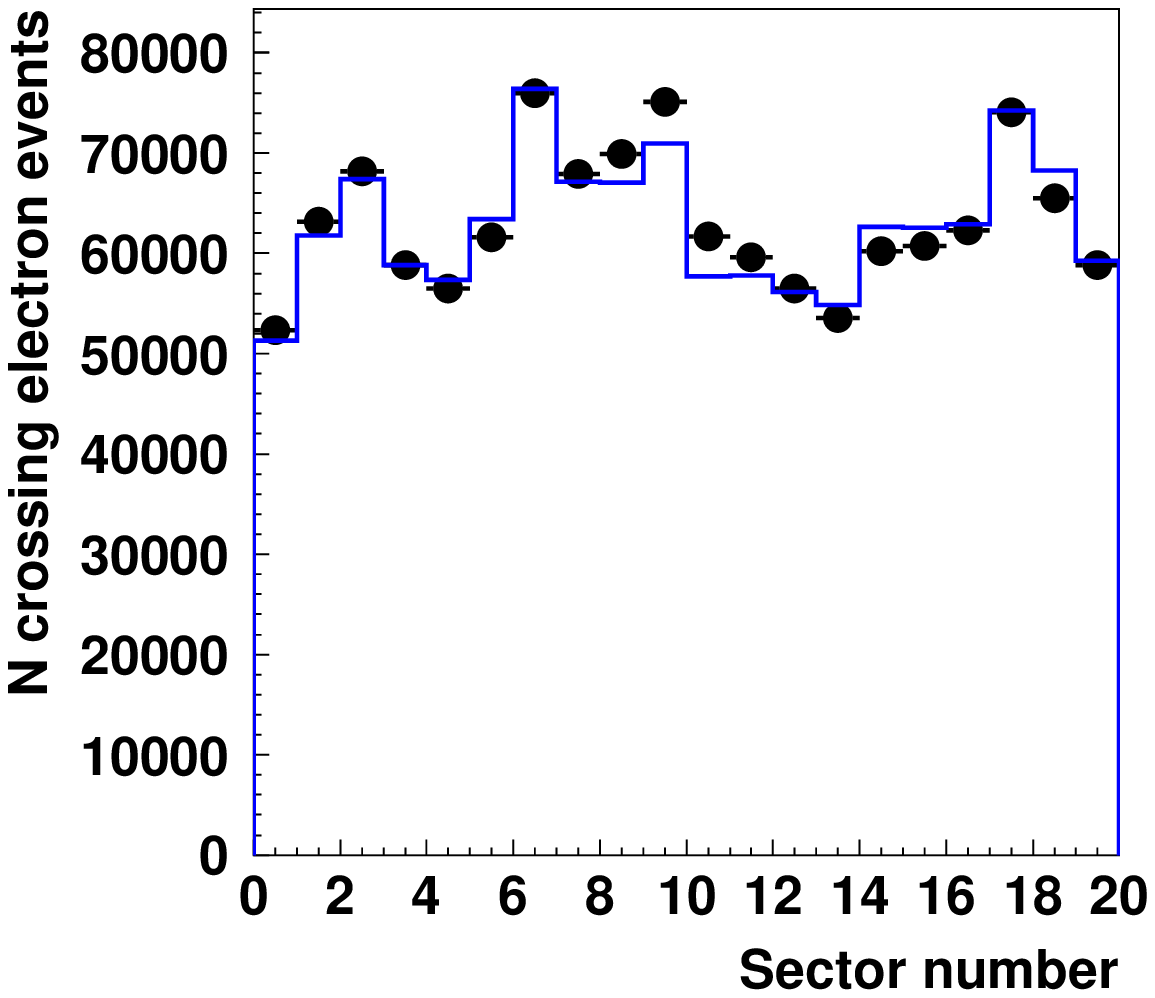}
\caption{Two distributions of the number of events per sector 
(left $e\gamma$-external and right crossing electrons). The data are given by black 
dots and the MC simulations are shown with a solid line.}
\label{fig:exbg_sector}
\end{center}
\end{figure}

The total number of crossing electron and $e\gamma$-external events and their observed 
energy distributions are well reproduced for the whole detector. 
Nevertheless, one may expect that the background may vary 
from one sector to another due to possible inhomogeneities of the detector materials.
The distribution of the number of observed events by sector is not uniform but the pattern is 
repeated by Monte Carlo calculations, see Fig.~\ref{fig:exbg_sector}.
It indicates that the non-uniformity is mainly due to the detector's acceptance.
A small number of dead PMTs and Geiger cells that vary from 
sector-to-sector accounts for this. 
The difference between the number of observed and expected 
crossing electron  and $e\gamma$-external events
per sector does not exceed  10\%.

\subsection {Test of the model with $ee\gamma$-external events}

\begin{figure}[htp]
\begin{center}
\includegraphics[width=0.41\textwidth]{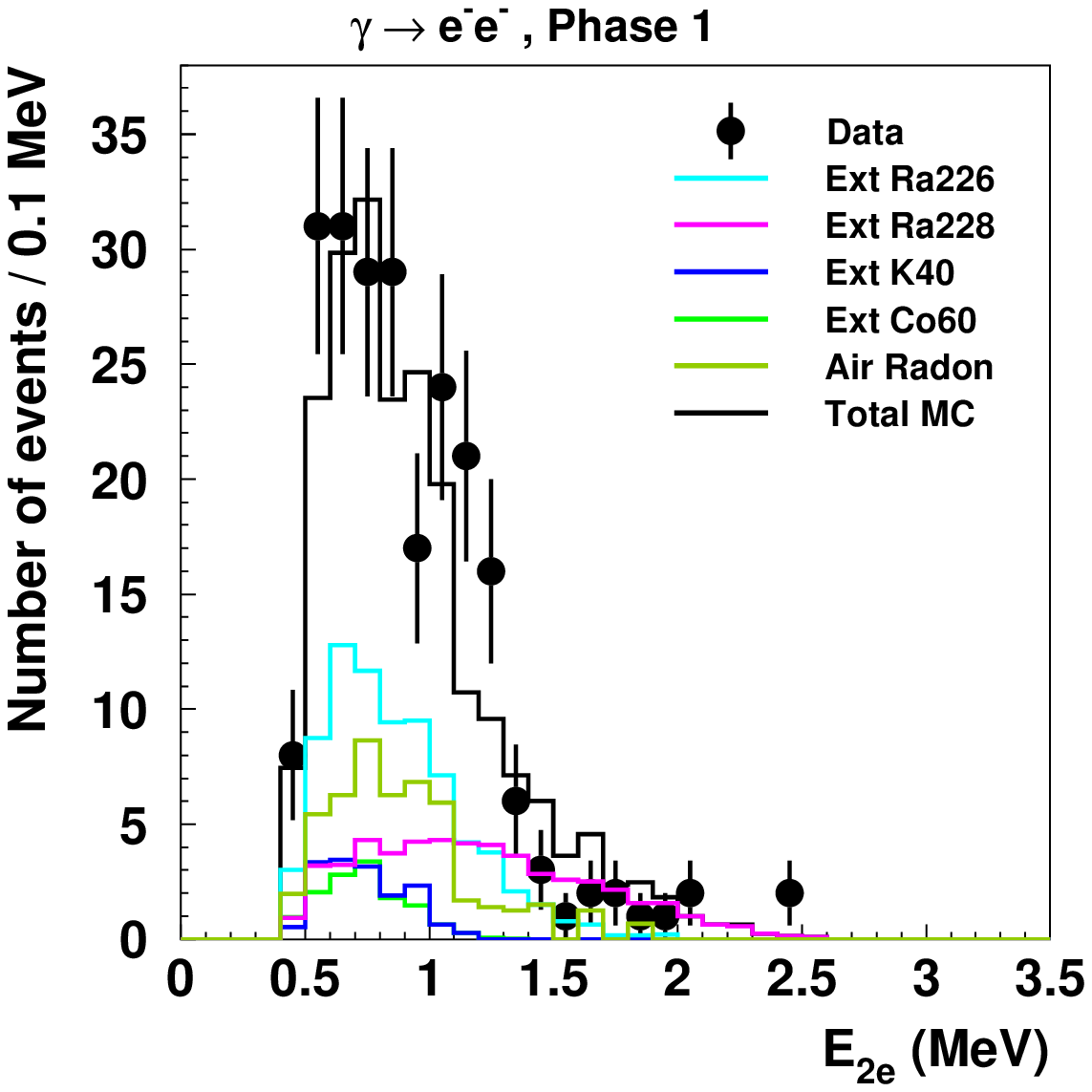}
\includegraphics[width=0.41\textwidth]{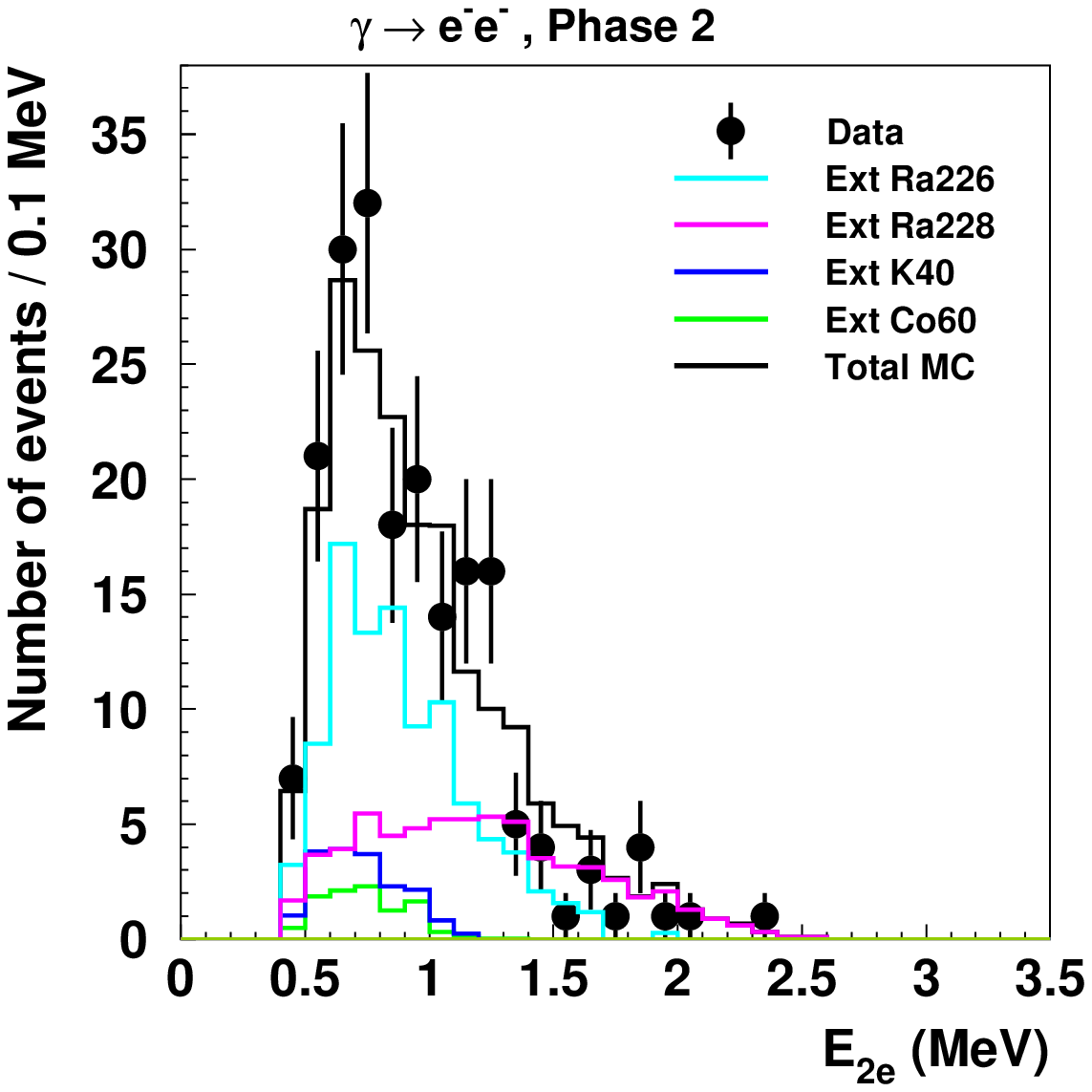}\\
\end{center}
\caption[$e^-e^-$]{Energy sum distribution of the two electrons in 
$ee\gamma$-external events compared 
to the results obtained with the external background model.}
\label{fig:gammaemem}
\end{figure}

This type of event is similar to the $e\gamma$-external one, where an incoming $\gamma$-ray
deposits some part of its energy in the calorimeter and hits a foil,
but here two electrons are emitted 
from the foil due to double Compton scattering in the foil or
due to a M{\o}ller scattering of a single Compton electron.
The probability for a  $\gamma$-ray to produce two electrons in the foil is about 
three orders of magnitude lower than for a single Compton electron.
So the statistics for this channel are rather poor.
Requiring the $\gamma$-ray energy deposit E$_\gamma$~$>$~200~keV
the total number of events observed
for 969 days of data collection is 420 compared to 409 events expected according to the 
external background model. The distribution of the energy sum of the two electrons in these
events is shown separately for Phase~1 and Phase~2 in Fig.~\ref{fig:gammaemem}.
The total number of observed 2e$^-$ 
events produced by detected external $\gamma$-rays as well 
as their energy sum distribution are well reproduced by 
MC calculations.

\section{Radioactivity inside the source foils}
\label{sect:inbg}
\subsection{Measurements of the internal $^{214}$Bi activity} 
The internal $^{214}$Bi contamination 
of the source foils is measured with BiPo decays 
through the detection of e$\alpha$ events.
The energy loss of  alpha particles in the foil is significantly 
larger compared to the energy loss in the gas.
As a consequence the $\alpha$-track length depends on the $^{214}$Po
decay location.
Therefore the delayed track length distribution was used to
measure the internal impurities of the source foils.

As in the case of radon activity measurements it is assumed that 
the $^{214}$Bi is deposited on the wire surfaces.
The possibility of a deposition on the foil surface is also considered.
Phase~2 data is used to minimize the 
contribution from the radon in the tracking detector.
Only events from the source foil (see Fig.~\ref{fig:ev_examples}a) are used.
The selected events are divided into four groups according to the
delayed track location with respect to the foil (inner or outer side)
and to the electron track (on the same side of the foil or 
on the opposite side). 

\begin{figure}[htb]
\centerline{
\parbox[t]{14.0cm}{\epsfxsize14.cm\epsffile{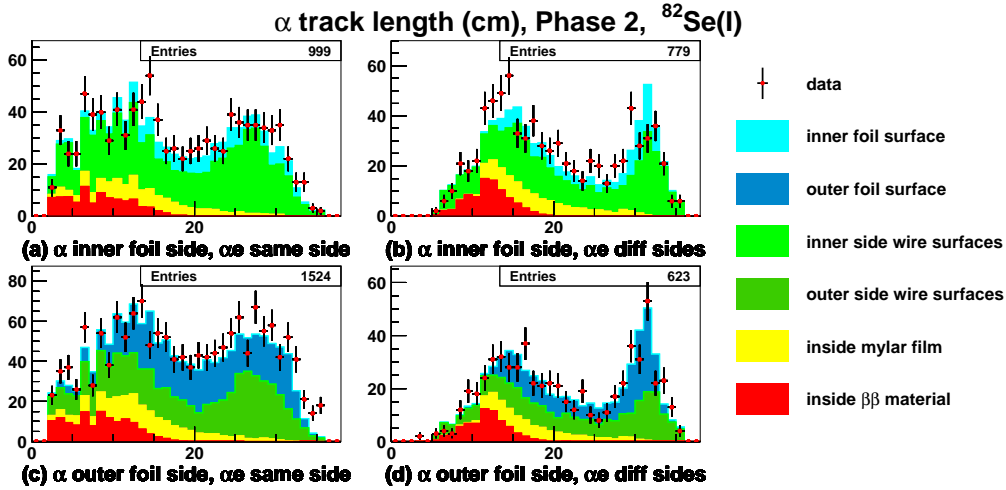}}
}
\caption{Delayed track length distributions for the 
  events of Phase~2 with vertices on the selenium
  source foils is shown by points with error bars. The
  MC simulations are shown with histograms. 
  The results for events with both the $\alpha$ track and the
  fast track (e$^-$) on the  same 
  side of the source foil are shown in plots~(a) and (c) while
  those having the $\alpha$ track and the electron on different foil sides 
  are shown in plots~(b) and (d).} 
\label{fig:bi214_int}
\end{figure}

An example of delayed track length distributions is shown for 
the selenium
foils of the first production\footnote{Selenium foils are from 
  two separate enrichment tasks, which are called $^{82}$Se(I) and
  $^{82}$Se(II).}, 
Fig.~\ref{fig:bi214_int}. These are fit with five
contributions corresponding to the different locations
of the $^{214}$Bi. 
The five are from inside the $\beta\beta$ material, on the two source 
surfaces and in the wire surfaces on both sides of the foil.

Except for the metallic molybdenum ($^{100}$Mo(m)) and the copper
foils, the NEMO~3 
source foils are made of thin mylar films
sandwiching the active $\beta\beta$ material. 
The mylar films are from three
different productions and their activities were measured with HPGe
detectors before foil fabrication.
The results of these measurements are given in column A2 of
Table~\ref{tab:tabinbg}. After the source foil fabrication 
the foil activity was again measured. Results are shown in column A4 of
Table~\ref{tab:tabinbg}. For most of the measurements only
limits could be achieved. 

To make the fit the $^{214}$Bi activity inside the mylar films
covering the foils (if any) is fixed to the value measured with the HPGe detector.
The measurement of the internal $^{214}$Bi contamination of each 
NEMO~3 source material, using this method, is given in column A1 of
Table~\ref{tab:tabinbg}. When only a limit on the mylar activity is available
two results are given. The first is obtained with the limit value, 
the second with the mylar activity set to zero.

\begin{table}
\vspace*{0.2cm}
\caption{Measurements of 
  $^{214}$Bi activity of the source foils, mBq/kg.
  A1 - results of the fit for the foil activity excluding the contribution from mylar.
  A2 - mylar activity measured with HPGe detectors, 
  A3 - total foil activity including mylar calculated from A1 and A2,
  A4 - foil activity measured with HPGe.
  The following
  notations are used for different source foil types:
  $^{100}$Mo(m) --- for the metallic molybdenum,
  $^{100}$Mo(c) --- for the composite molybdenum,
  $^{82}$Se(I) --- for the selenium sample of the first enrichment and 
  $^{82}$Se(II) --- for the second.  }   
\label{tab:tabinbg}
{\scriptsize
\begin{center}
\begin{tabular}{|l||l|l|l|l|}
\hline\hline
Source foil& A1 & A2 & A3 & A4 \\ \hline\hline
$^{100}$Mo(m)&$<$0.1                &-          &$<$0.1          &$<$0.39     \\ \hline
$^{100}$Mo(c)&$<$0.1; 0.30.0$\pm$0.07 &$<$0.67    &$<$0.15; 0.27$\pm$0.07     &$<$0.34    \\ \hline
$^{82}$Se(I) &1.0$\pm$0.13             &1.7$\pm$0.5&1.1$\pm$0.17 &$<$4.2      \\ \hline
$^{82}$Se(II)&0.4$\pm$0.15             &1.7$\pm$0.5&0.53$\pm$0.18&1.2$\pm$0.5 *\\ \hline
$^{96}$Zr    &6.4$\pm$2.; 7.8$\pm$2.  &$<$0.67    &5.5$\pm$1.7; 6.5$\pm$1.7  &$<$16.7     \\ \hline
$^{150}$Nd   &2.7$\pm$0.4              &3.3$\pm$0.5&2.8$\pm$0.4  &$<$3.3      \\ \hline
$^{130}$Te   &$<$0.1                   &3.3$\pm$0.5&0.39$\pm$0.06; 0.48$\pm$0.06&$<$0.67     \\ \hline
Te(nat)      &0.13$\pm$0.1             &1.7$\pm$0.5&0.28$\pm$0.14&$<$0.17     \\ \hline
$^{116}$Cd   &0.6$\pm$0.15; 0.67$\pm$0.13 &$<$1.0  &0.65$\pm$0.13; 0.59$\pm$0.13&$<$1.7 \\ \hline
Cu           &$<$0.1                   &-          &$<$0.1       &$<$0.12     \\ \hline 
\hline
\multicolumn{5}{l}{* samples of $^{82}$Se(I) and $^{82}$Se(II) having a combined mass of 800~g} \\ 
\multicolumn{5}{l}{have been measured. The $^{82}$Se(II) was not measured separately.} \\
\end{tabular} 
\end{center}
}
\end{table}
Column A3 shows the
total $^{214}$Bi activity of the source foils calculated from the
values of columns A1 and A2. 
When only a limit is known for the mylar contamination by $^{214}$Bi
there is a systematic uncertainty on the total foil activity
reflected by the difference between two numbers in the column A3.
When a limit is obtained on the internal foil material activity A1
(the case of $^{130}$Te) the first number in the column A3 corresponds to 
A1=0, the second is for the limit value of A1.
There is a good agreement
between the results obtained from  NEMO~3 data and those from the HPGe detectors.

\subsection{$^{208}$Tl inside the source foils}
The presence of a small quantity 
of $^{208}$Tl from the $^{232}$Th decay chain inside a source foil
is the origin of the most troublesome
background  for the neutrinoless double beta decay search. 
Therefore the radiopurity goals of molybdenum and selenium foils
were very high~\cite{ARN05}. The measurements
using HPGe detectors could not reach the required sensitivity and in
most cases only  limits on $^{208}$Tl activity were set. 
Here the measurements of $^{208}$Tl foil contamination performed with NEMO~3 
data are presented.

Events of $e\gamma\gamma$ and $e\gamma\gamma\gamma$ topologies are used.
The event selection criteria 
are similar to those described
in section~\ref{sect:thoron_sel} but in this case the event vertex
has to be on the source foil. All likely backgrounds have been taken
into account. The most important one is due to the thoron and radon in the detector
volume.
The two event topologies give consistent results and the activities evaluated with Phase 1 and Phase 2
data are in a good agreement.
The $^{208}$Tl activity in the $\beta\beta$ source foils and in the copper foils
based on the total available statistic are shown in Table~\ref{tab:tl208_int}.

The results of HPGe detector measurements are also shown in Table~\ref{tab:tl208_int}
for comparison.
\begin{table}[tbh]
\caption{Number of observed events  (N), signal-to-background ratio
(S/B), signal efficiency ($\varepsilon$) and results of the 
  measurements of $^{208}$Tl activity of the source foils 
   compared to the  
HPGe measurements.}
\label{tab:tl208_int}

\vspace*{0.1cm}
{\scriptsize
\begin{center}
\begin{tabular}{|l||r|r|r||r|r|}
\hline \hline
$\beta\beta$ material& N&S/B&$\varepsilon$ (\%)&A (mBq/kg)&A$_{HPGe}$ (mBq/kg)\\ \hline \hline
$^{100}$Mo(m) &666  &2.4 &1.7 & 0.11$\pm$0.01 &$<$0.13;$<$0.1;$<$0.12* \\
$^{100}$Mo(c) &1628 &1.7 &1.8 & 0.12$\pm$0.01 &$<$0.17       \\
$^{82}$Se (I) &446  &2.0 &2.0 & 0.34$\pm$0.05 &$<$0.670      \\
$^{82}$Se (II)&507  &3.4 &1.9 & 0.44$\pm$0.04 &0.4$\pm$0.13** \\
$^{48}$Ca     &42   &4.1 &1.4 & 1.15$\pm$0.22 &$<$2.         \\
$^{96}$Zr     &158  &7.9 &1.8 & 2.77$\pm$0.25 &$<$10.;$<$5.*  \\
$^{150}$Nd    &1002 &39.4&1.8 & 9.32$\pm$0.32 &10.$\pm$1.7    \\
$^{130}$Te    &448  &1.1 &2.0 & 0.23$\pm$0.05 &$<$0.5        \\
$^{nat}$Te    &495  &1.9 &1.8 & 0.27$\pm$0.04 &$<$0.08        \\
$^{116}$Cd    &196  &0.7 &1.7 & 0.17$\pm$0.05 &$<$0.83;$<$0.5* \\
Cu            &66   &0.6 &1.5 & 0.03$\pm$0.01 &$<$0.033       \\ \hline \hline
\multicolumn{6}{l}{* different foil samples have been measured}\\
\multicolumn{6}{l}{** samples of $^{82}$Se(I) and $^{82}$Se(II) having a combined mass of 800~g}\\ 
\multicolumn{6}{l}{have been measured.The $^{82}$Se(II) was not measured separately.}\\
\end{tabular}
\end{center}
}
\end{table}
With the exception of natural
tellurium  there exists a good agreement between the two measurements.
Due to the large data acquisition time  NEMO~3  achieves better 
sensitivity.

\subsection{Measurement of the internal activity of $\beta$ emitters}  
The beta activity of a foil is measured by events with a single electron
track. 
The sources of radioactivity considered above are not sufficient to explain
the single electron data. An 
additional source of electrons in the foils is required.
The results of the foil radiopurity measurements
performed with HPGe detectors before the foil installation in NEMO~3 are used 
to decide upon the list of contaminants in a given foil.
This list typically includes $^{40}$K and isotopes from 
$^{238}$U and $^{232}$Th decay chains.
The internal $^{226}$Ra and $^{228}$Ra activities are fixed 
from the results obtained for $^{214}$Bi and $^{208}$Tl  described above.
Their daughters are assumed to be in equilibrium.  
A possible foil surface pollution by $^{210}$Pb is also considered.

HPGe measurements made for the copper foils have revealed that the copper is
quite clean.
Upper limits of 5~mBq/kg and 8~mBq/kg on the  contamination 
by $^{234m}$Pa and $^{40}$K  respectively were attained.
Here the internal activities of these beta
emitters are determined by a fit to the single electron energy spectra
obtained from Phase~2.

\begin{figure}[htb]
\begin{center}
\includegraphics[width=0.33\textwidth]{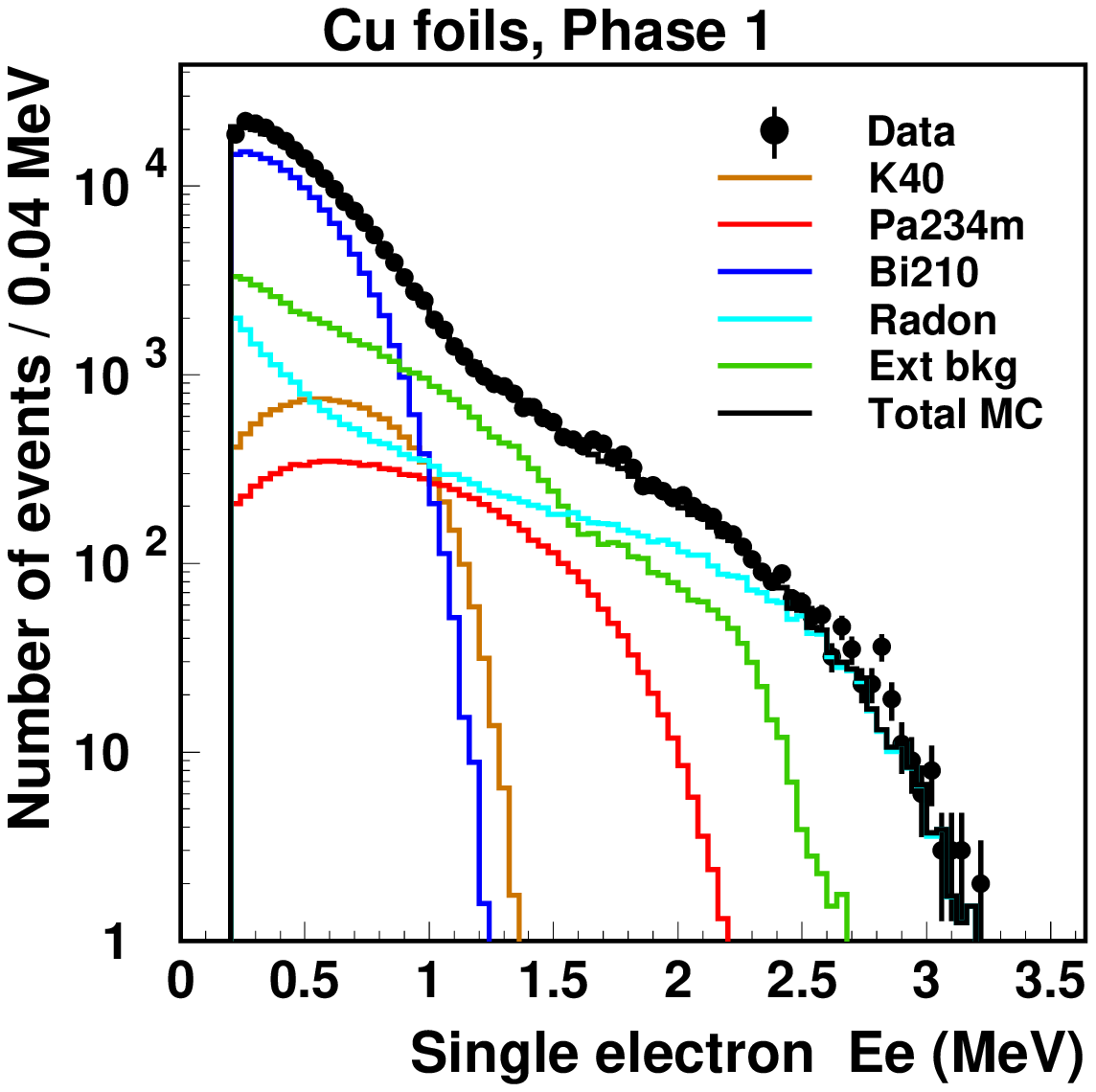}
\includegraphics[width=0.33\textwidth]{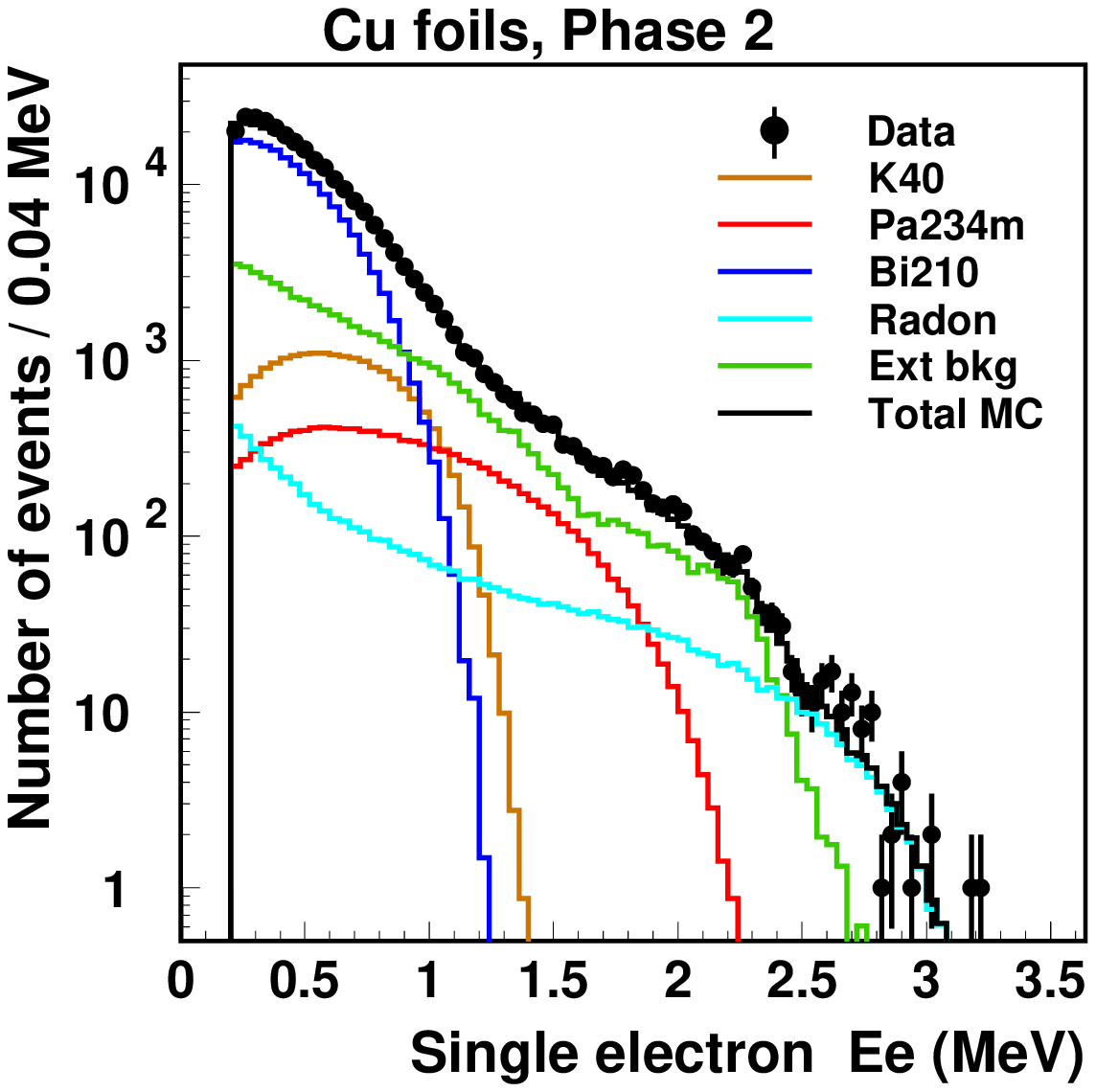}
\end{center}
\caption[1e]{Single electron energy spectra for copper,
Phase~1 and Phase~2 data.}
\label{fig:fig_1e}
\end{figure}
The results of the fit are shown in
Fig.~\ref{fig:fig_1e} demonstrating a good agreement with the data. 
An activity of $1.5 \pm 0.1$ mBq/kg is found for $^{234m}$Pa 
and  $3.7 \pm 0.1$ mBq/kg for $^{40}$K. 
No pollution on the copper foil surface by $^{210}$Pb 
was evident. In Fig.~\ref{fig:fig_1e} the $^{210}$Bi contribution corresponds to
the wire surface contamination.

The same method provides the determination of the contaminant 
activities in the $\beta\beta$ source foils. 
The "single electron" channel is the most appropriate for pure beta emitters.
For isotopes decaying  with copious $\gamma$-ray emission such as
$^{60}$Co and $^{207}$Bi, the $e\gamma$-internal channel is used.
\section{Test of the background model}
\label{sect:testmodel}
The highly radiopure copper foils (621 g) occupy one of the 20 sectors in  
NEMO~3. They are used to measure the external background. 
The internal $e\gamma$ and 2e$^-$ channels are used to compare the data 
with the background model described above.
\subsection{Internal $e\gamma$ events from the copper foils}
\begin{figure}[htb]
\begin{center}
\includegraphics[width=0.32\textwidth]{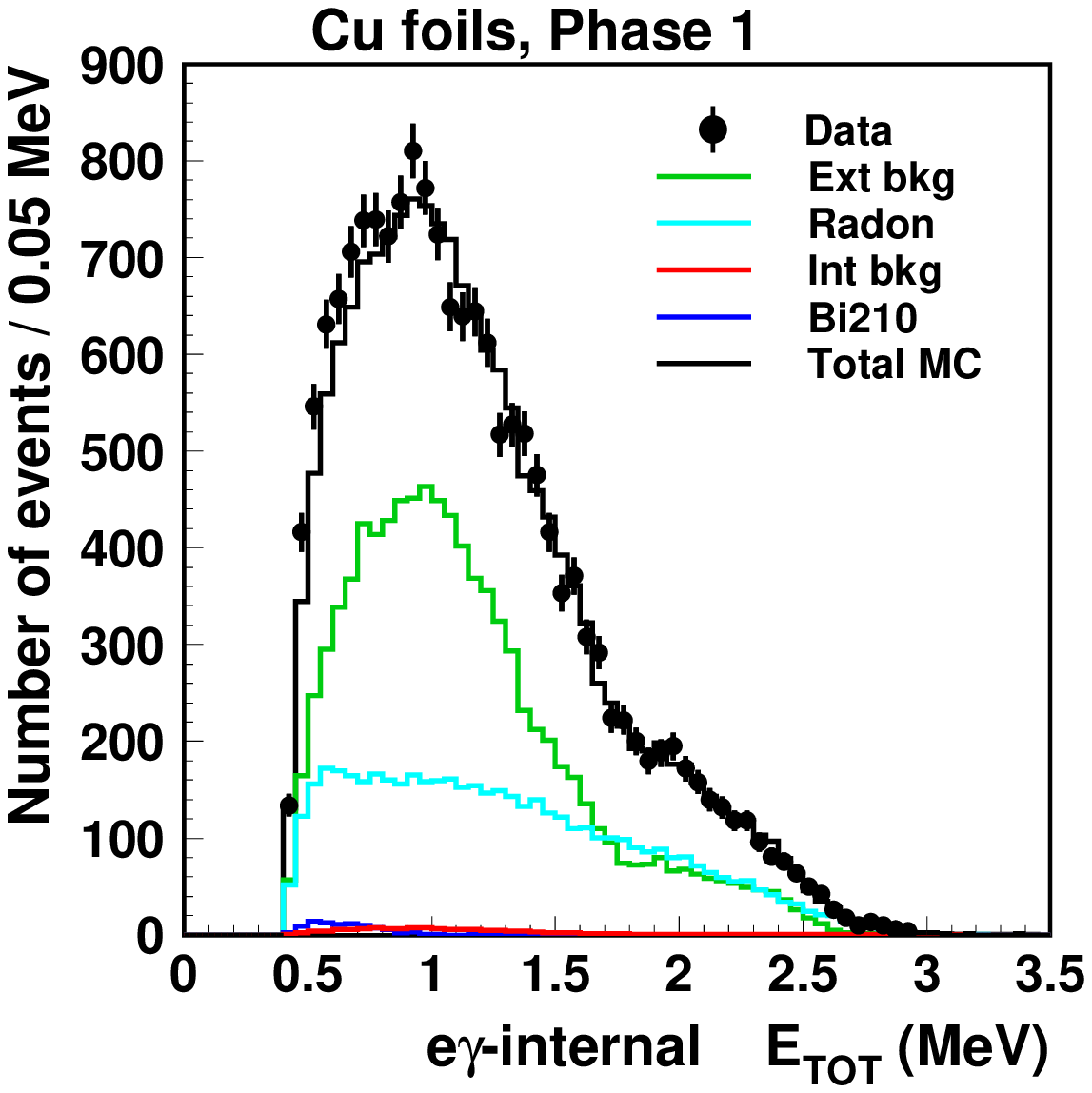}
\includegraphics[width=0.32\textwidth]{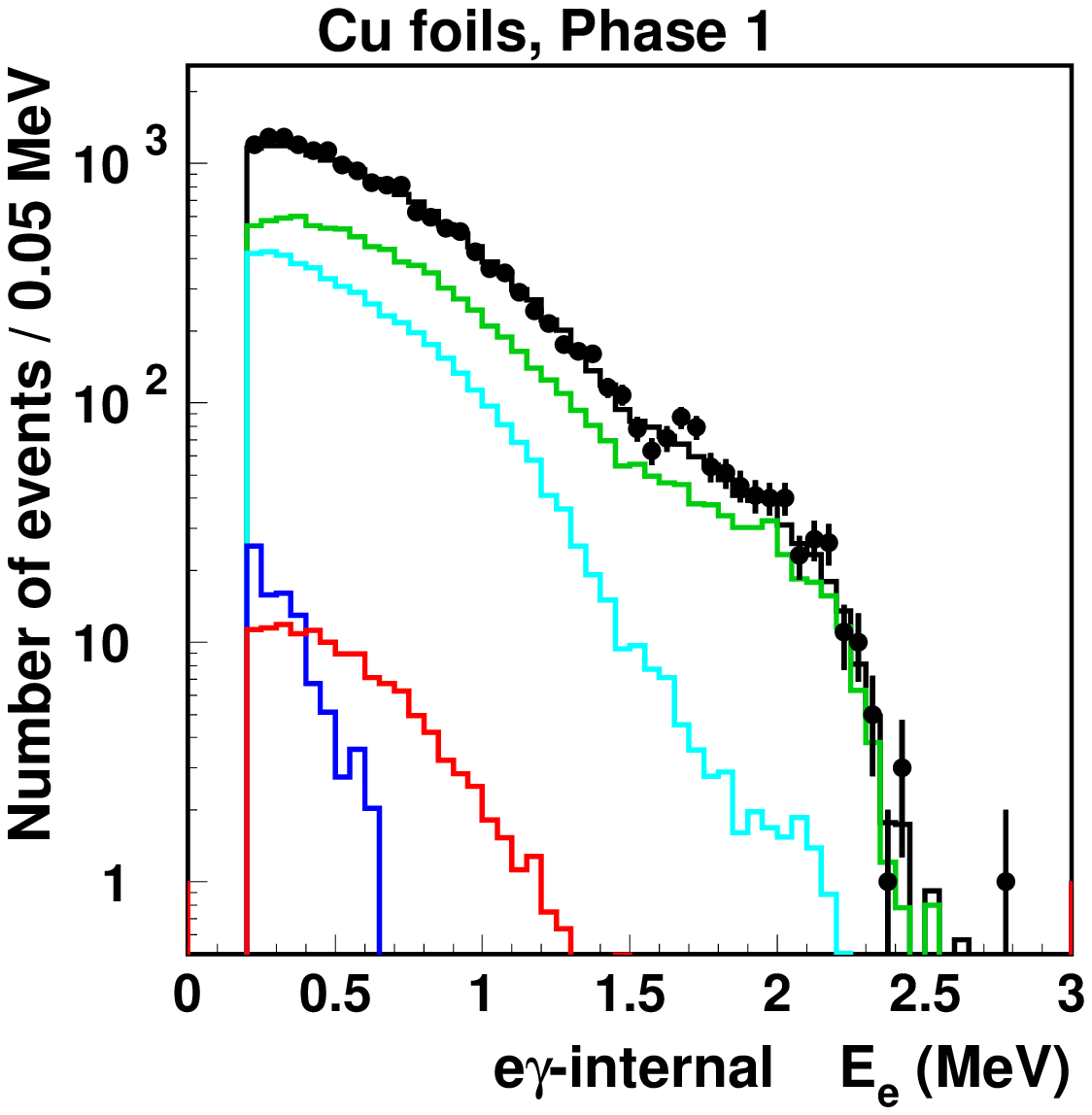}
\includegraphics[width=0.32\textwidth]{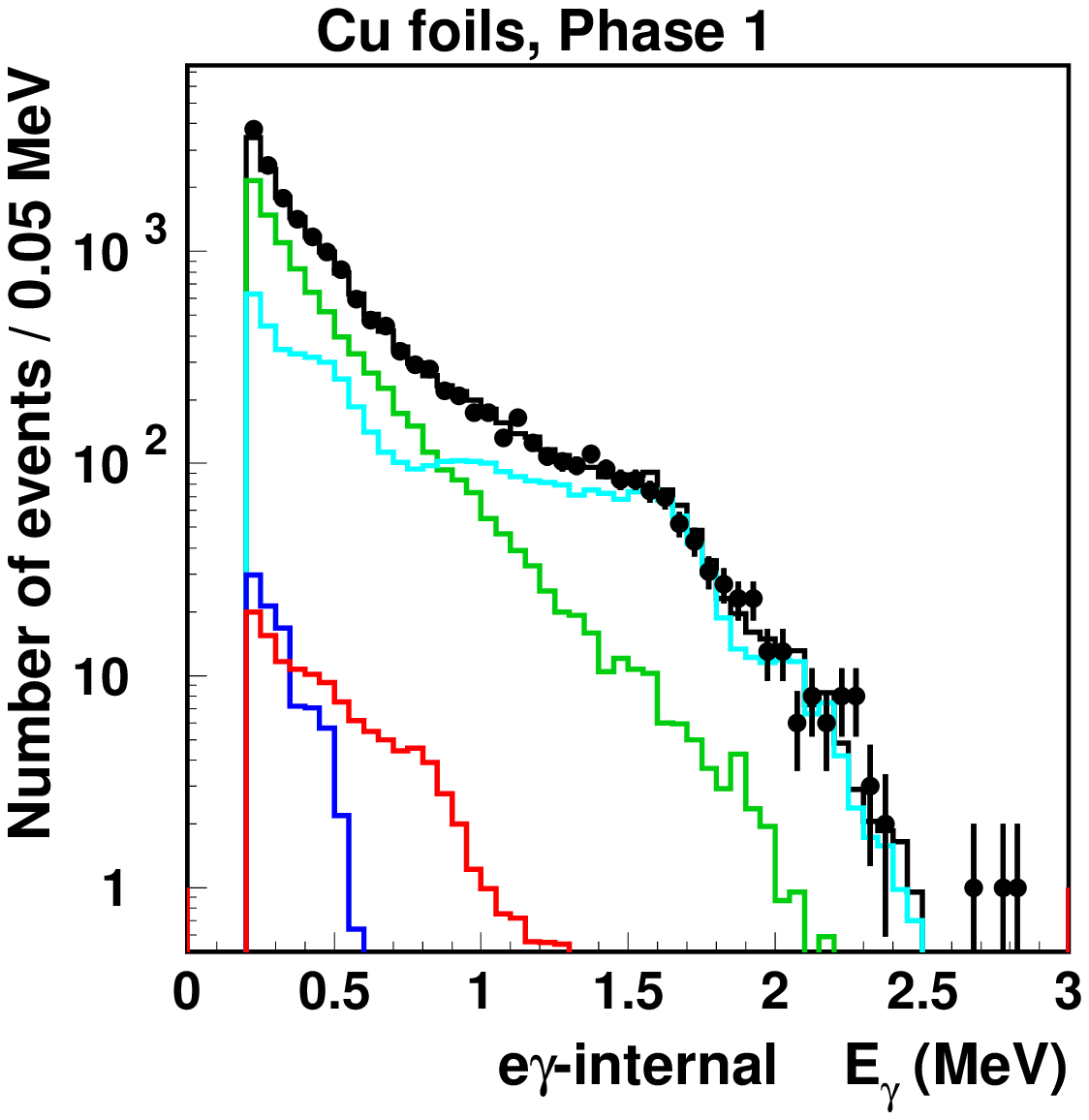}\\
\includegraphics[width=0.32\textwidth]{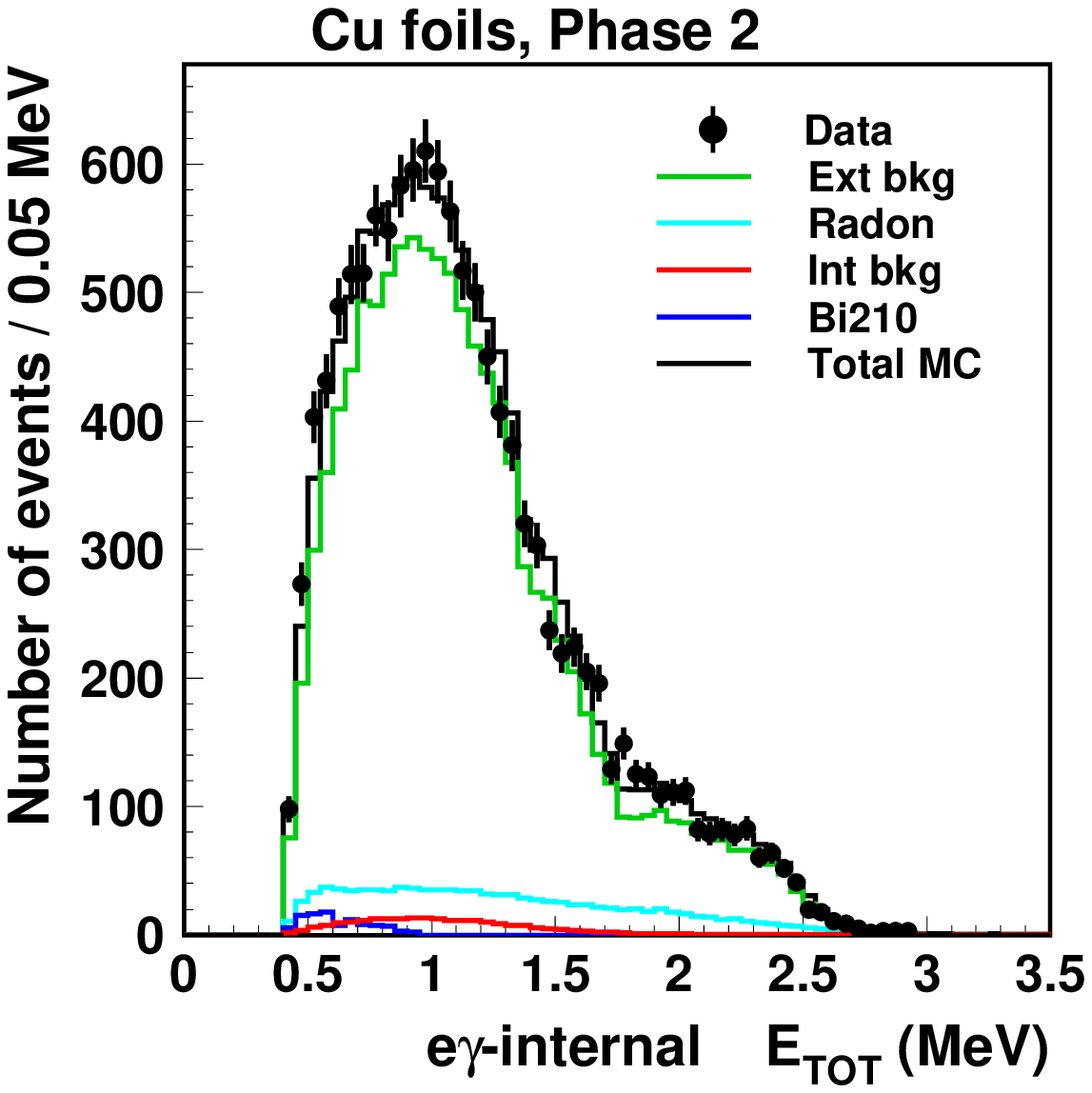}
\includegraphics[width=0.32\textwidth]{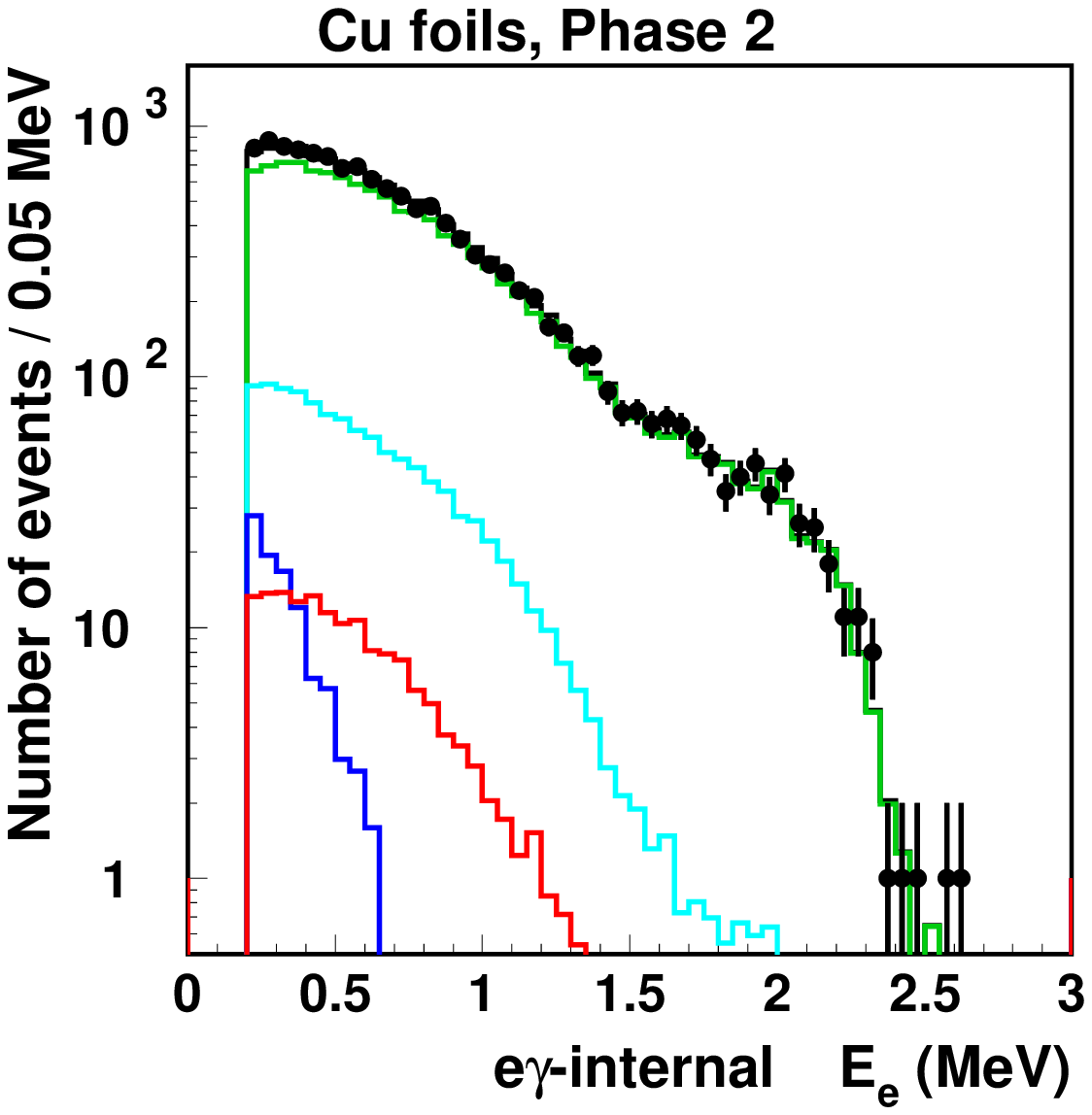}
\includegraphics[width=0.32\textwidth]{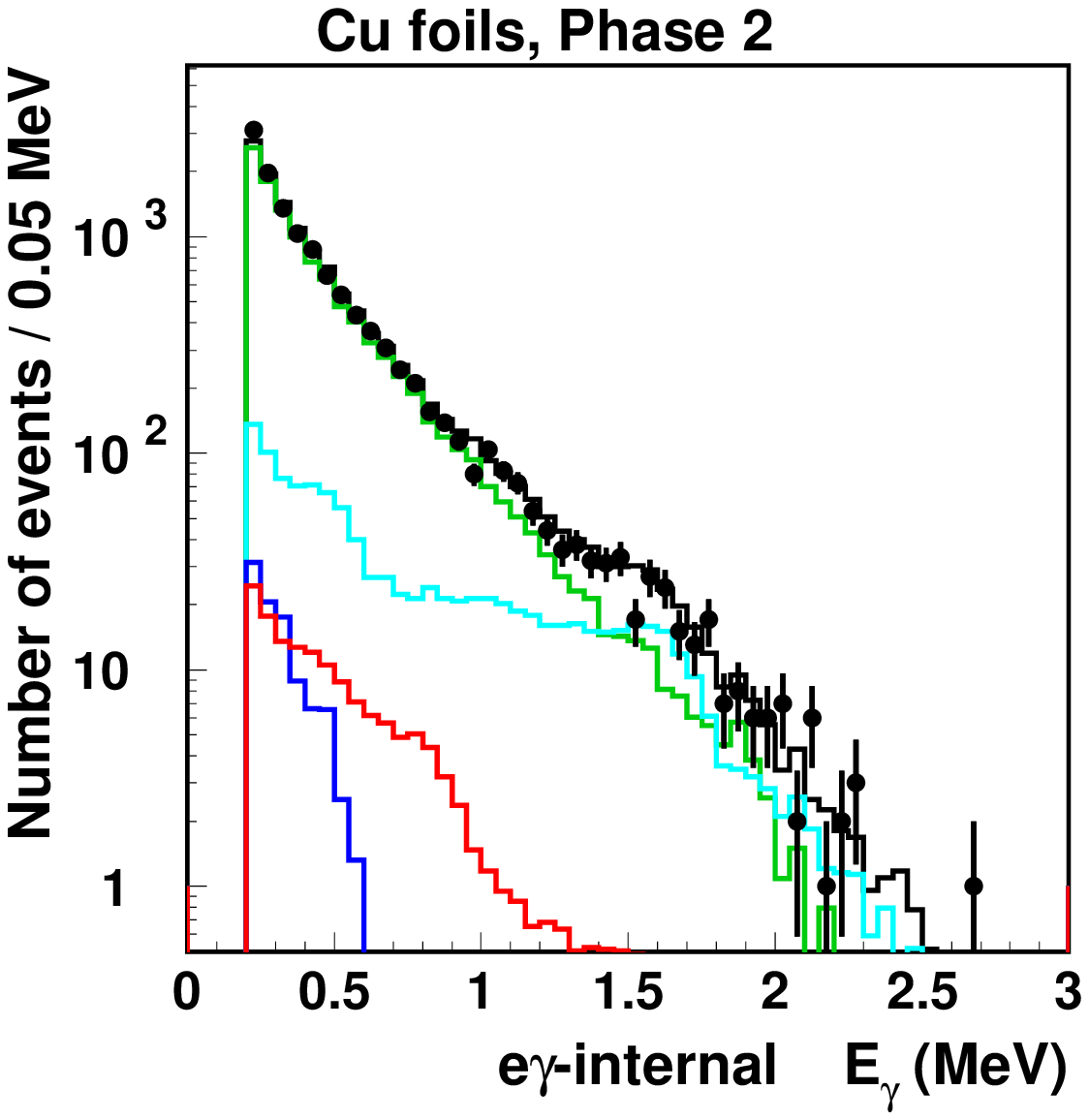}\\
\hfill
\end{center}
\caption[eg-int]{The distribution of the energy sum of the e$^-$ and $\gamma$-ray, the 
single e$^-$ energy and the single $\gamma$-ray energy of 
internal $e^-\gamma$ events from copper foils for Phase~1 and Phase~2 data.}
\label{fig:egint}
\end{figure}

In Fig.~\ref{fig:egint} some distributions of internal $e\gamma$ events 
coming from the copper 
foils are compared with the MC simulation based on the  background model. 
One can see that the  total number of events and the energy spectra 
are closely reproduced.

The external background is dominant and represents almost 90\% of all 
internal $e\gamma$ events in Phase~2.
In Phase~1 the radon produces a noticeable contribution, and in the energy sum 
region below 1 MeV
there are slightly more events than expected. The difference  is less than 2\%
of the total number of  events. 
In Phase 2 the radon contribution is 
significantly lower and the model describes the data very well even at low energies.
A similar problem at low energies for the Phase~1 data is also observed in the 2e$^-$ channel
and may have the same origin.

The internal background contribution in the $e\gamma$ channel is negligible so
the data  confirms that the copper foils are clean.
\subsection{Two electron events from the copper foils}
Copper is not a source of double beta decay. That
allows one to check the validity of the background model in the 2e$^-$ channel
where double beta decay is searched for. 
To study this channel the events are selected by requiring two reconstructed
electron tracks emitted from the foil with the correct curvature and a common vertex in 
the foil.
The energy of each electron measured in the calorimeter is required to be 
greater than 200 keV.
Each track must hit an isolated scintillator block 
and no additional PMT signals are allowed to avoid events with $\gamma$-rays.
The event is also recognized as internal by the time-of-flight
difference of the two electrons.
 
\begin{figure}[htb]
\begin{center}
\includegraphics[width=0.32\textwidth]{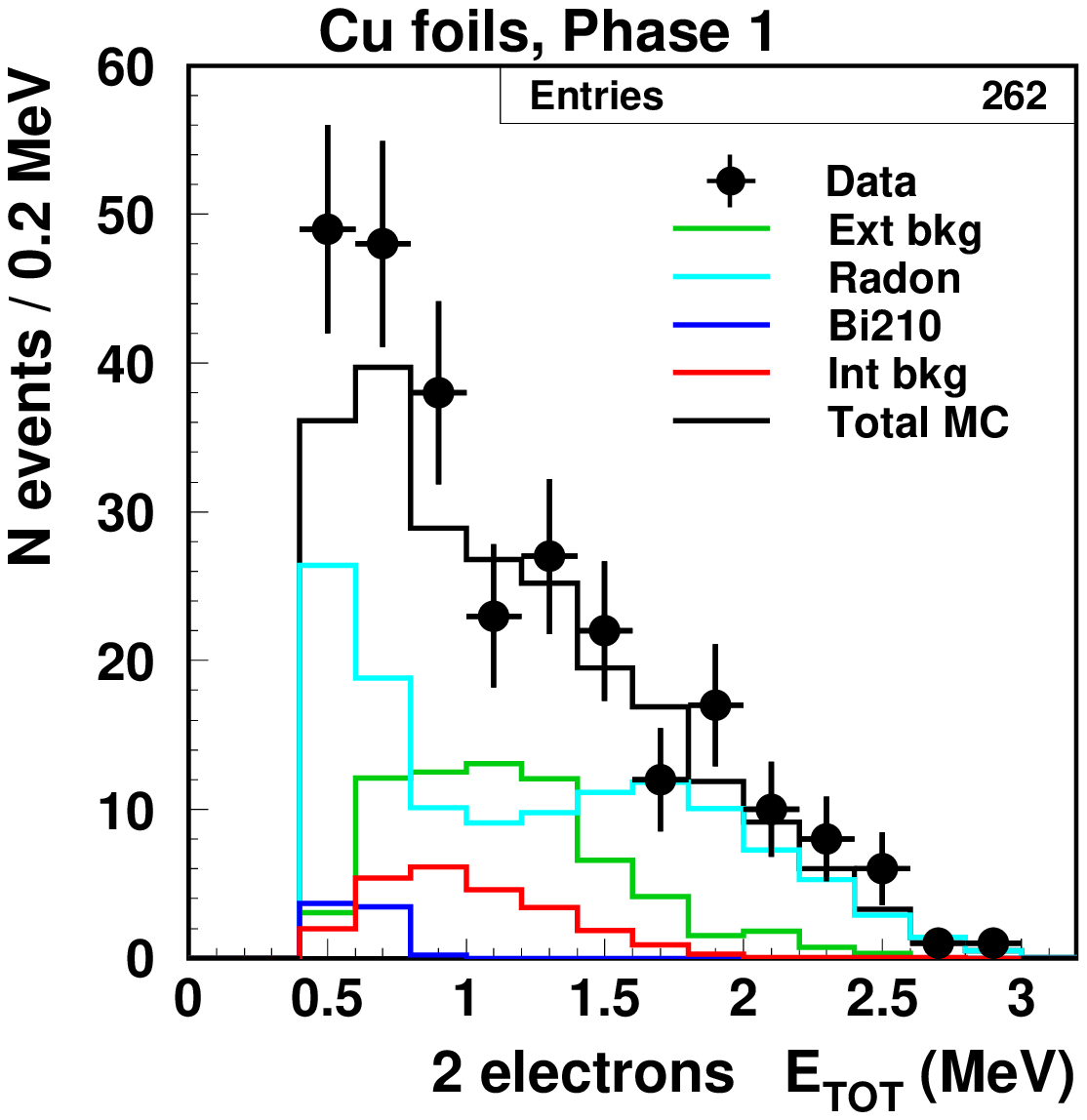}
\includegraphics[width=0.32\textwidth]{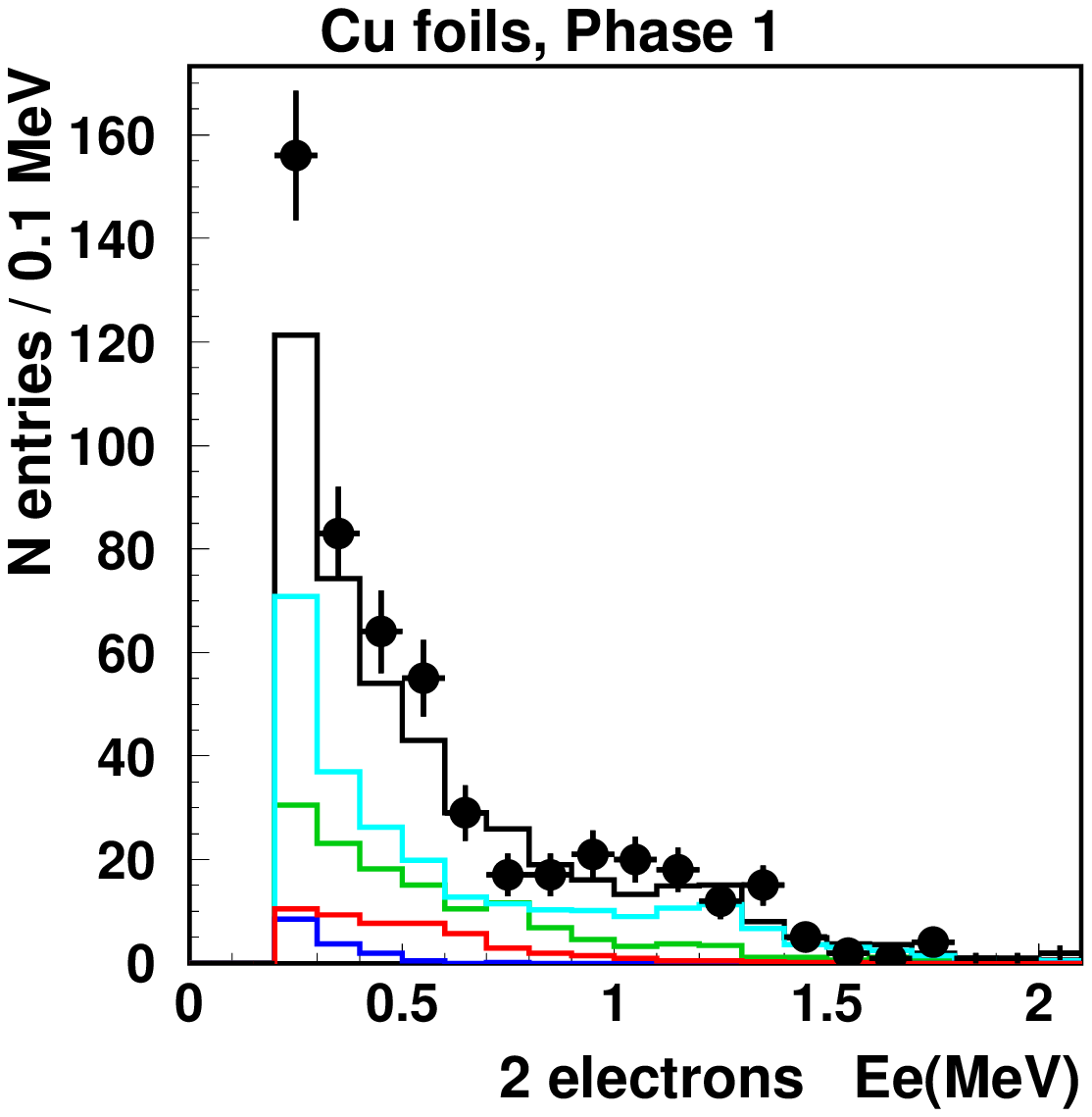}
\includegraphics[width=0.32\textwidth]{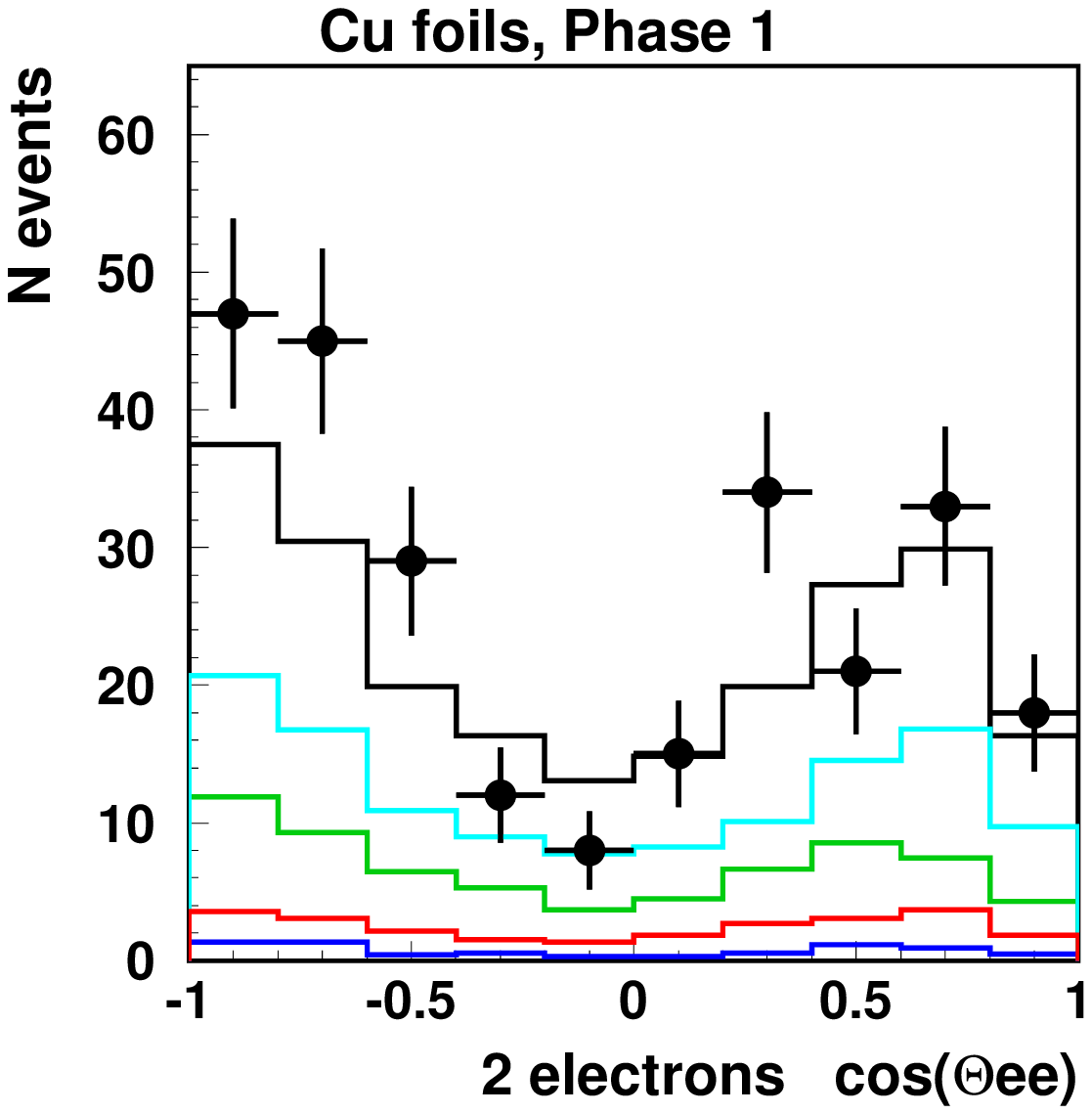}\\
\hfill
\includegraphics[width=0.32\textwidth]{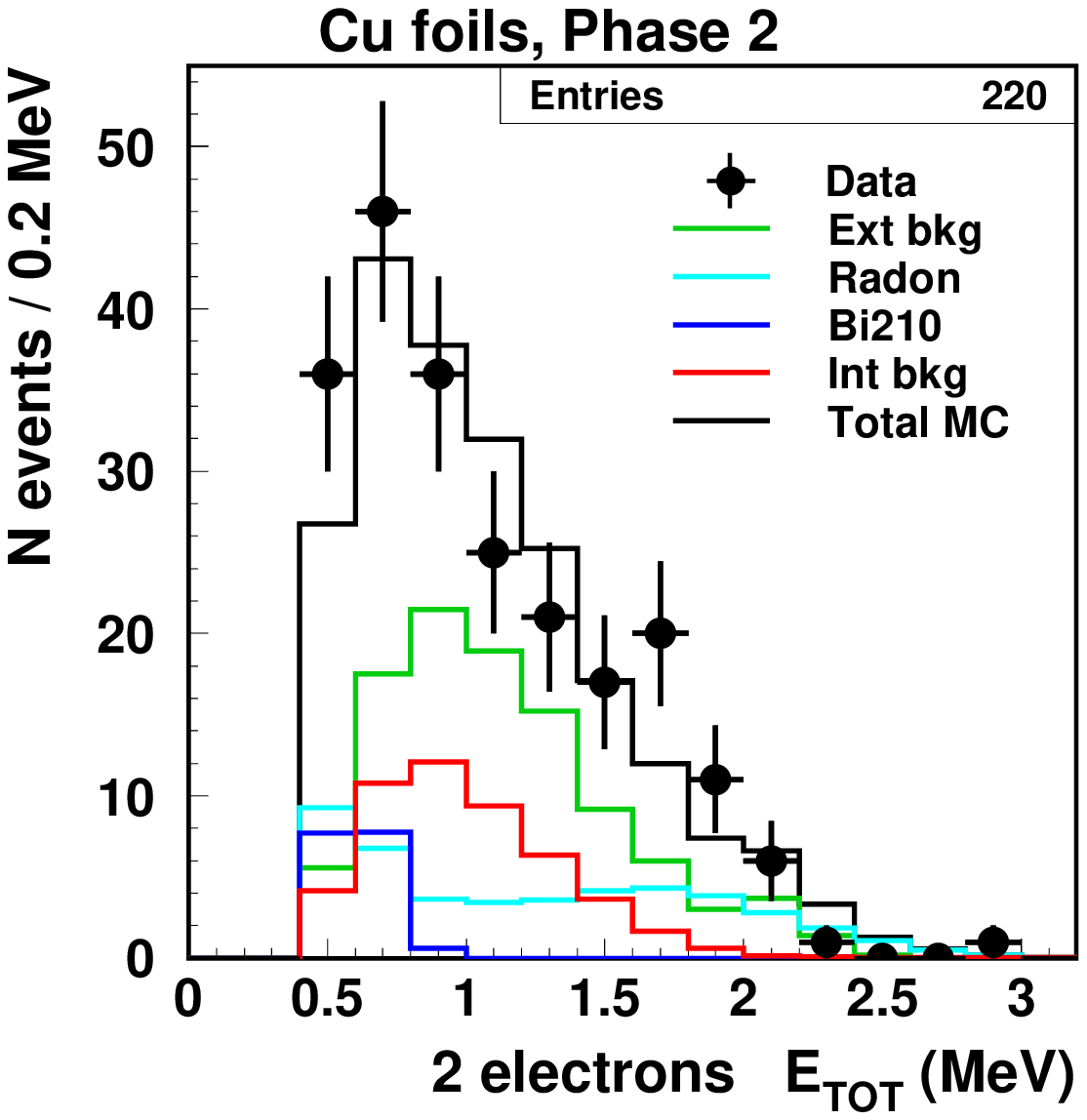}
\includegraphics[width=0.32\textwidth]{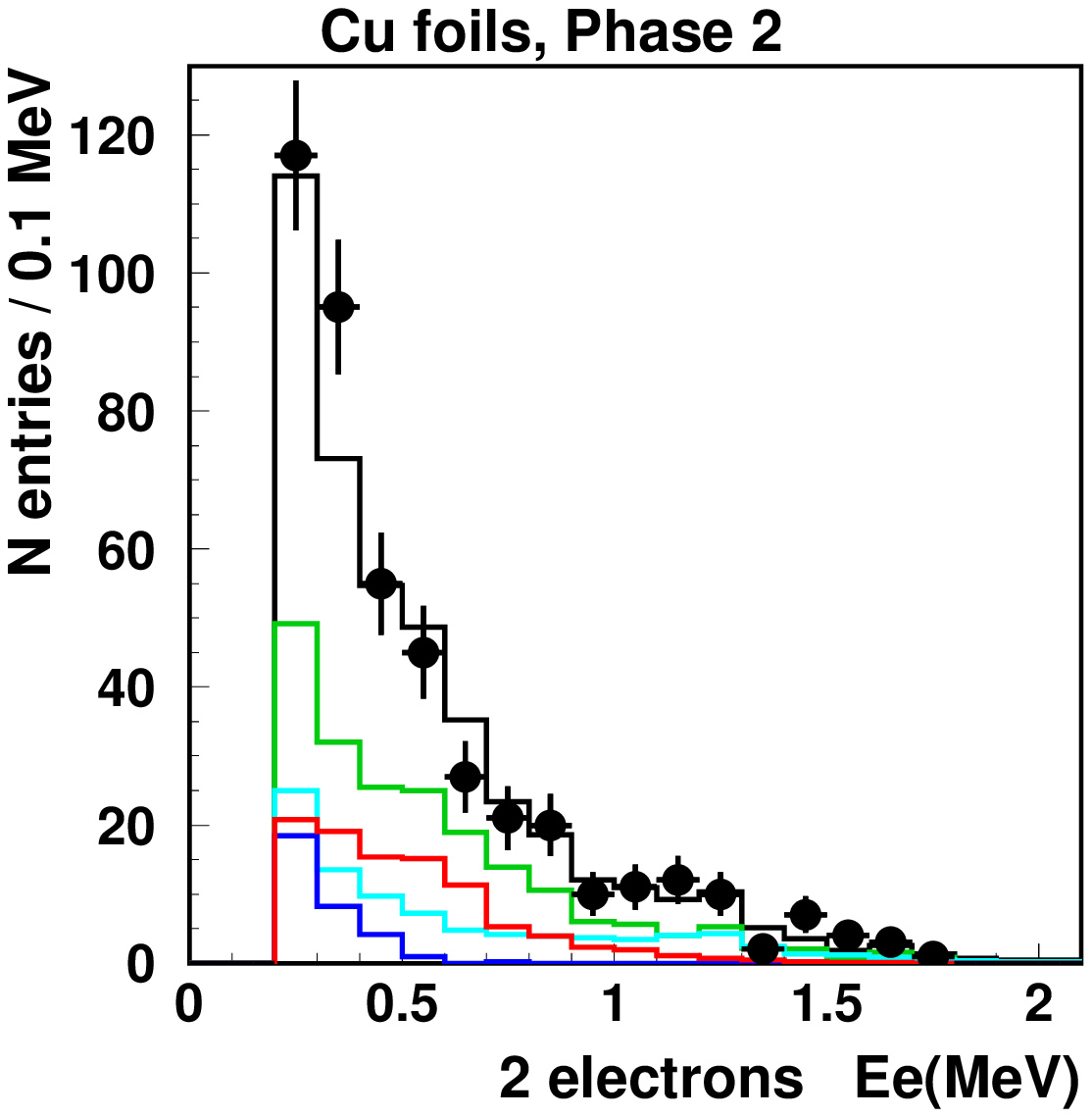}
\includegraphics[width=0.32\textwidth]{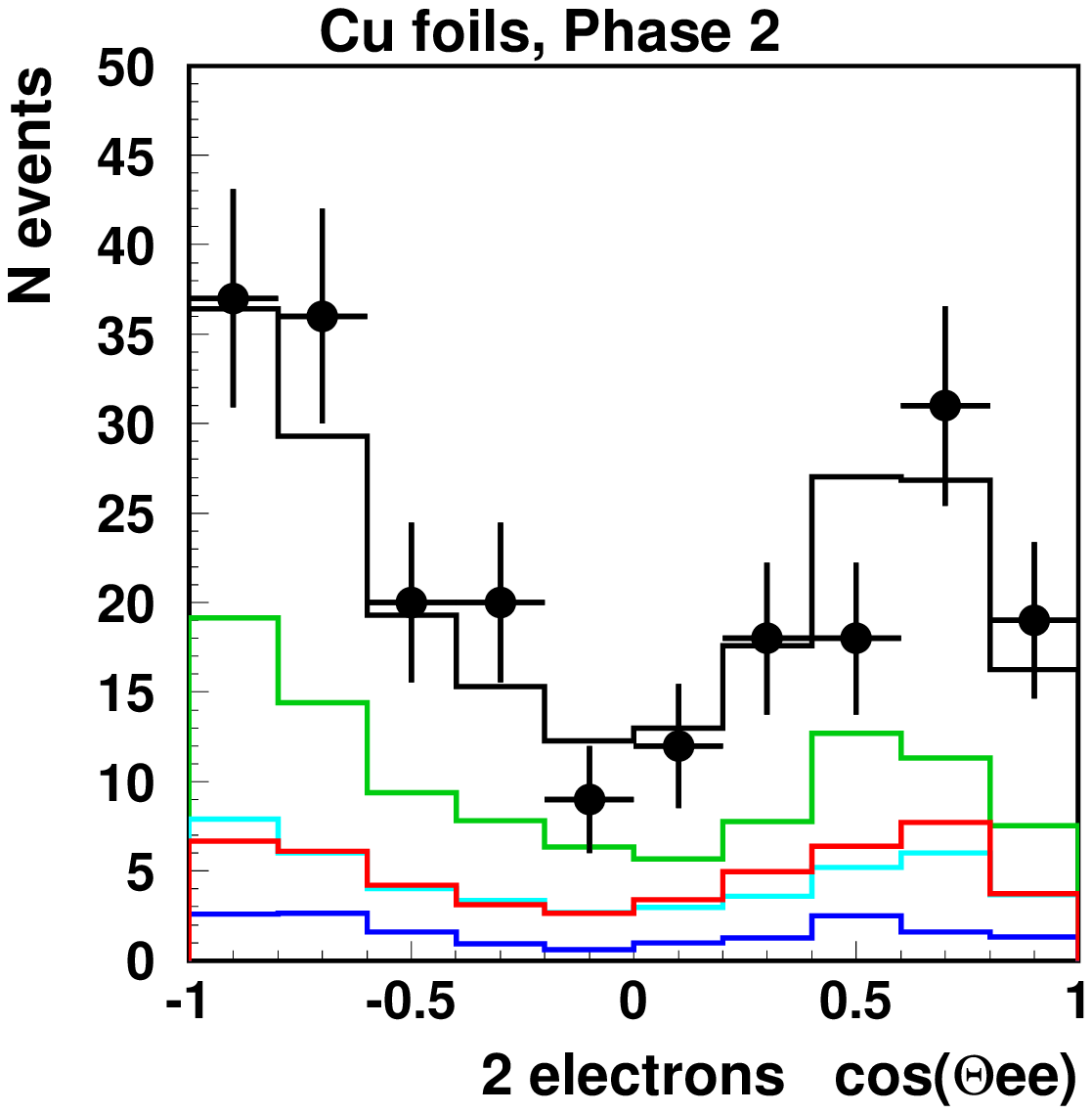}\\
\end{center}
\caption[bb-test]{Background model prediction compared to the data for 2e$^-$ 
events from the copper foils.}
\label{fig:bb-test}
\end{figure}

In Fig.~\ref{fig:bb-test} the  distributions of the energy sum of the two electrons, 
the single electron energies and angular correlation
of two-electron events coming from the copper foils are compared to the 
prediction of the background model.
In Phase 2 the total number of events (220 observed with 213 expected for 788 days 
of data acquisition) have energy and angular distributions in  good agreement 
with the MC predictions.
In Phase 1, where the radon activity level is six times higher, there are 262 events observed 
for the 374 days of acquisition, which is 15\% higher than the MC prediction. 
The excess of events is observed in the energy sum distribution below 1 MeV.
The difference is relatively small and is apparently due to radon.

The radon contribution is provided by $^{214}$Pb ($Q_{\beta}$=1.023~MeV) and 
$^{214}$Bi ($Q_{\beta}$=3.272~MeV) decays simulated  
according to the model based on ~\cite{Akovali1995}. 
The observed discrepancy at low energies seems to indicate some imperfection in the  
model for $^{214}$Pb and can be due to the shape of the beta spectra.
Shapes of beta spectra corresponding to allowed decays
were generated, while the beta transitions are mainly of the 
first non unique forbidden type. A deviation from the allowed shape
for this type of transition may be large ~\cite{Gor},\cite{Behr}, however it is not 
known for $^{214}$Pb and $^{214}$Bi.

Apart from the problem at low energies in  Phase~1 the background model 
successfully reproduces the 2e$^-$ events produced in the copper foils due to the 
external and internal radioactive sources.
\section{Summary}

The methods of background measurements in the double beta decay experiment NEMO~3
are presented. 
The background is classified as internal and external according to its origin.
Internal backgrounds come from the source foils.
The external background is subdivided into two groups. 
The first is due to radioactive sources inside the tracking volume, and the second 
results from the radioactivity outside of the tracking volume.
All these backgrounds were estimated from the data with events of 
various topologies and summarized below.
\begin{description}
\item{-}
The external background component due to the presence of radon in the 
tracking chamber is measured using
events with a detected electron accompanied by a delayed $\alpha$-particle track. 
This topology results from the beta decay of $^{214}$Bi followed by the alpha 
decay of $^{214}$Po.
\item{-}
The thoron is measurable with the events of $e\gamma\gamma$ and $e\gamma\gamma\gamma$ 
topologies. The method is based on the detection of events with the signal of a 
2.615~MeV $\gamma$-ray, typical of the $^{208}$Tl beta decay.
\item{-}
The presence of $^{210}$Pb on the wires is measured with a single electron
starting from a wire. 
\item{-}
The external $\gamma$-ray flux coming from outside of the tracking detector volume 
is measured with a crossing electron  and $e\gamma$-external events.
\item{-}
The internal background due to $^{214}$Bi and $^{208}$Tl decays inside the foil is 
measured in a manner similar to that used for radon and thoron measurements 
but requires the event vertex to be located on the foil.
\item{-}
The use of single electron events without detected $\gamma$-rays and coming from 
a source foil allows one to measure the activity of beta emitters inside the foils.
\end{description}

Very pure copper foils are used for the study of the 
external background with internal  $e\gamma$ events.
The validity of the background model is successfully verified with 2e$^-$ events
coming from the copper foils.
 
It has been demonstrated  that with the NEMO~3 detector  the
backgrounds are measurable from the experimental data. 
In particular the activities of the two most troublesome background sources 
for $\beta\beta 0\nu$
decay,  $^{214}$Bi and $^{208}$Tl, are measured with the 
adequate precision.

\section*{Acknowledgements}
The authors would like to thank the Modane Underground Laboratory staff 
for their technical assistance in running the experiment. Portions of this 
work were supported by
grants from RFBR (no 06-02-16672, 06-02-72553) and by the Russian 
Federal Agency for Atomic Energy.
We acknowledge support by the Grant Agencies of the Czech Republic (MSM
6840770029, LA305, LC07050).

\end{document}